\documentclass[journal]{IEEEtran} \newcommand{\figsize}{3.5}
\usepackage{graphicx,psfrag}
\usepackage{url}
\usepackage[usenames]{color}
\usepackage[cmex10]{amsmath}
\usepackage{amsfonts,amssymb,latexsym,cite,color,multirow,rotating}
\usepackage{algorithm,algorithmic}
\usepackage{bbm}

 \newcommand{\putTable}[3]{\begin{table}[t]
  			    \centering
		            #3
     			    \caption{#2}
     			    \label{tab:#1}
			    \vspace{-6mm}
			  \end{table} }
 \newcommand{\putFrag}[4]{\begin{figure}[t]
                            \centering
                            #4
			    \includegraphics[width=#3in]{figures/#1.eps}
            		    \caption{#2}
           		    \label{fig:#1}
                          \end{figure} }
 \newcommand{\putFragSpace}[4]{\begin{figure}[t]
                            \centering
                            #4
			    \includegraphics[width=#3in]{figures/#1.eps}
			    \vspace{-5mm}
            		    \caption{#2}
           		    \label{fig:#1}
                          \end{figure} }

 \renewcommand{\hat}{\widehat}
 
 \newcommand{\defn}{\triangleq}
 
 \newcommand{\uvec}[1]{\ensuremath{\underline{\boldsymbol{#1}}}}
 
 \newcommand{\ovec}[1]{\ensuremath{\overline{\boldsymbol{#1}}}}
 \newcommand{\hvec}[1]{\ensuremath{\boldsymbol{\Hat{#1}}}}
    
 \renewcommand{\vec}[1]{\ensuremath{\boldsymbol{#1}}}
 \newcommand{\mat}[1]{\ensuremath{\begin{bmatrix}#1\end{bmatrix}}}

 \newcommand{\norm}[1]{\ensuremath{\| #1 \|}}
 \newcommand{\mc}[1]{\ensuremath{\mathcal{#1}}}
 \newcommand{\st}{{~\text{s.t.}~}}

 \newcommand{\Real}{{\mathbb{R}}}
 
 \newcommand{\Int}{{\mathbb{Z}}}

 \newcommand{\tran}{^\textsf{T}}

 \newcommand{\giv}{\,|\,}
 \newcommand{\biggiv}{\,\big|\,}

 \DeclareMathOperator{\E}{E}
 \DeclareMathOperator{\var}{var}

 \DeclareMathOperator{\tr}{tr}

 \DeclareMathOperator*{\argmax}{arg\, max}
 \DeclareMathOperator*{\argmin}{arg\, min}

 \renewcommand{\eqref}[1]{(\ref{eq:#1})}
 
 \newcommand{\Figref}[1]{Figure~\ref{fig:#1}}
 
 \newcommand{\figref}[1]{Fig.~\ref{fig:#1}}
 
 \newcommand{\tabref}[1]{Table~\ref{tab:#1}}
 \newcommand{\secref}[1]{Sec.~\ref{sec:#1}}

 \newcommand{\appref}[1]{Appendix~\ref{app:#1}}

 \newcommand{\SNR}{\textsf{SNR}}
 \newcommand{\NMSE}{\textsf{NMSE}}

 \newcommand{\SAD}{{\textsf{SAD}}}

 \newcommand{\X}{\textsf{x}}
 \newcommand{\vX}{\textsf{\textbf{\textit{x}}}}
 \newcommand{\mX}{\textsf{\textbf{\textit{X}}}}
 \newcommand{\A}{\textsf{a}}
 \newcommand{\Ao}{\overline{\textsf{a}}}
 \newcommand{\mA}{\textsf{\textbf{\textit{A}}}}
 
 \newcommand{\D}{\textsf{d}}
 \newcommand{\vD}{\textsf{\textbf{\textit{d}}}}
 \newcommand{\mD}{\textsf{\textbf{\textit{D}}}}
 \renewcommand{\S}{\textsf{s}}
 \newcommand{\So}{\overline{\textsf{s}}}
 \newcommand{\Su}{\underline{\textsf{s}}}
 \newcommand{\mS}{\textsf{\textbf{\textit{S}}}}
 \newcommand{\mSo}{\overline{\textsf{\textbf{\textit{S}}}}}
 \newcommand{\mSu}{\underline{\textsf{\textbf{\textit{S}}}}}
 \newcommand{\Y}{\textsf{\textit{y}}}
 \newcommand{\Yo}{\overline{\textsf{\textit{y}}}}
 
 \newcommand{\mY}{\textsf{\textbf{\textit{Y}}}}
 \newcommand{\mYo}{\overline{\textsf{\textbf{\textit{Y}}}}}
 \newcommand{\Z}{\textsf{\textit{z}}}
 \newcommand{\Zo}{\overline{\textsf{\textit{z}}}}
 
 \newcommand{\mZ}{\textsf{\textbf{\textit{Z}}}}
 \newcommand{\mZo}{\overline{\textsf{\textbf{\textit{Z}}}}}

 \newcommand{\EE}{\textsf{e}}
 \newcommand{\mE}{\textsf{\textbf{\textit{E}}}}
 \newcommand{\vE}{\textsf{\textbf{\textit{e}}}}

 \newcommand{\R}{\textsf{r}}
 \newcommand{\Q}{\textsf{q}}
 
 \newcommand{\msg}[2]{\Delta^{#1}_{#2}}

 \newcommand{\BIGAMP}{\textbf{\textsf{BiGAMP}}}
 \newcommand{\MRF}{\textbf{\textsf{MRF}}}
 \newcommand{\GausMark}{\textbf{\textsf{GaussMarkov}}}
 \newcommand{\EM}{\textbf{\textsf{EM}}}

\begin{document}
\setlength{\arraycolsep}{0.4mm}
 \title{Hyperspectral Unmixing via Turbo Bilinear Approximate Message Passing}
	 \author{Jeremy Vila, \emph{Student Member, IEEE}, Philip Schniter, \emph{Fellow, IEEE} and Joseph Meola, \emph{Member, IEEE}%
	 \thanks{This work has been supported in part by NSF grants IIP-0968910, CCF-1018368, and CCF-1218754, and by an allocation of computing time from the Ohio Supercomputer Center. Portions of this work were presented at the 2013 SPIE Defense, Security, and Sensing symposium \cite{Vila:SPIE:13}.}
	 \thanks{J. Vila and P. Schniter are with the Department of Electrical and Computer Engineering, The Ohio State University, Columbus, OH 43210 USA (vila.2@osu.edu; schniter@ece.osu.edu; phone 614.247.6488; fax 614.292.7596).} 
         \thanks{J.\ Meola is with the Multispectral Sensing Division, Air Force Research Laboratory, Wright-Patterson AFB, OH 45433 USA (joseph.meola@wpafb.af.mil).}%
		}
 \date{\today}
 \maketitle

\begin{abstract}
The goal of hyperspectral unmixing is to decompose an electromagnetic spectral dataset measured over $M$ spectral bands and $T$ pixels into $N$ constituent material spectra (or ``endmembers'')  with corresponding spatial abundances.  
In this paper, we propose a novel approach to hyperspectral unmixing based on loopy belief propagation (BP) that enables the exploitation of spectral coherence in the endmembers and spatial coherence in the abundances.
In particular, we partition the factor graph into spectral coherence, spatial coherence, and bilinear subgraphs, and pass messages between them using a ``turbo'' approach.
To perform message passing within the bilinear subgraph, we employ the bilinear generalized approximate message passing algorithm (BiG-AMP), a recently proposed belief-propagation-based approach to matrix factorization.
Furthermore, we propose an expectation-maximization (EM) strategy to tune the prior parameters and a model-order selection strategy to select the number of materials $N$.
Numerical experiments conducted with both synthetic and real-world data show favorable unmixing performance relative to existing methods.  
\end{abstract}


\section{Introduction} \label{sec:intro}

In \emph{hyperspectral unmixing} (HU), the objective is to jointly estimate the spectral signatures and per-pixel abundances of the $N$ materials present in a scene, given measurements across $M$ spectral bands at each of $T = T_1 \times T_2$ pixels.
Often, a \emph{linear mixing model}\cite{Johnson:Geo:83,Bioucas:JSTAEO:12} is assumed, in which case the measurements $\vec{Y}\in\Real^{M\times T}$ are modeled as 
\begin{equation}
\label{eq:LMM}
\vec{Y} = \vec{SA} + \vec{W},
\end{equation}
where the $n$th column of $\vec{S} \in \Real_+^{M \times N}$ represents the spectrum (or ``endmember'') of the $n$th material,
the $n$th row of $\vec{A} \in \Real_+^{N \times T}$ represents the spatial abundance of the $n$th material, and $\vec{W}$ represents noise.  
Both $\vec{S}$ and $\vec{A}$ must contain only non-negative (NN) elements, and
each column of $\vec{A}$ must obey the simplex constraint (i.e., NN and sum-to-one).
Recently, nonlinear mixing models have also been considered (e.g., \cite{Heylen:JSTAEO:14,Dobigeon:SPM:14}), although such models lie outside of the scope of this paper.

Traditionally, hyperspectral unmixing is a two-step procedure, consisting of \emph{endmember extraction} (EE) to recover the endmembers followed by \emph{inversion} to recover the abundances.
Many EE algorithms leverage the ``\emph{pure pixel}'' assumption: for each material, there exists at least one observed pixel containing only that material (i.e., all columns of the $N\times N$ identity matrix can be found among the columns of $\vec{A}$).
Well-known examples of pure-pixel-based EE algorithms include N-FINDR\cite{Winter:SPIE:99} and
VCA\cite{Nasc:TGRS:05}.
The existence of pure pixels in HU is equivalent to ``\emph{separability}'' in the problem of non-negative matrix factorization (NMF), where the goal is to find $\vec{S} \in \Real_+^{M\times N}$ and $\vec{A} \in \Real_+^{N\times T}$ matching a given $\vec{Z}=\vec{S}\vec{A}$. 
There, separability has been shown to be sufficient for the existence of unique factorizations\cite{Donoho:NIPS:03} and polynomial-time solvers\cite{Arora:STOC:12}, with a recent example being the FSNMF algorithm from\cite{Gillis:TPAMI:14}.
In HU, however, the limited spatial-resolution of hyperspectral cameras implies that the pure-pixel assumption does not always hold in practice.
With ``mixed pixel'' scenarios in mind, algorithms such as Minimum Volume Simplex Analysis (MVSA)\cite{Li:IGARSS:08} and Minimum Volume Enclosing Simplex (MVES)\cite{Chan:TSP:09} attempt to find the minimum-volume simplex that contains the data $\vec{Y}$. 

In the inversion step, the extracted endmembers in $\hvec{S}$ are used to recover the simplex-constrained abundances in $\vec{A}$. 
Often this is done by solving\cite{Heinz:TGRS:01,Heylen:TGRS:11} 
\begin{equation}
\label{eq:FCLS}
\hvec{A} = \argmin_{\vec{A} \geq 0} \norm{\vec{Y} - \hvec{S}\vec{A}}_F^2 \st \vec{1}\tran_N\vec{A} = \vec{1}\tran_T ,
\end{equation} 
where $\vec{1}_N$ denotes the $N\times 1$ vector of ones,
which is usually referred to as \emph{fully constrained least squares} (FCLS).

Real-world hyperspectral datasets can contain significant structure beyond non-negativity on $s_{mn}$ and simplex constraints on $\{a_{nt}\}_{n=1}^N$.
For example, the abundances $\{a_{nt}\}_{n=1}^N$ will be \emph{sparse} 
if most pixels contain significant contributions from only a small subset of the $N$ materials.
Also, the abundances $\{a_{nt}\}_{t=1}^T$ will be \emph{spatially coherent} if the presence of a material in a given pixel makes it more likely for that same material to exist in neighboring pixels. 
Likewise, the endmembers $\{s_{mn}\}_{m=1}^M$ will be \emph{spectrally coherent} if the radiance values are correlated across frequency. 

Various unmixing algorithms have been proposed to leverage these additional structures.
For example, given an endmember estimate $\hvec{S}$, the 
SUnSAL algorithm\cite{Iordache:TGRS:11} estimates sparse abundances $\vec{A}$ using $\ell_1$-regularized least-squares (LS),
and the SUnSAL-TV algorithm\cite{Iordache:TGRS:12} adds total-variation (TV) regularization\cite{Chambolle:JMIV:04} to also penalize changes in abundance across neighboring pixels (i.e., to exploit spatial coherence).
SUnSAL and SUnSAL-TV can be categorized as unmixing algorithms, rather than inversion algorithms, since their $\ell_1$-regularization supports the use of large (i.e., $N>M$) and scene-independent endmember libraries for $\hvec{S}$. 
However, there are limitations on the size of the library $\hvec{S}$, and it can be difficult to determine suitable choices for the $\ell_1$ and TV regularization weights.

Traditional NMF techniques have also been enhanced to account for spectral and spatial coherence.
For instance, the $\ell_{1/2}$ NMF (L$\tfrac{1}{2}$NMF) \cite{Qian:TGRS:11} algorithm promotes sparse abundances by adding $\ell_{1/2}$ regularization to the traditional NMF formulation.
The algorithms in \cite{Yuan:TGRS:15,Lu:TGRS:14,Lu:TGRS:13} then expand on this idea by adding additional regularizations to promote coherent abundances.
For example, the Substance Dependence constrained NMF (SDSNMF) \cite{Yuan:TGRS:15}, which was shown in \cite{Yuan:TGRS:15} to perform the best out of \cite{Yuan:TGRS:15,Lu:TGRS:14,Lu:TGRS:13}, employs a sparse pixel-by-pixel weighting matrix that accounts for similarities in the abundances across in the scene. 
Additional coherence-promoting NMF techniques include a method based on hierarchical rank-2 decompositions \cite{Gillis:TGRS:15}, a method that promotes both abundance separability and coherence \cite{Liu:TGRS:11}, and a piece-wise spectral/spatial smoothness constrained method \cite{Jia:TGRS:09}. 

Bayesian approaches to hyperspectral unmixing 
have also been proposed.
For example, the Bayesian Linear Unmixing (BLU) algorithm\cite{Dobigeon:TSP:09} employs priors that enforce NN constraints on the endmembers and simplex constraints on the per-pixel abundances, and returns either (approximately) minimum mean-square error (MMSE) or maximum a posteriori (MAP) estimates using 
Gibbs sampling. 
The Spatially Constrained Unmixing (SCU)\cite{Mittelman:TSP:12} algorithm, an extension of BLU, furthermore exploits spatial coherence using a hierarchical Dirichlet-process prior.   
Both BLU and SCU have been shown to outperform N-FINDR and VCA-plus-FCLS under certain conditions\cite{Mittelman:TSP:12}, but at the cost of several orders-of-magnitude increase in runtime.

\begin{figure*}
\centering
  \newcommand{\sz}{0.75}
  \psfrag{x_jl}[b][B][\sz]{$\underline{\S}_{mn}$}
  \psfrag{y_il}[b][Bl][\sz]{$h_{mt}$}
  \psfrag{d_ik}[b][Bl][\sz]{$\A_{nt}$}
  \psfrag{e_jl}[b][B][\sz]{$\EE_{mn}$}
  \psfrag{s_ik}[b][B][\sz]{$\D_{nt}$}
  \psfrag{pX}[b][Bl][\sz]{$f_{mn}$}
  \psfrag{pD}[b][Bl][\sz]{$g_{nt}$}
  \psfrag{Spectral}[b][B][0.5]{\sf Spectral Coherence}
  \psfrag{BIGAMP}[b][B][0.5]{\sf Bilinear Structure}
  \psfrag{SPATIAL}[b][B][0.5]{\sf Spatial Coherence}
  \psfrag{j}[B][Bl][\sz]{$n$}
  \psfrag{l}[B][Bl][\sz]{$m$}
  \psfrag{k}[B][Bl][\sz]{$n$}
  \psfrag{i}[B][Bl][\sz]{$t$}
  \includegraphics[width=14cm]{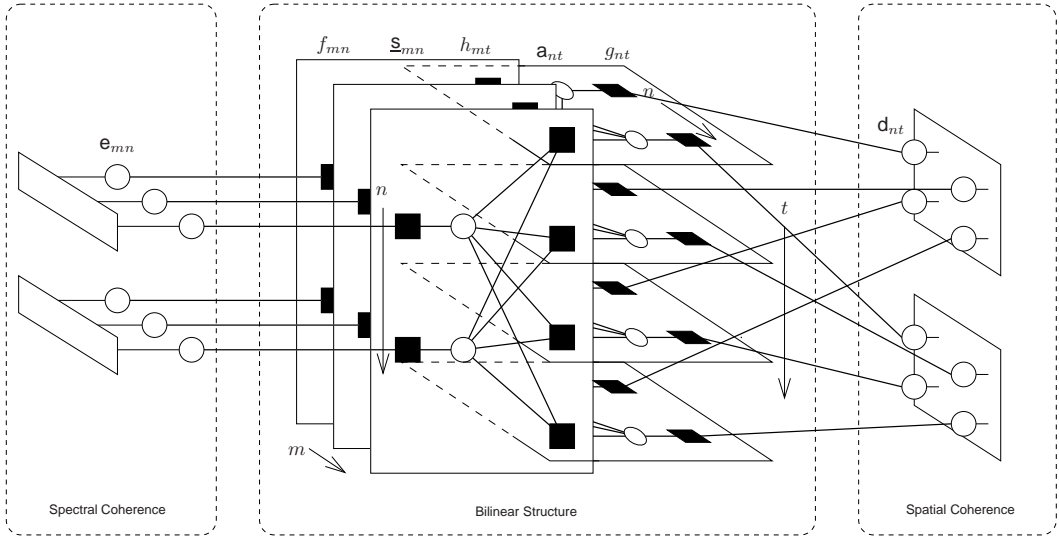}
  \caption{The factor graph for HUT-AMP for the toy-problem dimensions $M=3$, $N=2$, and $T=4$.  Circles represent random variables and dark squares represent pdf factors.  Each elongated bar in the left subgraph conglomerates the factors associated with an $M$-variable Markov chain (detailed in \figref{GAUSSMARK}), while each square in the right subgraph conglomerates the factors associated with a $T_1\times T_2$-pixel Markov random field (detailed in \figref{MRF1}).}
  \label{fig:HSIAMP}
\end{figure*}

In this paper, we propose a novel empirical-Bayesian approach to HU that is based on loopy belief propagation (LBP)\cite{Murphy:UAI:99}.
Our approach, referred to as HU turbo-AMP (HUT-AMP), simplifies the intractable task of LBP on the entire factor graph (see \figref{HSIAMP}) by partitioning it into three subgraphs: one that models spectral coherence (using $N$ Gauss-Markov chains), one that models spatial coherence (using $N$ binary Markov Random Fields (MRFs)), and one that models the NN bilinear structure of \eqref{LMM}.  
While the first two subgraphs yield inference problems that are handled efficiently by standard methods\cite{Bouman:ICIP:95,Li:Book:09}, the third does not.
Thus, to perform efficient inference on the latter subgraph, we apply the recently proposed Bilinear Generalized Approximate Message Passing (BiG-AMP) algorithm\cite{Parker:TSP:14a}.
BiG-AMP can be interpreted as an extension of approximate message passing (AMP) techniques\cite{Donoho:PNAS:09,Donoho:ITW:10a,Rangan:ISIT:11}, originally proposed for the linear observation models that arise in compressive sensing, to bilinear models like \eqref{LMM}.
To merge BiG-AMP-based inference with Markov-chain and MRF-based inference, we leverage the ``turbo AMP'' approach first proposed in\cite{Schniter:CISS:10} and subsequently applied to 
joint channel-estimation and decoding\cite{Schniter:JSTSP:11,Nassar:TSP:14}, 
compressive image retrieval\cite{Som:TSP:12,Som:ICML:11}, and
compressive video retrieval\cite{Ziniel:TSP:13b}, 
all with state-of-the-art results.
In formulating our statistical model, we treat the parameters of the prior distributions as deterministic unknowns and estimate them from the data using the expectation-maximization (EM) algorithm, building on the NN sparse reconstruction work in\cite{Vila:TSP:14}.
As such, our approach can be classified as \emph{empirical Bayesian} \cite{Efron:Book:10}.
Lastly, when the number of materials $N$ is unknown, we show how it can be accurately estimated using a classical model-order selection (MOS) strategy\cite{Stoica:SPM:04}.
The resulting algorithm has the following desirable features: 1) it requires no tuning parameters, 2) it exploits both spectral and spatial coherence, and 3) it uses a computationally efficient inference procedure. 

We evaluate the performance of our proposed technique, in comparison to several recently proposed methods, through a detailed numerical study that includes both synthetic and real-world datasets.
The results, presented in \secref{results}, suggest that HUT-AMP yields an excellent combination of unmixing performance and computational complexity.

Regarding novel contributions to HU models, we believe that our work (first presented in \cite{Vila:SPIE:13}) is the first to use either of the following:
i) Gauss-Markov chains to model spectral coherence in endmembers,
ii) Bernoulli truncated-Gaussian mixtures to model abundance amplitudes.
As for novel contributions to inference methodology, we believe that our work is the first to combine \emph{any} of the following methods, and in fact we combine all four of them: 
i) compressed sensing with non-negative Bernoulli-Gaussian-mixture priors whose parameters are learned via EM \cite{Vila:TSP:14},
ii) turbo compressed sensing that combines AMP with Markov-chain inference and learns the parameters via EM \cite{Ziniel:TSP:13b},
iii) turbo compressed sensing that combines AMP with MRF inference and learns the parameters via EM \cite{Som:ICML:11},
iv) bilinear AMP \cite{Parker:TSP:14a}.

\emph{Notation}:
  For matrices, we use boldface capital letters like $\vec{A}$, and we use $\vec{A}\tran$, $\tr(\vec{A})$, and $\norm{\vec{A}}_F$ to denote the transpose, trace, and Frobenius norm, respectively.
  For vectors, we use boldface small letters like $\vec{x}$, and we use $\norm{\vec{x}}_p=(\sum_n |x_n|^p)^{1/p}$ to denote the $\ell_p$ norm, with $x_n=[\vec{x}]_n$ representing the $n^{th}$ element of $\vec{x}$.
  We use $\vec{1}_N$ to denote the $N\times 1$ vector of ones.
  Deterministic quantities are denoted using serif typeface (e.g., $x,\vec{x},\vec{X}$), while random quantities are denoted using san-serif typeface (e.g., $\X,\vX,\mX$).
  For random variable $\X$, we write the probability density function (pdf) as $p_{\X}(x)$, the expectation as $\E\{\X\}$, and the variance as $\var\{\X\}$.
  For a Gaussian random variable $\X$ with mean $m$ and variance $v$, we write the pdf as $\mc{N}(x;m,v)$ 
  and, for the special case of $\mc{N}(x;0,1)$, we abbreviate the pdf as $\varphi(x)$ and write the complimentary cdf as $\Phi_c(x)$.
  Finally, we use $\delta(x)$ (where $x\in\Real$) to denote the Dirac delta distribution and $\delta_n$ (where $n\in\Int$) to denote the Kronecker delta sequence.

\section{Signal and Observation Models}\label{sec:models}

\subsection{Background on BiG-AMP}\label{sec:BiGAMP}

As described in the introduction, a distinguishing feature of our approach is the use of BiG-AMP\cite{Parker:TSP:14a} for bilinear inference.
We begin by overviewing BiG-AMP, since its operating assumptions affect the construction of our statistical model.

Consider the problem of estimating the elements of the matrices $\ovec{S}\in\Real^{M\times N}$ and $\ovec{A}\in\Real^{N\times T}$ from a noisy observation $\ovec{Y}\in\Real^{M\times T}$ of the hidden bilinear form $\ovec{Z}\defn\ovec{S}\ovec{A}\in\Real^{M\times T}$. 
(Our use of overbar notation will become clear in the sequel.)
Suppose that the elements of both $\ovec{S}$ and $\ovec{A}$ can be modeled as \emph{independent} random variables $\So_{mn}$ and $\Ao_{nt}$ with known prior pdfs $p_{\So_{mn}}(\cdot)$ and $p_{\Ao_{nt}}(\cdot)$, respectively, with $\So_{mn}$ being zero-mean.
Suppose also that the likelihood function of $\ovec{Z}$ is known and \emph{separable}, i.e., of the form
\begin{equation}
p_{\mYo|\mZo}(\ovec{Y}|\ovec{Z})
=\prod_{m=1}^M\prod_{t=1}^T p_{\Yo_{mt}|\Zo_{mt}}(\overline{y}_{mt}|\overline{z}_{mt}) 
\label{eq:likelihood} .
\end{equation}
Finally, suppose that the dimensions $M,N,T$ are sufficiently large.
In this case, approximations of the marginal posterior pdfs of $\So_{mn}$, $\Ao_{nt}$, and $\Zo_{mt}$ can tractably be computed using loopy belief propagation (LBP)\cite{Murphy:UAI:99}, and in particular using an approximation of the sum-product algorithm (SPA)\cite{Kschischang:TIT:01} known as BiG-AMP\cite{Parker:TSP:14a}.
More precisely, BiG-AMP approximates the marginal posterior pdf of $\So_{mn}$ as
\begin{equation}
p_{\So_{mn}|\mYo}(\overline s_{mn}|\ovec{Y}) 
= \frac{p_{\So_{mn}}\!(\overline s_{mn})\mc{N}(\overline s_{mn};\hat{q}_{mn},\nu^q_{mn})}
  {\int p_{\So_{mn}}\!(\overline s_{mn}')\mc{N}(\overline s_{mn}';\hat{q}_{mn},\nu^q_{mn}) \text{d}s_{mn}'} ,
\end{equation}
where the parameters $\hat{q}_{mn}$ and $\nu^q_{mn}$ are iteratively updated at each BiG-AMP iteration;
similar approximations are made for the marginal posteriors of $\Ao_{nt}$ and $\Zo_{mt}$.
BiG-AMP also computes the means and variances of these approximate marginal posteriors at each iteration, yielding approximate MMSE estimates of $\So_{mn}$, $\Ao_{nt}$, and $\Zo_{mt}$, as well as approximations of their corresponding MSEs.
For many priors of interest (e.g, the ones used in this paper), these means and variances can be computed in closed form.

In the big picture, BiG-AMP can be understood as a recent generalization of the AMP methods \cite{Donoho:PNAS:09,Donoho:ITW:10a,Rangan:ISIT:11} from linear to bilinear inference.
These AMP methods can be derived by starting with the SPA and applying i) central-limit-theorem arguments that approximate all messages as Gaussian and ii) Taylor-series approximations that reduce the number of messages.  
Under additional independence and sub-Gaussianity assumptions, these AMP methods can be analyzed in the large-system limit, where their behavior is fully characterized by a state evolution \cite{Bayati:TIT:11}.
When the state evolution has a unique fixed point, the posterior approximations computed by AMP are in fact Bayes-optimal in the large-system limit\cite{Bayati:TIT:11}. 
For finite-sized problems, the fixed points of AMP are known to coincide with the stationary points of a particular Bethe free energy approximation\cite{Rangan:ISIT:13,Krzakala:ISIT:14}.
For a more detailed description of how AMP methods fit into the larger family of variational Bayesian methods, we refer the reader to the recent tutorial\cite{Pereyra:JSTSP:15}.
For a detailed derivation of BiG-AMP, we refer the reader to \cite{Parker:TSP:14a}.

BiG-AMP's complexity is in general dominated by ten matrix multiplies (of the form $\ovec{S}\ovec{A}$) per iteration, although simplifications can be made in the case of Gaussian $p_{\Yo_{mt}|\Zo_{mt}}(\overline{y}_{mt}|\overline{z}_{mt})$ that reduce the complexity to three matrix multiplies per iteration\cite{Parker:TSP:14a}. 
Furthermore, when BiG-AMP's likelihood function and priors include unknown parameters $\vec{\Omega}$, expectation-maximization (EM) methods can be used to learn them, as described in\cite{Parker:TSP:14a}.
BiG-AMP was shown\cite{Parker:TSP:14b} to yield excellent performance on matrix completion, robust PCA, and dictionary learning problems, and here we show that it performs very well on the NMF and HU problems as well.

\subsection{Augmented Observation Model}\label{sec:obs}

We model the elements of the $m$th row of the additive noise matrix $\vec{W}$ in \eqref{LMM} as i.i.d zero-mean Gaussian with variance $\psi_m>0$.
Thus, the BiG-AMP marginal likelihoods take the form 
$p_{\Yo_{m}|\Zo_{m}}(\overline{y}_{mt}|\overline{z}_{mt})
= \mc{N}(\overline{y}_{mt};\overline{z}_{mt};\psi_m)$.
For now we treat $\vec{\psi}$ as known, but later (in \secref{EM}) we describe how it and other model parameters can be learned from $\vec{Y}$.

Leveraging the zero-mean property of the noise, we first perform mean-removal on the observations $\vec{Y}$.
In particular, we subtract the empirical mean 
\begin{align}
\mu 
&\defn \frac{1}{MT} \sum_{t=1}^T \sum_{m=1}^M y_{mt}  
= \frac{1}{MT} \vec{1}_M\tran\vec{Y}\vec{1}_T 
	\label{eq:mu}
\end{align}
from $\vec{Y}$ to obtain
\begin{align}
 \uvec{Y}
 &\defn \vec{Y} - \mu\vec{1}_M\vec{1}_T\tran 
 	\label{eq:LMMmr0} \\
 &= \underbrace{ \big(\vec{S} - \mu\vec{1}_M\vec{1}_N\tran\big) 
 	}_{\displaystyle \defn \uvec{S}} 
 	\vec{A} + \vec{W}
 	\label{eq:LMMmr} ,
\end{align}
where \eqref{LMMmr} employed \eqref{LMM} and $\vec{1}_N\tran\vec{A}=\vec{1}_T\tran$, the latter of which results from the simplex constraint on the columns of $\vec{A}$. 
It can then be shown (see \appref{mean}) that the elements of $\uvec{S}$ in \eqref{LMMmr} are approximately zero-mean.

To enforce the linear equality constraint $\vec{1}_N\tran\vec{A} \!=\! \vec{1}\tran_T$, we augment the observation model \eqref{LMMmr} into the form
\begin{equation}
\underbrace{ \mat{\uvec{Y}\\ \vec{1}\tran_T} }_{\displaystyle \defn \ovec{Y}}
= \underbrace{ \mat{\uvec{S} \\ \vec{1}\tran_N}}_{\displaystyle \defn \ovec{S}}
  \vec{A} + \underbrace{ \mat{\vec{W} \\ \vec{0}\tran_T} }_{\displaystyle \defn \ovec{W}} .
\label{eq:model}
\end{equation}
For the augmented model \eqref{model}, the likelihood function of $\ovec{Z}\defn \ovec{S}\vec{A}$ takes the form in \eqref{likelihood} with
\begin{align}
  p_{\overline \Y_{m}|\overline \Z_{m}}\!(\overline y_{mt}|\overline z_{mt}) 
  &\!=\! \underbrace{
    \begin{cases}
    \!\mc{N}(\overline y_{mt};\overline z_{mt},\psi_m)
    & m\!=\!1,\dots,M\\
    \!\delta(\overline{y}_{mt}\!-\overline z_{mt}) 
    & m\!=\!M\!+\!1.
  \end{cases}
  }_{\displaystyle \defn h_{mt}(\overline z_{mt})}
  \label{eq:pY_Z} 
\end{align}

We note that, ignoring spectral and spatial coherence, the model \eqref{model} is appropriate for the application of BiG-AMP, since the likelihood function $p_{\mYo|\mZo}(\ovec{Y}|\ovec{Z})$ is known (up to $\vec{\psi}$) and separable, and since the elements in $\ovec{S}$ and $\vec{A}$ can be treated as independent random variables with priors known up to a set of parameters, with those in $\ovec{S}$ being approximately zero-mean.
In the sequel, we describe how the model \eqref{model} can be extended to capture spectral and spatial coherence.
As we will see, this will be done through the introduction of additional variables that allow $\ovec{S}$ and $\vec{A}$ to be treated as \emph{conditionally} independent. 

\subsection{Endmember Prior}\label{sec:endm}

We desire a model that promotes spectral coherence in the endmembers, i.e., correlation among the (mean removed) spectral amplitudes $\{\Su_{mn}\}_{m=1}^M$ of each material $n$.
However, since BiG-AMP needs $\Su_{mn}$ to be independent, we cannot impose correlation on these variables directly.
Instead, we introduce an auxiliary sequence of correlated amplitudes $\{\EE_{mn}\}_{m=1}^M$ such that $\Su_{mn}$ are independent \emph{conditional on} $\EE_{mn}$.
In particular,
\begin{align}
p_{\mSu|\mE}(\uvec{S}|\vec{E})
&= \prod_{m=1}^M \prod_{n=1}^N p_{\Su|\EE}(\underline{s}_{mn}|e_{mn}) \\
p_{\Su|\EE}(\underline{s}_{mn}|e_{mn}) 
&= \underbrace{ \delta(\underline{s}_{mn}-e_{mn}) }_{
	\displaystyle \defn f_{mn}(s_{mn},e_{mn}) } ,
\label{eq:pS_E}
\end{align}
implying that $\EE_{mn}$ is merely a copy of $\Su_{mn}$.
To impart correlation within the auxiliary sequences $\{\EE_{mn}\}_{m=1}^M$, we model them as independent Gauss-Markov models
\begin{align}
p_\mE(\vec{E}) 
&= \prod_{n=1}^N 
	\underbrace{ p(e_{1n}) \prod_{m=2}^M p(e_{mn}|e_{m-1,n})
	}_{\displaystyle \defn p_{\vE_n}(\vec{e}_n)},
\end{align}
where $\vE_n\defn[\EE_{1n},\dots,\EE_{Mn}]\tran$, 
$\vec{e}_n\defn[e_{1n},\dots,e_{Mn}]\tran$, and 
\begin{align}
p(e_{1n})
&= \mc{N}(e_{mn};\kappa_n,\sigma^2_n)
\label{eq:speccorr1} \\
p(e_{mn}|e_{m-1,n})
&= \mc{N}\big(e_{mn}; (1\!-\!\eta_n)e_{m-1,n} \!+\! \eta_n \kappa_n, \eta_n^2 \sigma^2_n\big)
\label{eq:speccorr2} .
\end{align}
In \eqref{speccorr1}-\eqref{speccorr2}, 
$\kappa_n \in \Real$ controls the mean of the $n$th process, 
$\sigma^2_n$ controls the variance, 
and $\eta_n \in [0,1]$ controls the correlation.
The resulting factor graph is illustrated in \figref{GAUSSMARK}. 

We note that the model \eqref{speccorr1}-\eqref{speccorr2} does not explicitly enforce non-negativity in $s_{mn}$ because, for simplicity, we have omitted the constraint $\underline{s}_{mn}\geq -\mu$. 
Enforcement of $\underline{s}_{mn}\geq -\mu$ could be accomplished by replacing the pdfs in \eqref{speccorr1}-\eqref{speccorr2} with truncated Gaussian versions, but the computations required for inference would become much more tedious.
In our experience, this tedium is not warranted: with practical HU datasets,\footnote{Throughout our numerical experiments, the proposed inference method never produced a negative estimate of $\S_{mn}$.} it suffices to enforce non-negativity in $\vec{A}$ and keep $\vec{Y}\approx\vec{SA}$. 

\putFrag{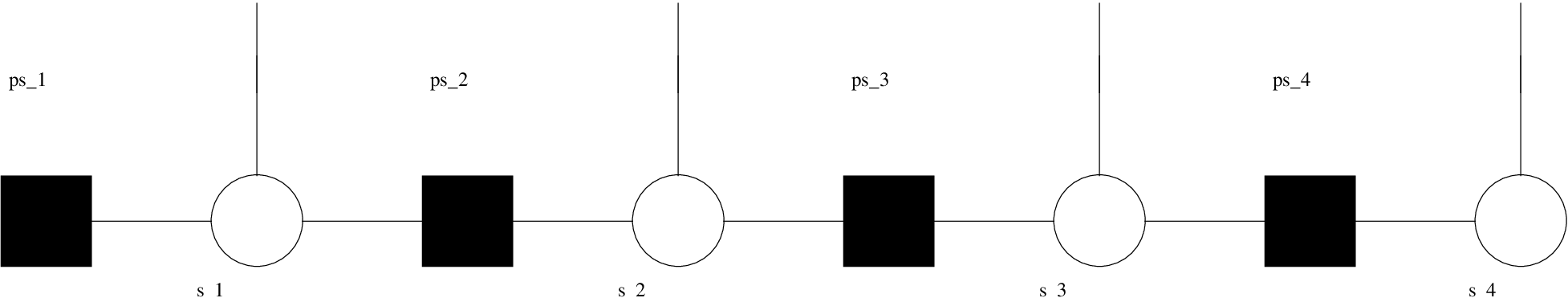}
{Factor graph for the stationary first-order Gauss-Markov chain used to model coherence in the spectrum of the $n^{\text{th}}$ endmember, shown here for $M=4$ spectral bands. 
Incoming messages from BiG-AMP flow downward into the $\EE_{mn}$ nodes, and outgoing messages to BiG-AMP flow upward from the $\EE_{mn}$ nodes.
}
{2.25}
{
\newcommand{\sz}{0.8}
\psfrag{s_1}[c][c][\sz]{$\EE_{1n}$}
\psfrag{s_2}[c][c][\sz]{$\EE_{2n}$}
\psfrag{s_3}[c][c][\sz]{$\EE_{3n}$}
\psfrag{s_4}[c][c][\sz]{$\EE_{4n}$}
\psfrag{ps_1}[c][r][\sz]{$p(e_{1n})$}
\psfrag{ps_2}[c][r][\sz]{$p(e_{2n}|e_{1n})$}
\psfrag{ps_3}[c][r][\sz]{$p(e_{3n}|e_{2n})$}
\psfrag{ps_4}[c][r][\sz]{$p(e_{4n}|e_{3n})$}
} 

\subsection{Abundance Prior}  \label{sec:abun}
We desire a model that promotes both sparsity and spatial coherence in the abundances $\A_{nt}$. 
To accomplish the latter, we impose structure on the \emph{support} of $\{\A_{nt}\}_{t=1}^T$ for each material $n$.
For this purpose, we introduce the support variables $\D_{nt} \in \{-1,1\}$, where $\D_{nt}=-1$ indicates that $\A_{nt}$ is zero-valued, and $\D_{nt} = 1$ indicates that $\A_{nt}$ is non-zero with probability $1$, which we will refer to as ``active.'' 
By modeling the abundances $\A_{nt}$ as independent \emph{conditional on} $\D_{nt}$, we comply with the independence assumptions of BiG-AMP.
In particular, we assume that 
\begin{align}
p_{\mA|\mD}(\vec{A}|\vec{D})
&= \prod_{n=1}^N\prod_{t=1}^T p_{\A_n|\D_n}(a_{nt} | d_{nt}) \\
p_{\A_n|\D_n}(a_{nt} | d_{nt}) 
&= \underbrace{
	\begin{cases}
	  \delta(a_{nt}) & d_{nt}=-1 \\
	  \zeta_n(a_{nt}) & d_{nt}=1 
	\end{cases}
   }_{ \displaystyle \defn g_{nt}(a_{nt},d_{nt}) }
\label{eq:pA_D},
\end{align}
where $\zeta_n(\cdot)$ denotes the pdf of $\A_{nt}$ when active. 
Essentially, we employ a Bernoulli-$\zeta_n(\cdot)$ distribution for the $n$th material.

We then place a Markov random field (MRF) prior on the support of the $n$th material, $\vD_n\defn[\D_{n1},\dots,\D_{nT}]\tran$:
\begin{align}
p_\mD(\vec{D}) 
&= \prod_{n=1}^N p_{\vD_n}(\vec{d}_n)  \\
p_{\vD_n}(\vec{d}_n) 
&\propto \exp\Bigg(\sum_{t=1}^T 
	\Bigg(\frac{1}{2} \sum_{i \in \mc{D}_{t}} \beta_n d_{ni} - \alpha_n\Bigg)
	d_{nt} \Bigg),
\label{eq:Ising}
\end{align}
where $\mc{D}_{t}\subset\{1,\dots,T\}\setminus t$ denotes the neighbors of pixel $t$.
Roughly speaking, larger $\beta_n$ yields higher spatial coherence and larger $\alpha_n$ yields higher sparsity.
For simplicity, we adopt a neighborhood structure corresponding to the classical Ising model\cite{Bouman:ICIP:95}, as illustrated by the factor graph in \figref{MRF1}.

\putFrag{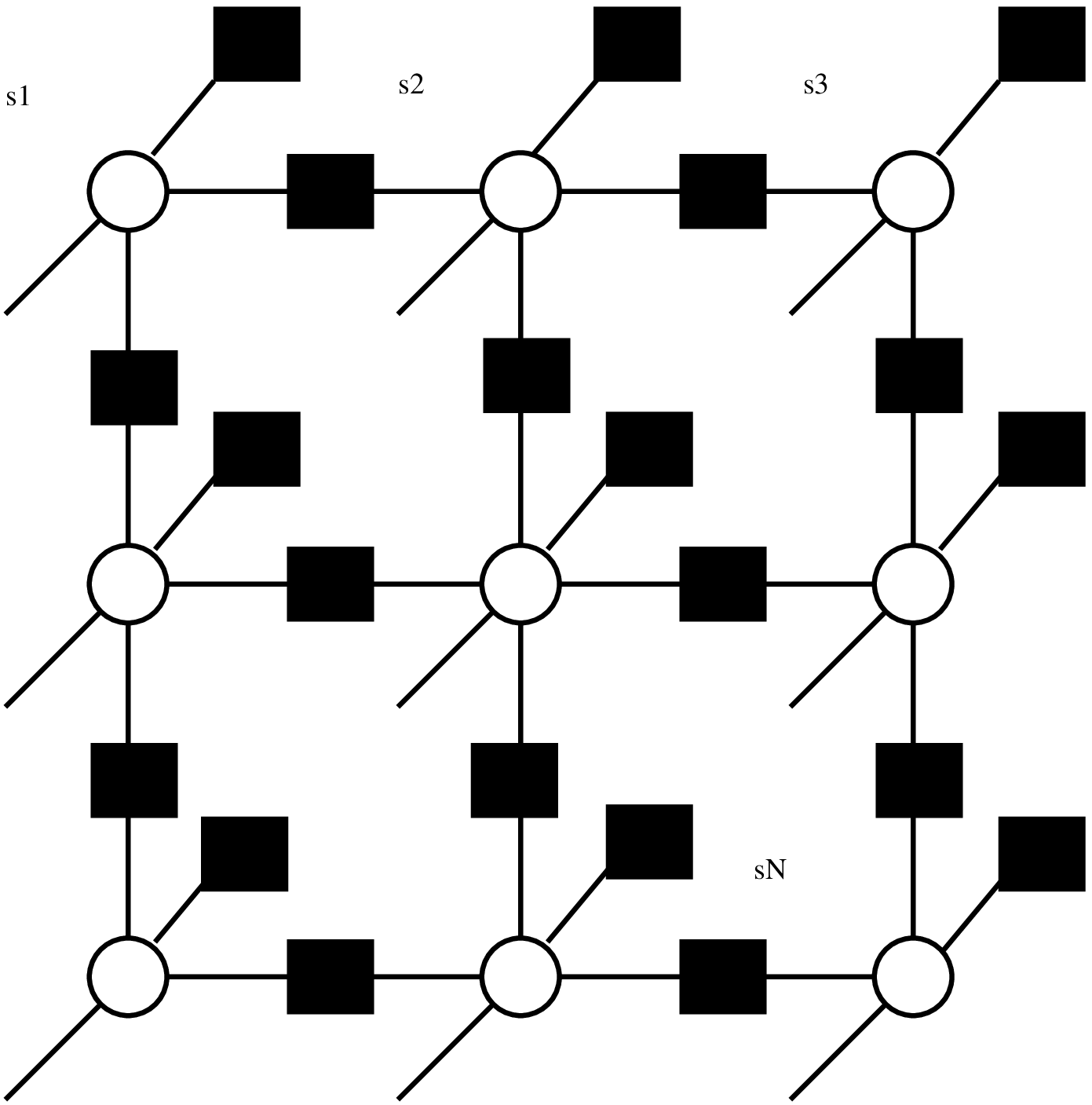}
{Factor graph for the Ising MRF used to model spatial coherence in the support of the $n$th abundance map, here for $T = 3 \times 3$ pixels.
Incoming messages from BiG-AMP flow diagonally upward into the $\D_{nt}$ nodes, and outgoing messages to BiG-AMP flow diagonally downward from the $\D_{nt}$ nodes.
}
{1.7}
{
\newcommand{\sz}{0.8}
\psfrag{s1}[l][c][\sz]{$\D_{n1}$}
\psfrag{s2}[lt][ct][\sz]{$\D_{n2}$}
\psfrag{s3}[lt][ct][\sz]{$\D_{n3}$}
\psfrag{sN}[lt][ct][\sz]{$\D_{nT}$}
}

As for the active abundances, we adopt a non-negative Gaussian mixture (NNGM) distribution for $\zeta_n(\cdot)$:
\begin{equation}
\zeta_n(a) 
= \sum_{\ell = 1}^L \omega^\A_{n\ell} \mc{N}_+(a;\theta^\A_{n\ell},\phi^\A_{n\ell}),  
\label{eq:nu}
\end{equation}
where $\omega^\A_{n\ell}\geq 0$ and $\sum_{\ell=1}^L\omega^\A_{n\ell}=1$.
In \eqref{nu}, $\mc{N}_+$ refers to the truncated Gaussian pdf 
\begin{equation}
\label{eq:NNG}
\mc{N}_+(x;\theta,\phi) \defn 
\begin{cases}
0 & x < 0 \\
\displaystyle 
\frac{\mc{N}(x;\theta,\phi)}{\Phi_c(\theta/\sqrt{\phi})} & x \geq 0
\end{cases},
\end{equation}
where $\theta\in\Real$ is a location parameter (but not the mean), $\phi>0$ is a scale parameter (but not the variance), and $\Phi_c(\cdot)$ is the complimentary cdf of the $\mc{N}(0,1)$ distribution.
In practice, we find that $L=3$ mixture components suffice, and we used this value throughout our numerical experiments in \secref{results}.
We use a NNGM prior based on its ability to faithfully model a wide range of distributions (including multi-modal ones) and the ease by which its parameters, $\{\omega^\A_{n\ell}, \theta^\A_{n\ell}, \phi^\A_{n\ell}\}$, can be accurately tuned using the EM method developed in \cite{Vila:TSP:14} and discussed in \secref{EM}.

We note that the abundance model described in this section treats the abundance coefficients as correlated across pixels but statistically independent across materials.
Meanwhile, the likelihood function described in \secref{obs} enforces a sum-to-one constraint across materials at each pixel. 
These statistical structures are then merged in the posterior.
An alternative approach that allows an abundance prior with correlation across pixels and sum-to-one across materials recently appeared in \cite{Altmann:14}.
Implementing this approach in conjunction with AMP is an interesting topic for future research. 

\section{The HUT-AMP Algorithm}  	\label{sec:algorithm}

\subsection{Message Passing and Turbo Inference}		\label{sec:turbo}

Our overall goal is to jointly estimate the (correlated, non-negative) endmembers $\vec{S}$ and (structured sparse, simplex-constrained) abundances $\vec{A}$ from noisy observations $\vec{Y}$ of the bilinear form $\vec{Z}=\vec{SA}$.
Using the mean-removed, augmented probabilistic models from \secref{models}, the joint pdf of all random variables can be factored as follows:
\begin{align}
\lefteqn{
p(\ovec{Y}, \ovec{S}, \vec{A}, \vec{E}, \vec{D}) 
}\nonumber\\
&= p(\ovec{Y} | \ovec{S},\vec{A})\,
   p(\ovec{S},\vec{E})\,
   p(\vec{A},\vec{D}) 
\label{eq:post} \\
& = p_{\mYo|\mZo}(\ovec{Y}|\ovec{S}\vec{A})\,
    p_{\mSo|\mE}(\ovec{S}|\vec{E})\,
    p_{\mE}(\vec{E})\,
    p_{\mA|\mD}(\vec{A}|\vec{D})\,
    p_{\mD}(\vec{D})\\
&= \Bigg(\prod_{m=1}^{M+1}\prod_{t=1}^T  
h_{mt}\Bigg(\sum_{n=1}^N \overline{s}_{mn}a_{nt}\Bigg)\Bigg)
\nonumber\\&\quad\times
\prod_{n=1}^N \Bigg(
\delta(\overline{s}_{M+1,n}-1)\,
p_{\vE_n}\!(\vec{e}_n) 
\prod_{m=1}^M f_{mn}(\overline{s}_{mn},e_{mn})
\nonumber\\&\qquad\qquad\times 
p_{\vD_n}\!(\vec{d}_n) \prod_{t=1}^Tg_{nt}(a_{nt},d_{nt})\Bigg),
\end{align}
yielding the factor graph in \figref{HSIAMP}.  
Due to the cycles within the factor graph, exact inference is NP-hard\cite{Cooper:AI:90}, and so we settle for approximate MMSE inference.

To accomplish approximate MMSE inference, we apply a form of loopy belief propagation that is inspired by the ``turbo decoding'' approach used in modern communications receivers\cite{McEliece:JSAC:98}.
In particular, after partitioning the overall factor graph into three subgraphs, as in \figref{HSIAMP}, we alternate between message-passing \emph{within} subgraphs and message-passing \emph{between} subgraphs.
In our case, BiG-AMP\cite{Parker:TSP:14a} is used for message-passing within the bilinear subgraph and standard methods from\cite{Bouman:ICIP:95,Li:Book:09} are used for message-passing within the other two subgraphs, which involve $N$ Gauss-Markov chains and $N$ binary MRFs, respectively.
Overall, our proposed approach can be interpreted as a bilinear extension of the ``turbo AMP'' approach first proposed in\cite{Schniter:CISS:10}.

\subsection{Messaging Between Subgraphs}		\label{sec:exch} 

For a detailed description of the message passing \emph{within} the Gauss-Markov, MRF, and BiG-AMP subgraphs, we refer interested readers to\cite{Bouman:ICIP:95},\cite{Li:Book:09}, and\cite{Parker:TSP:14a}, respectively.
We now describe the message passing \emph{between} subgraphs, which relies on the sum-product algorithm (SPA)\cite{Kschischang:TIT:01}.
In our implementation of the SPA, we assume that all messages are scaled to form valid pdfs (in the case of continuous random variables) or probability mass functions (pmfs) (in the case of discrete random variables), and we use $\msg{b}{c}(\cdot)$ to represent the message passed from node $b$ to node $c$.

As described in\cite{Kschischang:TIT:01}, the SPA message flowing out of a variable node along a given edge equals the (scaled) product of messages flowing into that node along its other edges. 
Meanwhile, the SPA message flowing out of a factor node along a given edge equals the (scaled) integral of the product of all incoming messages times the factor associated with that node. 
Finally, the SPA approximates the posterior of a given random variable as the (scaled) product of messages flowing into that random variable.

As discussed in \secref{BiGAMP}, a key property of BiG-AMP is that certain messages within its sub-graph are approximated as Gaussian.
In particular,
\begin{align}
\msg{\underline{\S}_{mn}}{f_{mn}}(\underline{s})
&= \mc{N}(\underline{s};\hat{q}_{mn}, \nu^q_{mn}) \label{eq:stof} \\
\msg{\A_{nt}}{g_{nt}}(a)
&= \mc{N}(a;\hat{r}_{nt}, \nu^r_{nt}) \label{eq:atog},
\end{align}
where the quantities $\hat{q}_{mn}, \nu^q_{mn}, \hat{r}_{nt}, \nu^r_{nt}$ are computed during the final iteration of BiG-AMP.
Thus, the SPA approximated posteriors on $\underline{\S}_{mn}$ and $\A_{nt}$ take the form
\begin{align}
p_{\underline{\S}_{mn}|\Q_{mn}}\!(\underline{s}\giv\hat{q}_{mn}; \nu^q_{mn}) 
&\propto \msg{f_{mn}}{\underline{\S}_{mn}}(\underline{s}) \mc{N}(\underline{s}; \hat{q}_{mn}, \nu^q_{mn}) 
\label{eq:gampposts} \\
p_{\A_{nt}|\R_{nt}}\!(a\giv\hat{r}_{nt}; \nu^r_{nt}) 
&\propto \msg{g_{nt}}{\A_{nt}}(a) \mc{N}(a; \hat{r}_{nt}, \nu^r_{nt}) 
\label{eq:gampposta} ,
\end{align}
where $\msg{f_{mn}}{\underline{\S}_{mn}}(\underline{s})$ and $\msg{g_{nt}}{\A_{nt}}(a)$ can be interpreted as priors, $\mc{N}(\underline{s}; \hat{q}_{mn}, \nu^q_{mn})$ and $\mc{N}(a; \hat{r}_{nt}, \nu^r_{nt})$ can be interpreted as likelihoods, and \eqref{gampposts} and \eqref{gampposta} can be interpreted as Bayes rule.
We will use these properties in the sequel.

First, we discuss the message-passing between the bilinear sub-graph and spectral-coherence sub-graph in \figref{HSIAMP}.
Given \eqref{pS_E}, \eqref{stof}, and the construction of the factor graph in \figref{HSIAMP}, the SPA implies that 
\begin{align}
\msg{f_{mn}}{\EE_{mn}}(e) 
&\propto \int f_{mn}(\underline{s},e) \, \msg{\underline{\S}_{mn}}{f_{mn}}(\underline{s}) \, d\underline{s} \\
&= \mc{N}(e; \hat{q}_{mn}, \nu^q_{mn})
\label{eq:ftoe} .
\end{align}
The messages in \eqref{ftoe} are used as inputs to the Gauss-Markov inference procedure.
By construction, the outputs of the Gauss-Markov inference procedure will also be Gaussian beliefs.
Denoting their means and variances by $\theta^{\underline{\S}}_{mn}$ and $\phi^{\underline{\S}}_{mn}$, respectively, we have that
\begin{align}
\msg{\EE_{mn}}{f_{mn}}(e) 
&\propto \mc{N}(e; \theta^{\underline{\S}}_{mn}, \phi^{\underline{\S}}_{mn}) \label{eq:etof} \\
\msg{f_{mn}}{\underline{\S}_{mn}}(\underline{s}) 
&= \int f_{mn}(\underline{s},e) \, \msg{\EE_{mn}}{f_{mn}}(e) \, de \\
&= \mc{N}(\underline{s}; \theta^{\underline{\S}}_{mn}, \phi^{\underline{\S}}_{mn}) \label{eq:ftos} .
\end{align}
When BiG-AMP is subsequently called for inference on the bilinear sub-graph, \eqref{ftos} is inserted into \eqref{gampposts}, i.e., $\msg{f_{mn}}{\underline{\S}_{mn}}(\cdot)$ acts as the prior on $\underline{\S}_{mn}$.

Next we discuss the message-passing between the bilinear sub-graph and the spatial-coherence sub-graph in \figref{HSIAMP}.
The SPA, together with the construction of the factor graph in \figref{HSIAMP}, imply
\begin{align}
\msg{g_{nt}}{\D_{nt}}(d)
&= \frac{\int g_{nt}(a,d) \, \msg{\A_{nt}}{g_{nt}}(a) \, da }
	{\sum_{d'=\pm 1} \int g_{nt}(a,d') \, \msg{\A_{nt}}{g_{nt}}(a) \, da} ,
	~d\in\pm 1.
\end{align}
Given \eqref{pA_D} and \eqref{atog}, we find that
\begin{align}
\lefteqn{
\int g_{nt}(a,d) \, \msg{\A_{nt}}{g_{nt}}(a) \, da 
}\nonumber\\
&= \begin{cases}
	\mc{N}(0;\hat{r}_{nt},\nu^r_{nt}) \, da & d=-1 \\
	\int \zeta_n(a) \, \mc{N}(a;\hat{r}_{nt},\nu^r_{nt}) \, da & d=1 \\
	\end{cases}
\end{align}
which implies
\begin{subequations}
\label{eq:gtod} 
\begin{align}
\msg{g_{nt}}{\D_{nt}}(d =+1)
&= \left(1 \!+\! 
	\frac{\mc{N}(0;\hat{r}_{nt},\nu^r_{nt})}
 	     {\int \zeta_n(a) \, \mc{N}(a;\hat{r}_{nt},\nu^r_{nt})}
	\right)^{-1} \label{eq:gtod1} \\
\msg{g_{nt}}{\D_{nt}}(d =-1)
&= 1-\msg{g_{nt}}{\D_{nt}}(d =+1) ,
\end{align}
\end{subequations}
where the fraction in \eqref{gtod1} is BiG-AMP's approximation of the likelihood ratio $p_{\mY|\D_{nt}}(\vec{Y}|-1)/p_{\mY|\D_{nt}}(\vec{Y}|+1)$.

The Bernoulli beliefs from \eqref{gtod} are used as inputs to the MRF-based support-inference procedure.
The outputs of the MRF inference procedure will also be Bernoulli beliefs of the form 
\begin{subequations}
\label{eq:dtog} 
\begin{align}
\msg{\D_{nt}}{g_{nt}}(d=+1) 
&= \pi_{nt} \\
\msg{\D_{nt}}{g_{nt}}(d=-1) 
&= 1-\pi_{nt} 
\end{align}
\end{subequations}
for some $\pi_{nt}\in(0,1)$.
The SPA and \eqref{pA_D} then imply that
\begin{align}
\msg{g_{nt}}{\A_{nt}}(a) 
&\propto \sum_{d=\pm 1} g_{nt}(a,d) \, \msg{\D_{nt}}{g_{nt}}(d) \\
&= (1-\pi_{nt})\delta(a) + \pi_{nt}\zeta_n(a)
\label{eq:gtoa}
\end{align}
for $\zeta_n(\cdot)$ defined in \eqref{nu}.
When BiG-AMP is subsequently called for inference on the bilinear sub-graph, \eqref{gtoa} is inserted into \eqref{gampposta}, i.e., $\msg{g_{nt}}{\A_{nt}}(\cdot)$ acts as the prior on $\A_{nt}$.

\subsection{EM Learning of the Prior Parameters}     		\label{sec:EM}
In practice, we desire that the parameters 
\begin{equation}
\vec{\Omega} 
= \Big\{\vec{\psi},\{\omega^\A_{n\ell},\theta^\A_{n\ell},\phi^\A_{n\ell}\}_{\forall n\ell}, \{\eta_n, \kappa_n, \sigma^2_n, \alpha_n, \beta_n\}_{\forall n} \Big\} 
\end{equation}
used for the assumed 
likelihood $p_{\Y_{mt}|\Z_{mt}}(y_{mt}|\cdot)$, NNGM abundance prior $\zeta_{n}(\cdot)$, Gauss-Markov chain $p_{\vE_n}(\cdot)$, and binary MRF $p_{\vD_n}(\cdot)$ are well tuned.
With this in mind, we propose an expectation-maximization (EM)\cite{Dempster:JRSS:77} procedure to tune $\vec{\Omega}$, similar to that used for the GAMP-based sparse-reconstruction algorithms in\cite{Vila:TSP:13} and\cite{Vila:TSP:14}.

To tune $\vec{\Omega}$, the EM algorithm\cite{Dempster:JRSS:77} iterates 
\begin{equation}
\vec{\Omega}^{i+1} =\argmax_{\vec{\Omega}} \E\big\{\ln p(\mE, \mA, \mD, \mYo;\vec{\Omega})\biggiv \ovec{Y};\vec{\Omega}^i\big\} 
\label{eq:EMmain}
\end{equation}
with the goal of increasing a lower bound on the true likelihood $p(\ovec{Y};\vec{\Omega})$ at each EM-iteration $i$.
In our case, the true posterior distribution used to evaluate the expectation in \eqref{EMmain} is NP-hard to compute, and so we use the SPA-approximated posteriors 
$\widehat{p_{\mE|\mYo}}(\vec{E}|\ovec{Y}) \propto \prod_{m,n} \msg{f_{mn}}{\EE_{mn}}(e_{mn}) \msg{\EE_{mn}}{f_{mn}}(e_{mn})$ from \eqref{ftoe}-\eqref{etof},
$\widehat{p_{\mD|\mYo}}(\vec{D}|\ovec{Y}) \propto \prod_{n,t} \msg{g_{nt}}{\D_{nt}}(d_{nt}) \msg{\D_{nt}}{g_{nt}}(d_{nt})$ from \eqref{gtod}-\eqref{dtog},
and
$\widehat{p_{\mA|\mYo}}(\vec{A}|\ovec{Y}) \propto \prod_{n,t} \msg{\A_{nt}}{g_{nt}}(a_{nt}) \msg{g_{nt}}{\A_{nt}}(a_{nt})$ from \eqref{atog} and \eqref{gtoa}.
Furthermore, since it is difficult to perform the maximization in \eqref{EMmain} jointly, we maximize $\vec{\Omega}$ one component at a time (while holding the others fixed), which is the well known ``incremental'' variant of EM\cite{Neal:Jordan:98}.

The resulting EM-update expressions for the noise and NNGM parameters $\vec{\psi},\omega^\A_{n\ell},\theta^\A_{n\ell},\phi^\A_{n\ell}$ can be found in\cite{Vila:TSP:14}, and
those for the Gauss-Markov chain parameters $\eta_n, \kappa_n, \sigma^2_n$ can be found in\cite{Ziniel:TSP:13b}.
They are all computed in closed-form using readily available quantities, and thus do not add significantly to the complexity of HUT-AMP.  
The update procedure for the binary MRF parameters $\alpha_n, \beta_n$ is described in\cite{Som:ICML:11} and uses gradient descent. 
Since a small number of gradient-descent iterations suffice, this latter procedure does not significantly increase the complexity of HUT-AMP.

\subsection{EM Initialization} 	\label{sec:inits}

Since the EM algorithm may converge to a local maximum of the likelihood, care must be taken when initializing the EM-learned parameters.
Below, we propose an initialization strategy for HUT-AMP that, based on our empirical experience, seems to work well.

We first initialize the endmembers $\vec{S}$. 
For this, we found it effective to use an off-the-shelf EE algorithm like VCA\cite{Nasc:TGRS:05} or FSNMF\footnote{\label{FSNMF}
With FSNMF (which was used for all of the experiments in \secref{results}), we found that it helped to post-process the observations to reduce the effects of noise.
For this, we used the standard PCA-based denoising approach described in \cite{Bioucas:JSTAEO:12}:
the signal subspace was estimated from the left singular vectors of $\vec{Y}$ after row-wise mean-removal, 
and the FSNMF-estimated endmembers were projected onto the signal subspace.}
\cite{Gillis:TPAMI:14} to recover $\hvec{S}^0$.
Then, as described in \eqref{LMMmr}, we subtract the observation mean $\mu$ from $\hvec{S}^0$ to obtain the initialization $\hvec{\underline{S}}^0$.

With the aid of $\hvec{\underline{S}}^0$, we next run BiG-AMP under
\begin{enumerate}
\item the trivial endmember prior 
\begin{equation}
\msg{f_{mn}}{\underline{\S}_{mn}}(\underline{s}) 
= \delta(\underline{s}-\hat{\underline{s}}_{mn}^0),
\label{eq:trivS}
\end{equation}
which essentially fixes the endmembers at $\hvec{\underline{S}}^0$,
\item
the agnostic NNGM abundance initialization from\cite{Vila:TSP:14}: 
\begin{equation}
\msg{g_{nt}}{\A_{nt}}(a)
= (1-\pi^0_{nt})\delta(a) + \pi^0_{nt}
\sum_{\ell=1}^L \omega^\A_{n\ell} \mc{N}_+(a;\theta^\A_{n\ell},\phi^\A_{n\ell})
\end{equation}
with $\{\omega^\A_{n\ell},\theta^\A_{n\ell},\phi^\A_{n\ell}\}_{\ell=1}^L$ set at the best fit to a uniform distribution on the interval $[0,1]$ and $\pi_{nt}^0=\frac{1}{2}$, and 
\item the noise variance initialization from\cite{Vila:TSP:14}:
\begin{equation}
\psi_m^0 = \frac{ \norm{\uvec{Y}}_F^2}{(\SNR_m^0+1)MT} \ \forall m, 
\end{equation}
where, without any prior knowledge of the true $\SNR_m\defn \E\{|\Z_{mt}|^2\}/\psi_m$, we suggest $\SNR_m^0\!=\!10$ dB.
\end{enumerate}
By running BiG-AMP under these settings, we initialize the messages $\msg{\underline{\S}_{mn}}{f_{mn}}(\cdot)$ and $\msg{\A_{nt}}{g_{nt}}(\cdot)$ from \eqref{stof}-\eqref{atog} and we also obtain an initial estimate of $\vec{A}$ from the mean of the approximate posterior \eqref{gampposta}, which we shall refer to as $\hvec{A}^0$.

Finally, we initialize the remaining parameters in $\Omega$.
Starting with the spectral coherence parameters,
we set the mean $\kappa_n^0$ and variance $(\sigma^2_n)^0$ at the empirical mean and variance, respectively, of the elements in the $n$th column of $\hvec{\underline{S}}^0$.
Then, we initialize the correlation $\eta_n$ as suggested in\cite{Ziniel:TSP:13b}, i.e., 
\begin{align}
\varphi_m^0 &= \frac{\norm{\uvec{y}_m}_2^2 - T \psi_m^0}{\big\|\hvec{A}^0\big\|_F^2} \\
\eta_n^0 &= 1 - \frac{1}{M-1} \sum_{m=1}^{M-1} \frac{ |\uvec{y}_m\tran\uvec{y}_{m+1}|}{\varphi_m^0  \big\|\hvec{A}^0\big\|_F^2 } \ \text{for} \ n=1,\dots,N,
\end{align}
where $\uvec{y}_m\tran$ denotes the $m$th row of $\uvec{Y}$.
Lastly, we initialize the spatial coherence parameters as suggested in \cite{Som:ICML:11}, i.e., $\beta_n^0 = 0.4$ and $\alpha_n^0 = 0.4$, since \cite{Som:ICML:11} shows these values to work well over a wide operating range.

\subsection{HUT-AMP Summary}	\label{sec:sched}

We now describe the scheduling of turbo-messaging and EM-tuning steps, which together constitute the HUT-AMP algorithm.  
Essentially, we elect to perform one EM update per turbo iteration,
yielding the steps tabulated in \tabref{HUTAMP}.
As previously mentioned, 
the ``{\small \BIGAMP}'' operation iterates the BiG-AMP algorithm to convergence as described in\cite{Parker:TSP:14a},
the ``{\small \GausMark}'' operation performs standard Gauss-Markov inference as described in\cite{Bouman:ICIP:95}, and 
the ``{\small \MRF}'' operation performs MRF inference via the belief-propagation method described in\cite{Li:Book:09}. 

\putTable{HUTAMP}
{HUT-AMP pseudocode for fixed number of materials $N$.}
{
\fbox{
\begin{minipage}{3.3in}
Definitions:\\ [1.5mm]
\begin{tabular}{l l}
$\vec{\Delta}^\mE_{\vec{F}} \defn \{\msg{\EE_{mn}}{f_{mn}}(\cdot)\}_{\forall mn}$ 
& $\vec{\Delta}^{\vec{F}}_\mE \defn \{\msg{f_{mn}}{\EE_{mn}}(\cdot)\}_{\forall mn}$ \\
$\vec{\Delta}^{\vec{F}}_{\underline{\mS}} \defn \{\msg{f_{mn}}{\underline{\S}_{mn}}(\cdot)\}_{\forall mn}$ 
&$\vec{\Delta}^{\underline{\mS}}_{\vec{F}} \defn \{\msg{\underline{\S}_{mn}}{f_{mn}}(\cdot)\}_{\forall mn}$ \\
$\vec{\Delta}^\mA_{\vec{G}} \defn \{\msg{\A_{nt}}{g_{nt}}(\cdot)\}_{\forall nt}$ 
&$\vec{\Delta}^{\vec{G}}_\mA \defn \{\msg{g_{nt}}{\A_{nt}}(\cdot)\}_{\forall nt}$ \\
$\vec{\Delta}^{\vec{G}}_\mD \defn \{\msg{g_{nt}}{\D_{nt}}(\cdot)\}_{\forall nt}$  
&$\vec{\Delta}^\mD_{\vec{G}} \defn \{\msg{\D_{nt}}{g_{nt}}(\cdot)\}_{\forall nt}$ \\ [2mm] \hline
\end{tabular}
\begin{algorithmic}[1]
\STATE Initialize $\vec{\Delta}^{\underline{\mS}}_{\vec{F}}$, $\vec{\Delta}^\mA_{\vec{G}}$, and $\vec{\Omega}^0$ as described in \secref{inits}.
\FOR{ $i=1,2,3,\dots$ }
\STATE convert $\vec{\Delta}^{\underline{\mS}}_{\vec{F}}$ to $\vec{\Delta}^{\vec{F}}_\mE$ via \eqref{stof} and \eqref{ftoe}
\STATE convert $\vec{\Delta}^\mA_{\vec{G}}$ to $\vec{\Delta}^{\vec{G}}_\mD$ via \eqref{atog} and \eqref{gtod}
\STATE $\vec{\Delta}^\mE_{\vec{F}} = \GausMark(\vec{\Delta}^{\vec{F}}_\mE, \vec{\Omega}^i)$
\STATE $\vec{\Delta}^\mD_{\vec{G}} = \MRF(\vec{\Delta}^{\vec{G}}_\mD, \vec{\Omega}^i)$
\STATE convert $\vec{\Delta}^\mE_{\vec{F}}$ to $\vec{\Delta}^{\vec{F}}_{\underline{\mS}}$ via \eqref{etof} and \eqref{ftos}
\STATE convert $\vec{\Delta}^\mD_{\vec{G}}$ to $\vec{\Delta}^{\vec{G}}_\mA$ via \eqref{dtog} and \eqref{gtoa}
\STATE $\vec{\Omega}^{i} = \EM(\vec{\Delta}^{\vec{F}}_\mE, \vec{\Delta}^\mE_{\vec{F}}, \vec{\Delta}^{\vec{G}}_\mD, \vec{\Delta}^\mD_{\vec{G}}, \vec{\Delta}^{\vec{G}}_\mA, \vec{\Delta}^\mA_{\vec{G}}, \vec{\Omega}^{i-1})$
\STATE $[\vec{\Delta}^{\underline{\mS}}_{\vec{F}},\vec{\Delta}^\mA_{\vec{G}}] = \BIGAMP(\vec{\Delta}^{\vec{F}}_{\underline{\mS}}, \vec{\Delta}^{\vec{G}}_\mA, \vec{\Omega}^{i})$  
\ENDFOR
\end{algorithmic}
\end{minipage}
}
}

\subsection{Selection of Model Order $N$} 					\label{sec:MOS}

In practice, the number of materials $N$ present in a scene may be unknown. 
Previous approaches such as the hyperspectral signal subspace identification by minimum error (HySime) \cite{Bioucas:TGRS:08}, and a Neyman-Pearson detection theory-based thresholding method (HFC) \cite{Chang:TGRS:04} directly address the problem of estimating the number of materials $N$.

As an alternative, we apply a standard penalized log-likelihood maximization\cite{Stoica:SPM:04} method to estimate $N$ from the observed data $\vec{Y}$. 
Specifically, we aim to solve
\begin{equation}
\label{eq:AICc}
\hat{N} = \argmax_{N} 2\ln p_{\mYo|\mZ}(\ovec{Y}|\underline{\hvec{S}}_N\hvec{A}_N;\hvec{\psi}_{\text{ML}}) - \gamma(N),
\end{equation}
where $\underline{\hvec{S}}_N$ and $\hvec{A}_N$ are the estimates of the mean-removed endmembers and abundances returned from $N$-material HUT-AMP, $\hvec{\psi}_{\text{ML}}$ is the ML estimate of the noise variance, and $\gamma(\cdot)$ is a penalty term.
As recommended in\cite{Parker:TSP:14b}, we choose $\gamma(\cdot)$ in accordance with the small-sample-corrected Akaike information criterion (AICc)\cite{Stoica:SPM:04}, i.e., $\gamma(N)=2\frac{MT}{MT-n(N)-1}n(N)$, where $MT$ is the number of scalar observations in $\vec{Y}$ and $n(N)$ is the number of scalar degrees-of-freedom (DoF) in our model, which depends on $N$.
In particular, $n(N)$ comprises $MN$ DoF from $\vec{S}$, $(N-1)T$ DoF from $\vec{A}$, and $5N+2NL+N(L-1)+M$ DoF from $\vec{\Omega}$.
Plugging the standard form of the ML estimate of $\psi$ (see, e.g.,\cite[eq. (7)]{Stoica:SPM:04}) into \eqref{AICc}, we obtain 
\begin{equation}
\label{eq:MOS}
\hat{N} \!=\! \argmax_{N} -MT\ln\!\left(\!\frac{\norm{\vec{Y} \!-\!\hvec{S}_N\hvec{A}_N}^2_F}{MT}\!\right) - \frac{2MT n(N)}{MT \!-\! n(N) \!-\!1} .
\end{equation}

To solve the maximization in \eqref{MOS}, we first run $N=2$ HUT-AMP to completion and compute the penalized log-likelihood.
We then increment $N$ by $1$, and compute the penalized log-likelihood again.  
If it increases, we increment $N$ by $1$ and repeat the procedure.
Once the penalized log-likelihood decreases, we stop the procedure and select the previous model order $N$, which is the local maximizer of the penalized log-likelihood.
We refer to the resulting procedure as ``HUT-AMP with model-order selection'' (HUT-AMP-MOS). 

We also note that a similar model-order selection strategy can be implemented to tune the number of NNGM components $L$ used in \eqref{nu}, and we refer interested readers to\cite{Vila:TSP:13} for more details.
We note, however, that the fixed choice $L=3$ was sufficient to yield the excellent results in \secref{results}.

\section{Numerical Results}  \label{sec:results}

In this section, we report the results of several experiments that we conducted to characterize the performance of our proposed methods on both synthetic and real-world datasets.

In these experiments, we compared the endmembers $\hvec{S}$ recovered from our proposed HUT-AMP and HUT-AMP-MOS\footnote{Matlab code can be found at \url{http://www.ece.osu.edu/~schniter/HUTAMP}.} unmixing algorithms to those recovered by the Bayesian unmixing algorithm SCU\cite{Mittelman:TSP:12}; the sparse NMF techniques L$\tfrac{1}{2}$NMF \cite{Qian:TGRS:11} and SDSNMF \cite{Yuan:TGRS:15}; and the endmember extraction (EE) algorithms VCA\cite{Nasc:TGRS:05}, FSNMF\cite{Gillis:TPAMI:14}, and MVSA\cite{Li:IGARSS:08}.

We also compared the abundances $\hvec{A}$ recovered by our proposed HUT-AMP and HUT-AMP-MOS unmixing algorithms to those recovered by SCU and SDSNMF, as well as those recovered by FCLS \eqref{FCLS} (implemented via Matlab's \texttt{lsqlin}) and SUnSAL-TV\cite{Iordache:TGRS:12} using the endmember estimates produced by VCA, FSNMF, and MVSA.
We note that SCU, SDSNMF, and SUnSAL-TV all exploit spatial coherence, and that SDSNMF is in fact L$\tfrac{1}{2}$NMF with additional mechanisms to exploit spatial coherence.

In all cases, algorithms were run using their authors' implementation and suggested default settings, unless noted otherwise.
The only exceptions are SDSNMF and L$\tfrac{1}{2}$NMF, which we implemented in MATLAB since their authors declined to provide source code. 
All algorithms (with the exception of HUT-AMP-MOS) were supplied the true number of materials $N$ in each experiment.
For SUnSAL-TV, the regularization weights for the $\ell_1$ and TV norms were hand-tuned, because cross-validation tuning was too computationally expensive given the sizes of the datasets.
For FSNMF, we used the PCA post-processing described in footnote~\ref{FSNMF} to reduce the effects of measurement noise, since this greatly improved its mean-squared estimation error.

\subsection{Pixel Purity versus Abundance Sparsity} \label{sec:mixed}

Our first experiment aims to assess EE performance as a function of pixel purity and abundance sparsity.
Our motivation stems from the fact that 
the proposed HUT-AMP algorithm aims to exploit \emph{sparsity} in the columns of the abundance matrix $\vec{A}$,
while
classical EE techniques like VCA and FSNMF aim to exploit the presence of \emph{pure pixels}, 
recalling the discussion in \secref{intro}. 
Thus, we are interested in seeing how these contrasting approaches fare under varying combinations of pixel purity and abundance sparsity.
We also compare against the minimum-volume-simplex approach from \cite{Li:IGARSS:08}, which is an alternative to both pixel purity and abundance sparsity.

We first constructed synthetic data consisting of 
$M\!=\!100$ spectral bands, $T\!=\!115$ spatial pixels, and $N\!=\!10$ materials. 
The endmember matrix $\vec{S}\in\Real_+^{M\times N}$ was drawn i.i.d such that $\S_{mn} \sim \mc{N}_+(0.5,0.05)$.
The abundance matrix $\vec{A}\in\Real_+^{N\times T}$ was generated as shown in \figref{NNMFdiag}, where $P$ of the columns of $\vec{A}$ were assigned (uniformly at random) to be pure pixels, and the remaining columns were drawn $K$-sparse on the simplex.
In particular, for each of these latter columns, the support was drawn uniformly at random, and the non-zero values $\{\underline{a}_k\}_{k=1}^{K}$ were drawn from a Dirichlet distribution, i.e., 
\begin{subequations} \label{eq:Dirichlet}
\begin{eqnarray}
&&p(\underline{a}_1,\dots,\underline{a}_{K-1}) 
= \begin{cases} \frac{\Gamma(\alpha K)}{\Gamma(\alpha)^K}\prod_{k=1}^K \underline{a}_k^{\alpha-1},
&\underline{a}_k\in [0,1] \\ 
0 & \text{else} \end{cases}\qquad \\
&&p(\underline{a}_K|\underline{a}_1,\dots,\underline{a}_{K-1})
= \delta(1-\underline{a}_1-\dots-\underline{a}_{K}) ,
\end{eqnarray}  
\end{subequations}
where $\Gamma(\cdot)$ denotes the gamma function, with concentration parameter $\alpha = 1$.
Finally, the observation matrix $\vec{Y}$ was created by adding white Gaussian noise $\vec{W}$ to $\vec{Z}=\vec{SA}$, where the noise variance $\psi$ was adjusted to achieve $\textsf{SNR}\!\defn\!\frac{1}{MT}\|\vec{Z}\|_F^2/\psi\!=\!80$~dB.

\putFrag{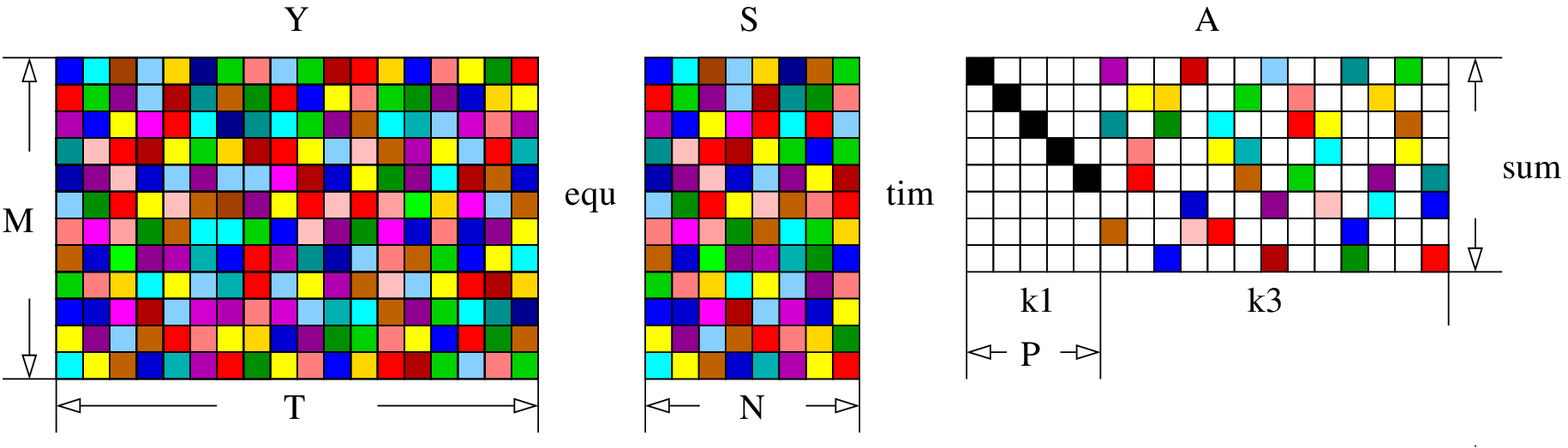}
{Illustration of the non-negative endmember matrix $\vec{S}$ and the $K$-sparse $P$-pure abundance matrix $\vec{A}$ for the first experiment.}
{3.5}
{
\newcommand{\sz}{0.8}
\psfrag{Y}[c][c][\sz]{$\vec{Y}$}
\psfrag{S}[c][c][\sz]{$\vec{S}$}
\psfrag{A}[c][c][\sz]{$\vec{A}$}
\psfrag{M}[c][c][\sz]{$M$}
\psfrag{T}[c][c][\sz]{$T$}
\psfrag{N}[c][c][\sz]{$N$}
\psfrag{P}[c][c][\sz]{$P$}
\psfrag{k1}[c][c][0.7]{\sf pure}
\psfrag{k3}[c][c][0.7]{$K\!=\!3$}
\psfrag{equ}[c][c][\sz]{$\vec{=}$}
\psfrag{tim}[c][c][\sz]{$\vec{\times}$}
\psfrag{sum}[c][c][0.65]{$\norm{\vec{a}}_1 \!=\! 1$}
}

\begin{figure}[t]
\newcommand{\pwid}{1.6 in}
\newcommand{\fwid}{1.6 in}
  \begin{minipage}{\pwid}
        \newcommand{\sz}{0.8}
 		\psfrag{Sparsity}[][][\sz]{Sparsity $K$}
 		\psfrag{Purepixels}[t][][\sz]{\sf Pixel purity $P$}
    \hspace{6 mm} (a) HUT-AMP\\
    \includegraphics[width=\fwid]{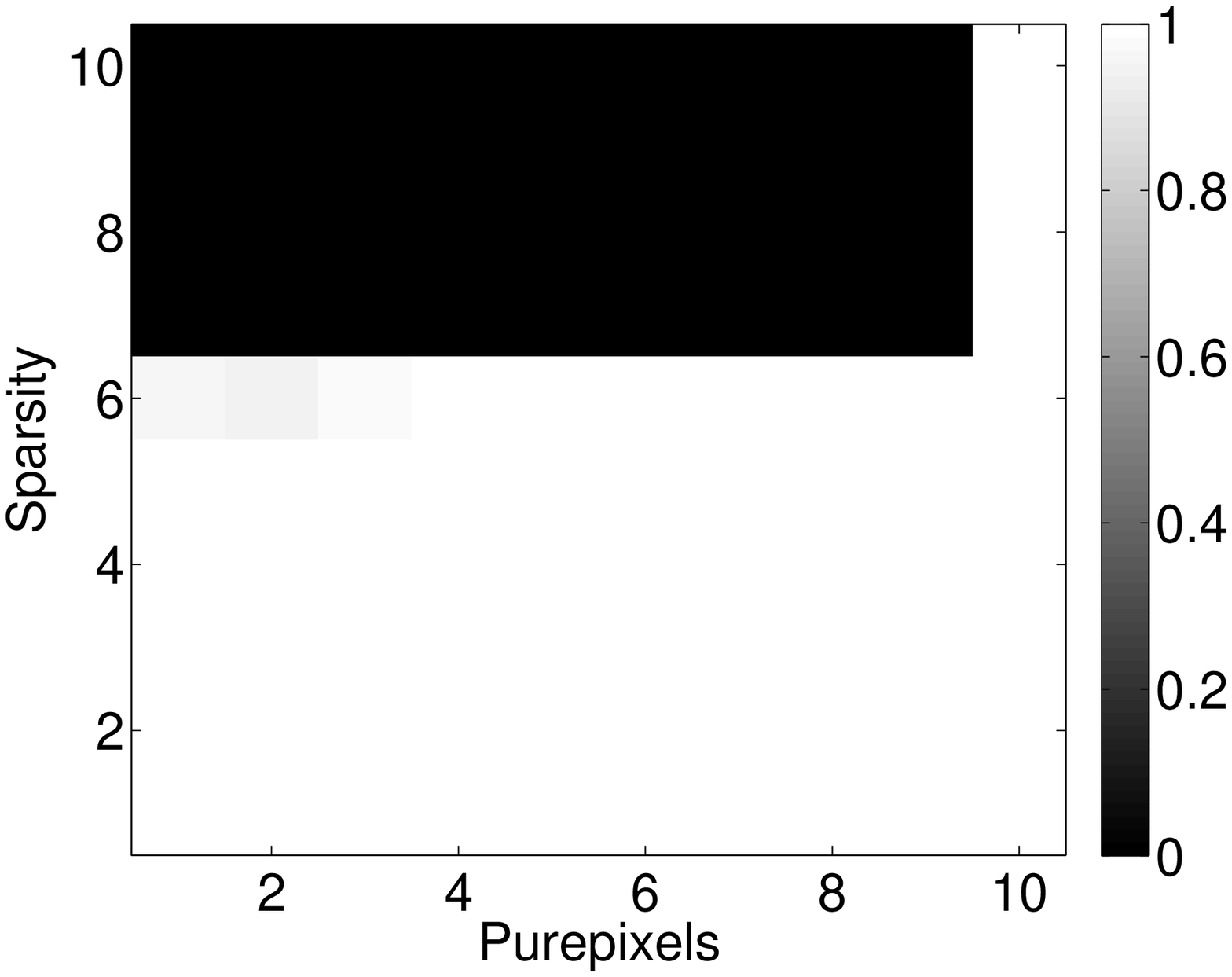}
  \end{minipage}
  \begin{minipage}{\pwid}
        \newcommand{\sz}{0.8}
 		\psfrag{Sparsity}[][][\sz]{Sparsity $K$}
 		\psfrag{Purepixels}[t][][\sz]{\sf Pixel purity $P$}
    \hspace{10 mm} (b) MVSA\\
    \includegraphics[width=\fwid]{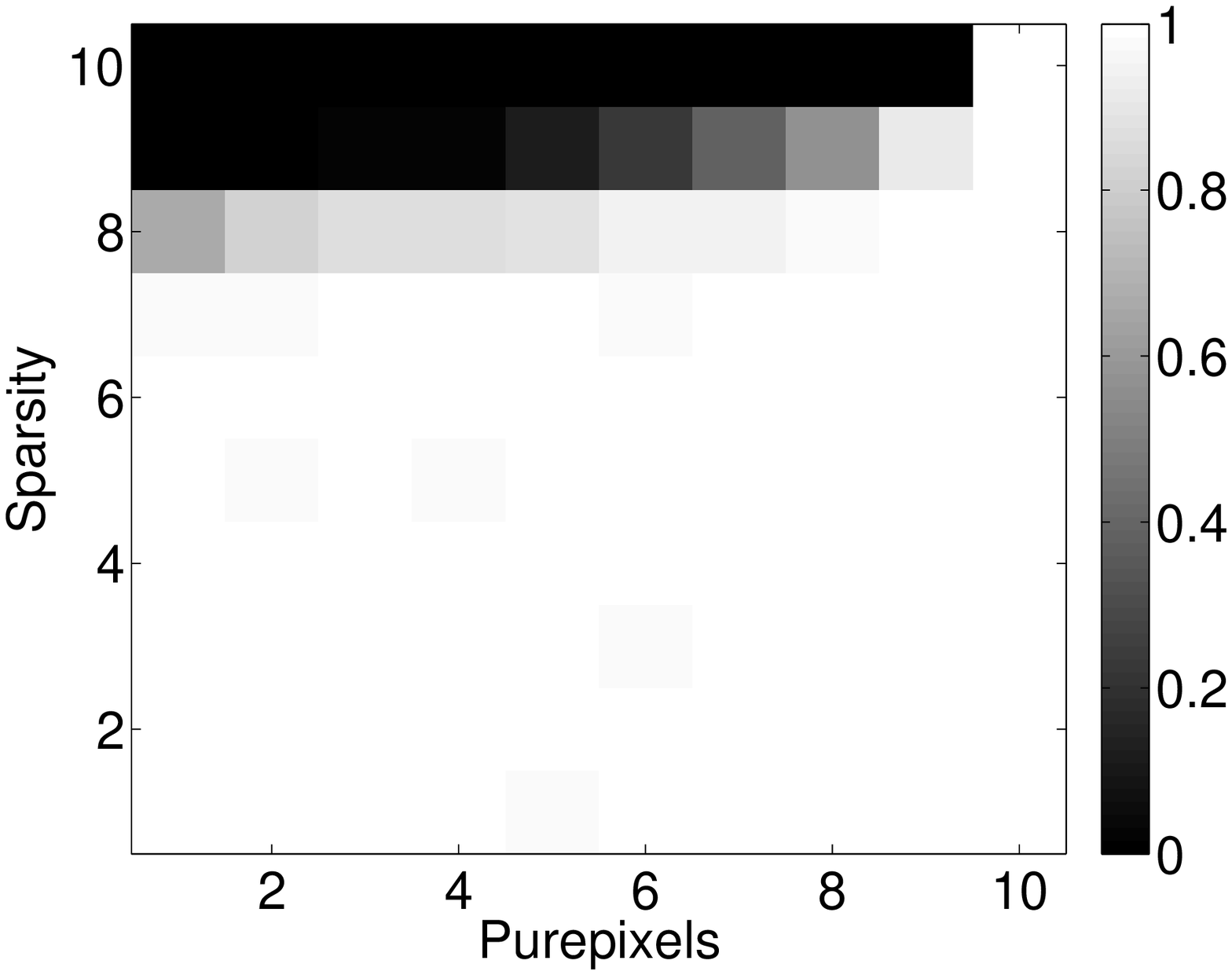}
  \end{minipage} \\[3mm]
  \begin{minipage}{\pwid}
           \newcommand{\sz}{0.8}
 		\psfrag{Sparsity}[][][\sz]{Sparsity $K$}
 		\psfrag{Purepixels}[t][][\sz]{\sf Pixel purity $P$}
    \hspace{12 mm} (c) VCA\\
    \includegraphics[width=\fwid]{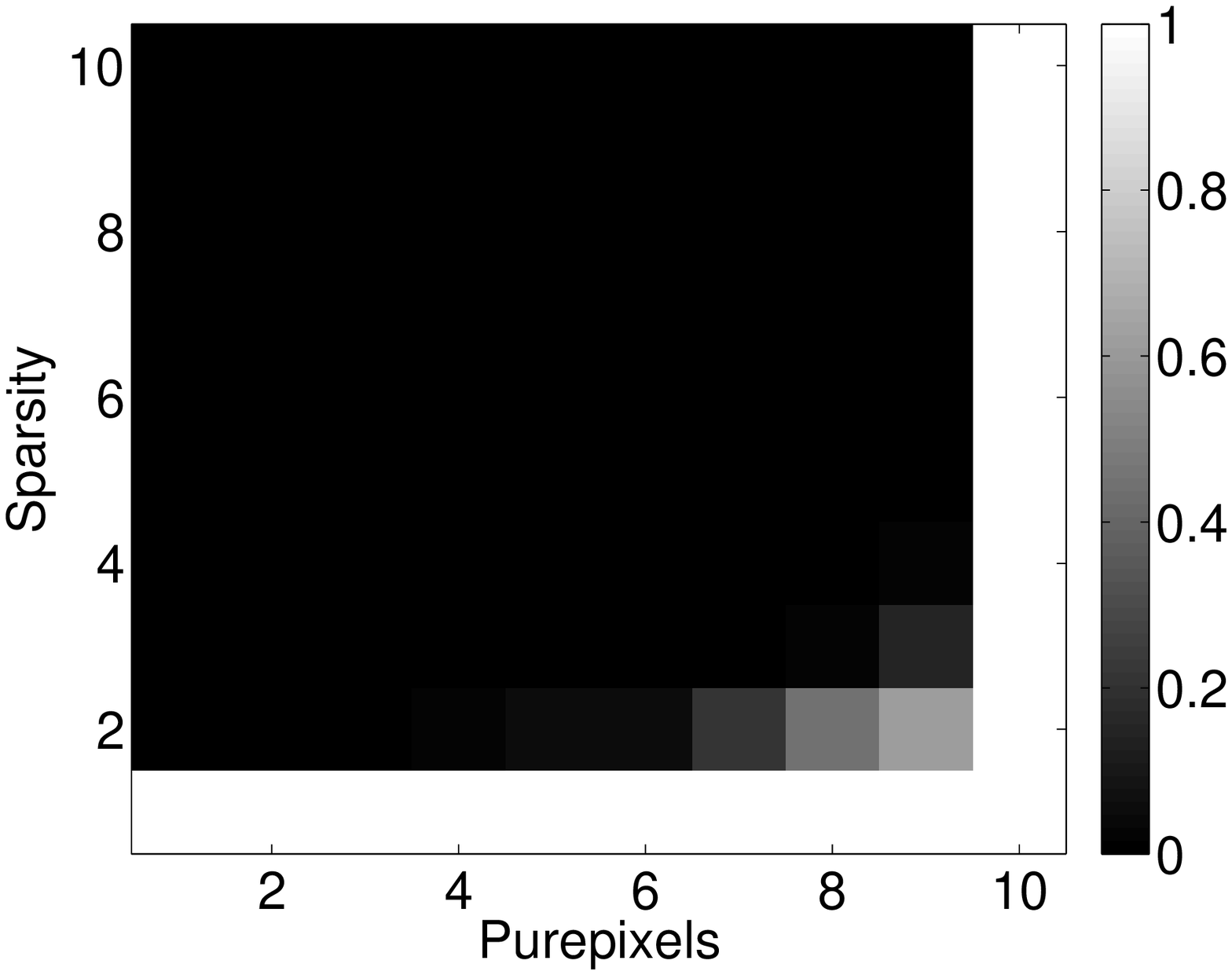}
  \end{minipage}
    \begin{minipage}{\pwid}
        \newcommand{\sz}{0.8}
 		\psfrag{Sparsity}[][][\sz]{Sparsity $K$}
 		\psfrag{Purepixels}[t][][\sz]{\sf Pixel purity $P$}
    \hspace{9 mm} (d) L$\tfrac{1}{2}$NMF\\
    \includegraphics[width=\fwid]{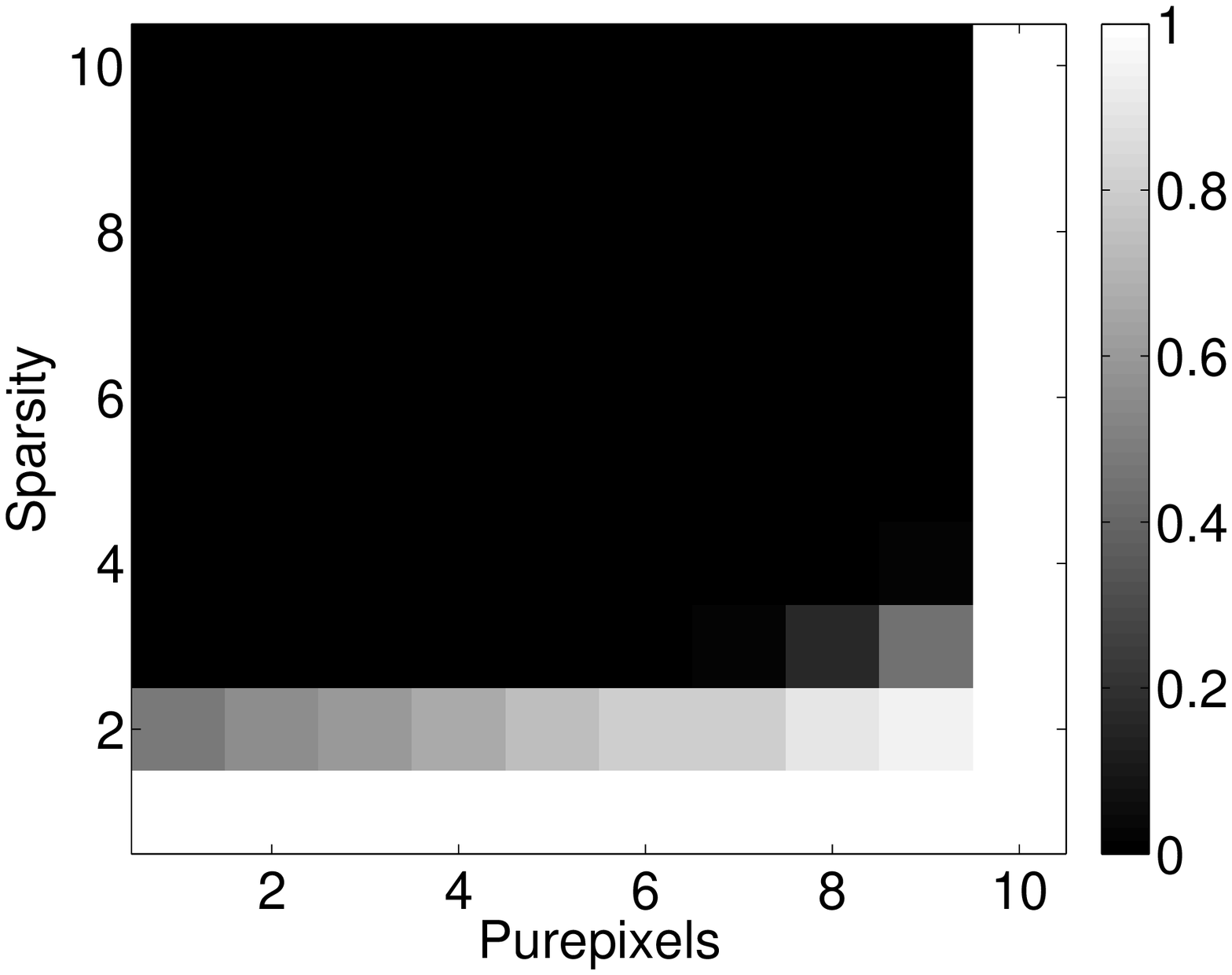}
  \end{minipage} \\[2 mm]
  \caption{First experiment: Average success rate for near-perfect recovery of i.i.d endmembers $\vec{S}$ and $K$-sparse and $P$-pure abundances $\vec{A}$ using (a) HUT-AMP, (b) MVSA, (c) VCA, and (d) L$\tfrac{1}{2}$NMF.}
  \label{fig:PTC}
\end{figure}

\Figref{PTC} shows empirical success probability averaged over $R = 100$ realizations, as a function of pixel purity $P$ and sparsity $K$, for the HUT-AMP, MVSA, VCA, and L$\tfrac{1}{2}$NMF algorithms.\footnote{%
Since our experimental findings into sparsity-versus-purity would be biased if the algorithms under test used different approaches to the exploitation of spatial and/or spectral coherence, we turn off the coherence-exploiting mechanisms in HUT-AMP and SDSNMF (reducing the latter to L$\tfrac{1}{2}$NMF) and compare to other algorithms that do not exploit spatial or spectral coherence: VCA and MVSA.}
It does not show FSNMF since its performance was indistinguishable from VCA's performance.
Here, a recovery was considered successful if $\NMSE_S \defn \norm{\vec{S} - \hat{\vec{S}}}^2_F/\norm{\vec{S}}^2_F <-40$~dB.  
As seen in \figref{PTC}(c) and \figref{PTC}(d), VCA and FSNMF were only successful for the $K\!=\!1$ and $P\!=\!10$ cases, i.e., the pure-pixel cases.  
L$\tfrac{1}{2}$NMF did slightly better, with successful recovery for $K \leq 2$.
HUT-AMP, on the other hand, was able to successfully recover the endmembers for $K \leq 6$-sparse abundances, even when there was only $P\!=\!1$ pure-pixels available.
We attribute HUT-AMP's improved performance to its exploitation of sparsity rather than pure pixels (as with VCA and FSNMF), and its ability to accurately learn the underlying sparsity rate.
Also, we conjecture that sparsity (i.e., $K\!>\!1$ and $P\!<\!N$) is more important in practice, since the spatial resolution of the hyperspectral sensors may not guarantee pixel-purity for all materials, while sparse abundances (i.e., $K\!\ll\!N$) are more likely to hold.
Finally, we note that, although MVSA performed remarkably well in this experiment, it performed relatively poorly for the experiments in \secref{pure} through \secref{Cuprite}.

\begin{figure}[t]
\newcommand{\pwid}{1.6 in}
\newcommand{\fwid}{1.6 in}
  \begin{minipage}{\pwid}
        \newcommand{\sz}{0.8}
 		\psfrag{Sparsity}[][][\sz]{Sparsity $K$}
 		\psfrag{Purepixels}[t][][\sz]{\sf Pixel purity $P$}
    \hspace{6 mm} (a) HUT-AMP\\
    \includegraphics[width=\fwid]{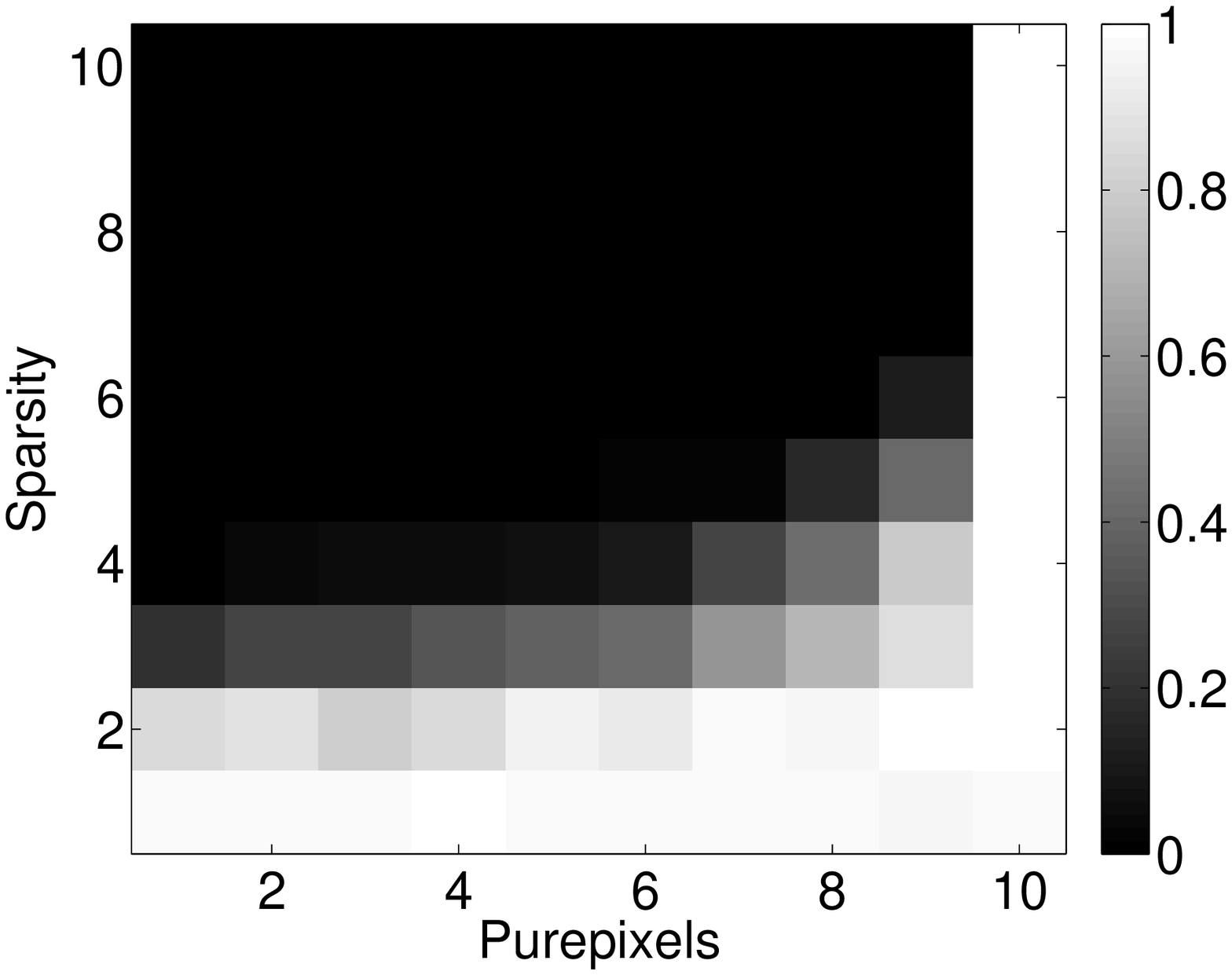}
  \end{minipage}
  \begin{minipage}{\pwid}
        \newcommand{\sz}{0.8}
 		\psfrag{Sparsity}[][][\sz]{Sparsity $K$}
 		\psfrag{Purepixels}[t][][\sz]{\sf Pixel purity $P$}
    \hspace{10 mm} (b) MVSA\\
    \includegraphics[width=\fwid]{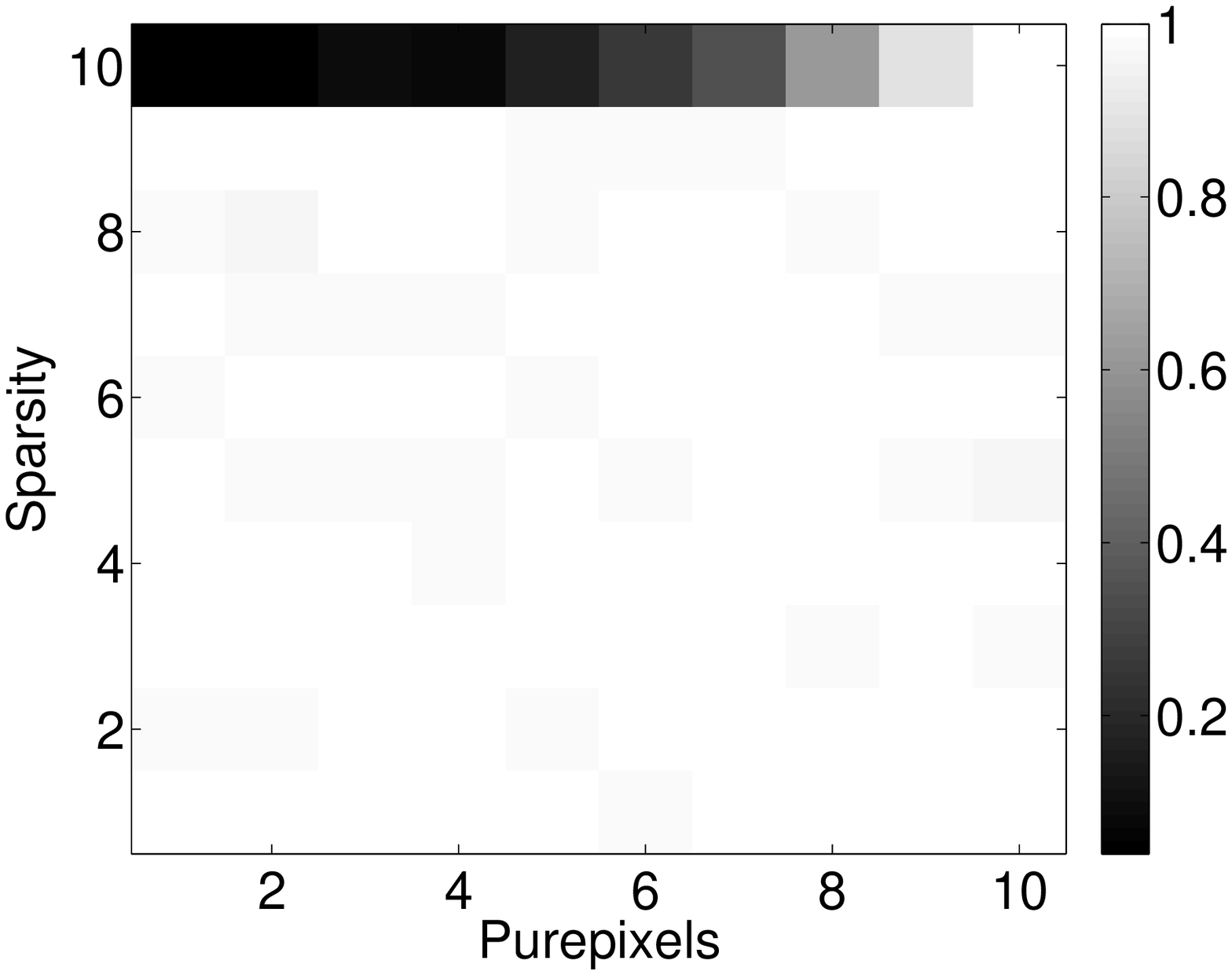}
  \end{minipage} \\[3mm]
  \begin{minipage}{\pwid}
           \newcommand{\sz}{0.8}
 		\psfrag{Sparsity}[][][\sz]{Sparsity $K$}
 		\psfrag{Purepixels}[t][][\sz]{\sf Pixel purity $P$}
    \hspace{12 mm} (c) VCA\\
    \includegraphics[width=\fwid]{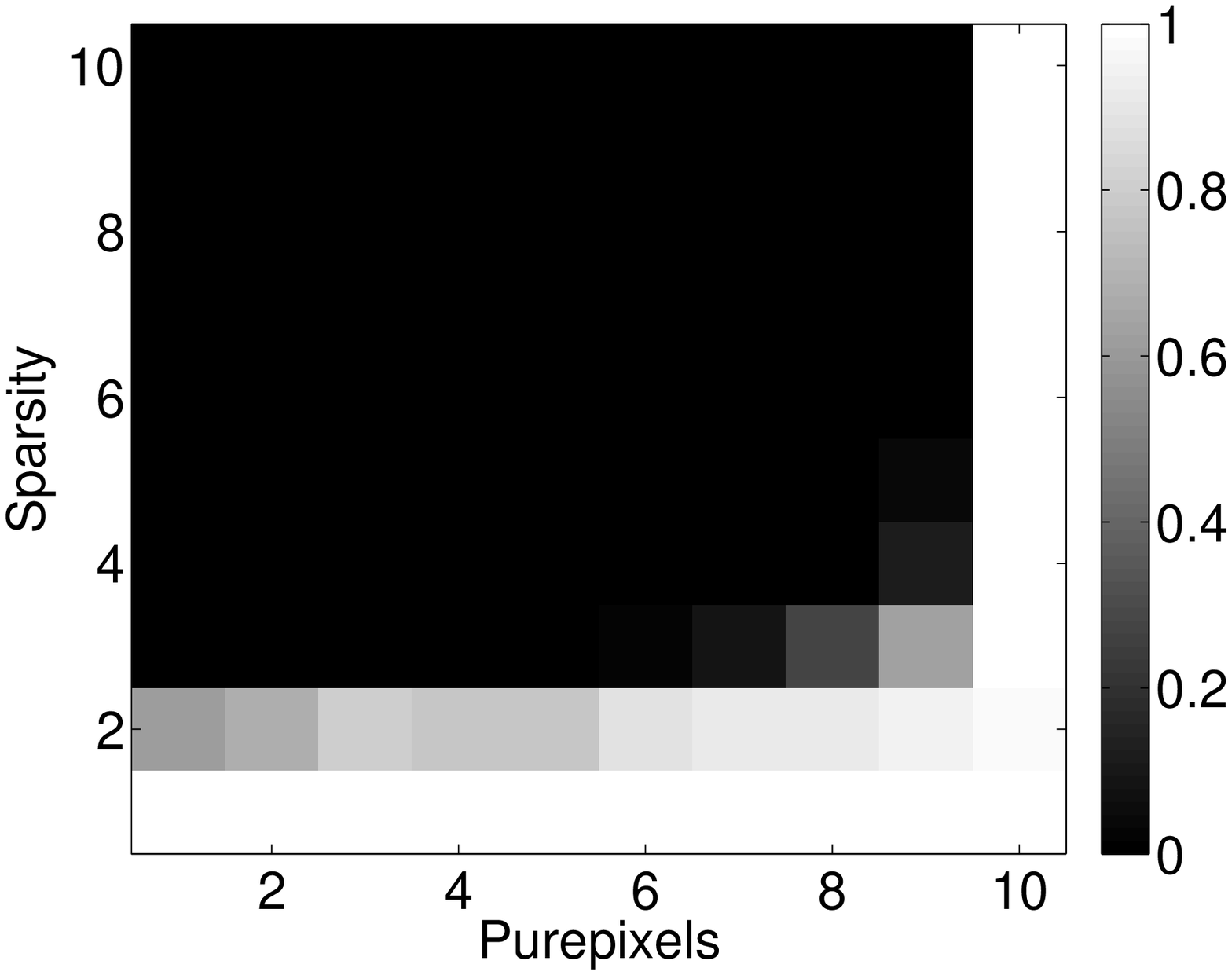}
  \end{minipage}
    \begin{minipage}{\pwid}
        \newcommand{\sz}{0.8}
 		\psfrag{Sparsity}[][][\sz]{Sparsity $K$}
 		\psfrag{Purepixels}[t][][\sz]{\sf Pixel purity $P$}
    \hspace{9 mm} (d)  L$\tfrac{1}{2}$NMF\\
    \includegraphics[width=\fwid]{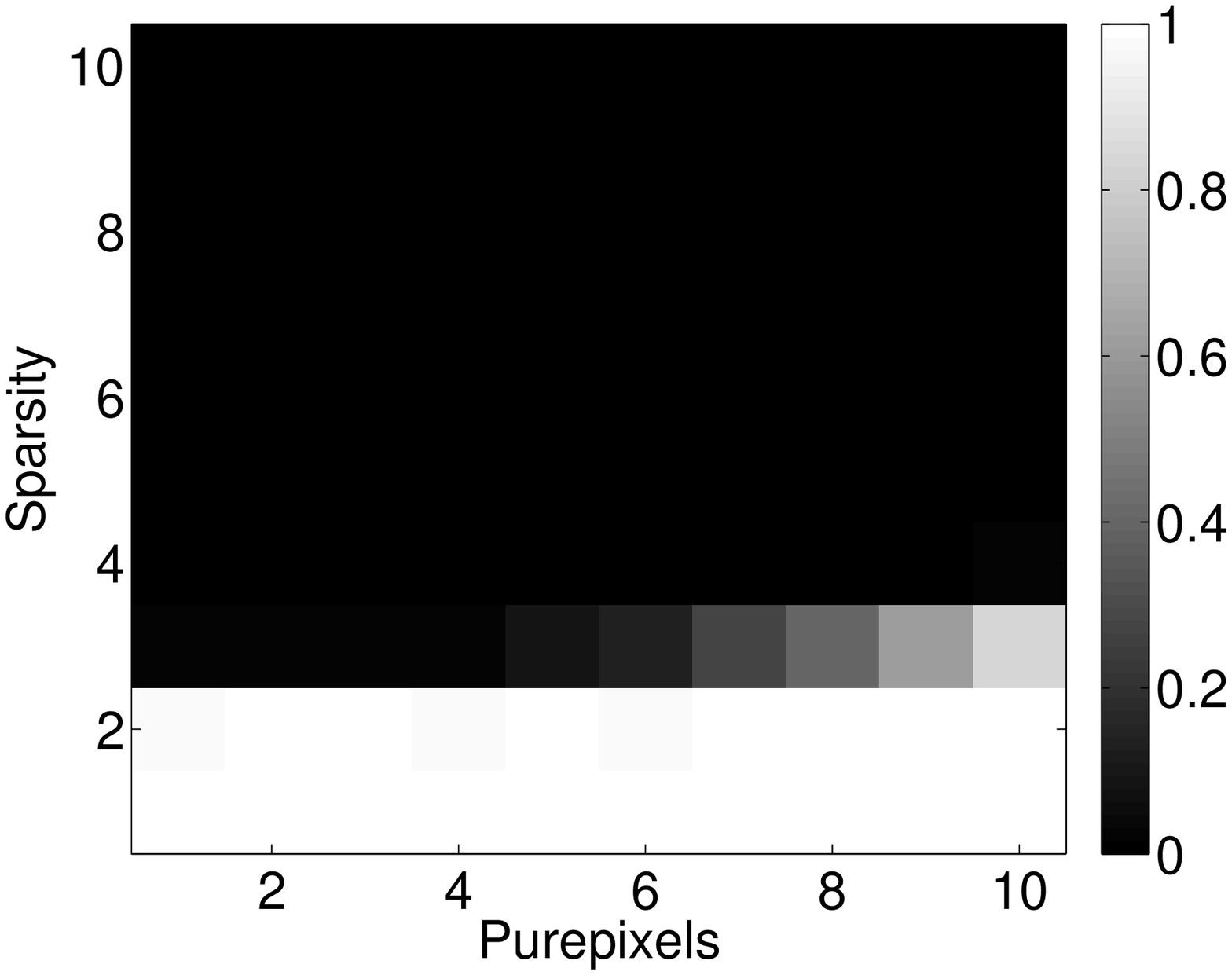}
  \end{minipage} \\[2 mm]
  \caption{Second experiment: Average success rate for near-perfect recovery of USGS endmembers $\vec{S}$ and $K$-sparse and $P$-pure abundances $\vec{A}$ using (a) HUT-AMP, (b) MVSA, (c) VCA, and (d) L$\tfrac{1}{2}$NMF.}
  \label{fig:USGSPTC}
\end{figure}

Next, we repeat the previous experiment at $\SNR\!=\! 60$~dB with $\vec{S}$ randomly selected from the USGS Digital Spectral Library splib06a,\footnote{See
\url{http://speclab.cr.usgs.gov/spectral.lib06/ds231/}}
which contains laboratory-measured reflectance values for various materials over $M \!=\! 224$ spectral bands.
In particular, for each Monte-Carlo realization we randomly select $N\!=\!10$ endmembers from the library such that $\min_{i \neq j} \SAD(\vec{s}_i,\vec{s}_j) \geq 15$~ degrees, for \emph{spectral angle distance}
\begin{equation}
\label{eq:SAD}
\SAD(\vec{s}_i,\vec{s}_j) \defn \arccos\left(\frac{\vec{s}_i\tran \vec{s}_j}{\norm{\vec{s}_i}_2\norm{\vec{s}_j}_2}\right)
\end{equation}
\Figref{USGSPTC} shows the empirical success probability for the HUT-AMP, MVSA, VCA, and L$\tfrac{1}{2}$NMF algorithms.
Although HUT-AMP's performance with USGS endmembers is not as good as with i.i.d.\ endmembers, it still outperformed VCA and L$\tfrac{1}{2}$NMF.
As before, MVSA has the best performance.

\begin{figure}[t]
\newcommand{\pwid}{1.6 in}
\newcommand{\fwid}{1.6 in}
  \begin{minipage}{\pwid}
        \newcommand{\sz}{0.8}
 		\psfrag{log10alpha}[][][\sz]{\sf concentration $\log_{10}\alpha$}
 		\psfrag{N}[t][][\sz]{\sf \# materials $N$}
    \hspace{6 mm} (a) HUT-AMP\\
    \includegraphics[width=\fwid]{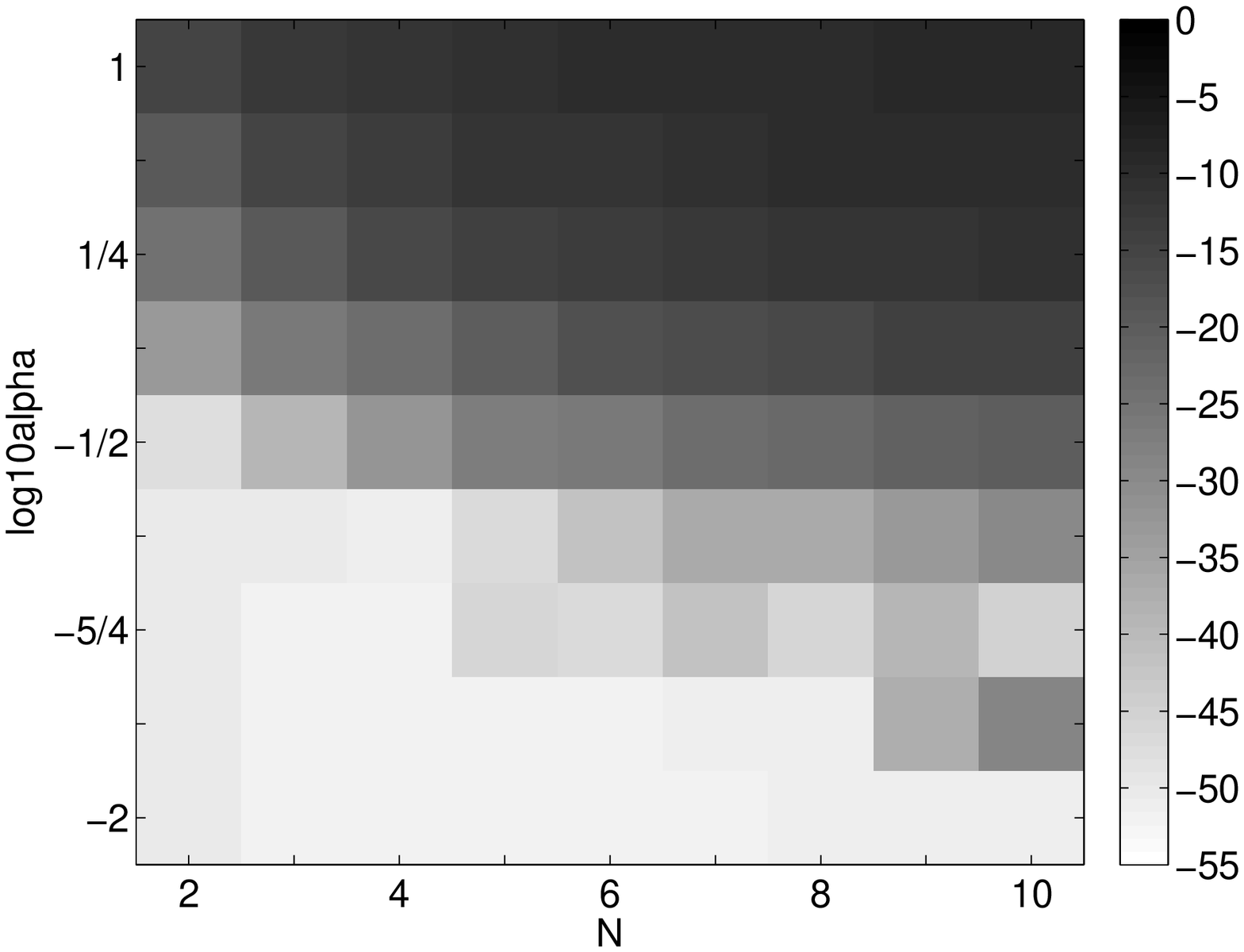}
  \end{minipage}
  \begin{minipage}{\pwid}
        \newcommand{\sz}{0.8}
 		\psfrag{log10alpha}[][][\sz]{\sf concentration $\log_{10}\alpha$}
 		\psfrag{N}[t][][\sz]{\sf \# materials $N$}
    \hspace{10 mm} (b) MVSA\\
    \includegraphics[width=\fwid]{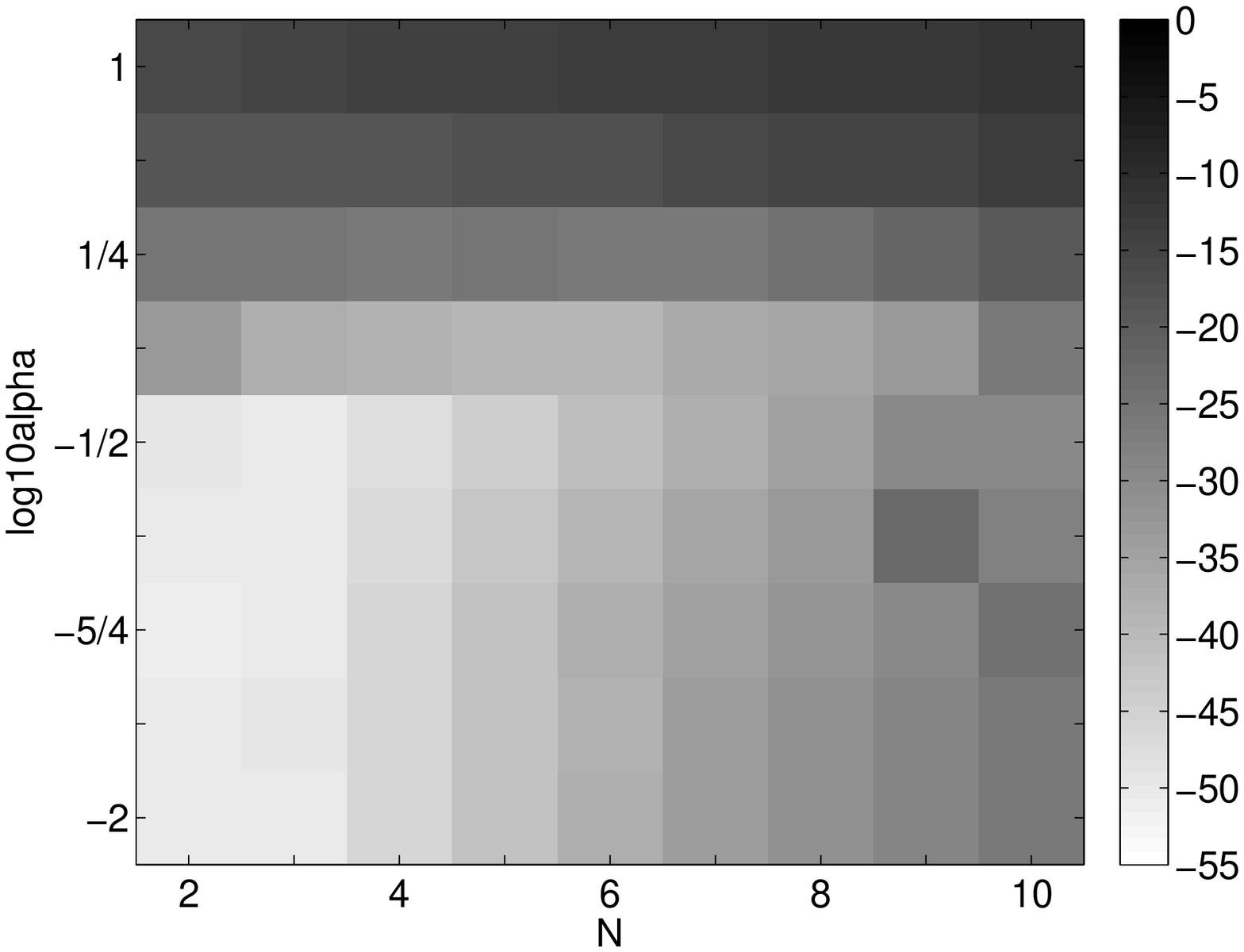}
  \end{minipage} \\[3mm]
  \begin{minipage}{\pwid}
           \newcommand{\sz}{0.8}
 		\psfrag{log10alpha}[][][\sz]{\sf concentration $\log_{10}\alpha$}
 		\psfrag{N}[t][][\sz]{\sf \# materials $N$}
    \hspace{12 mm} (c) VCA\\
    \includegraphics[width=\fwid]{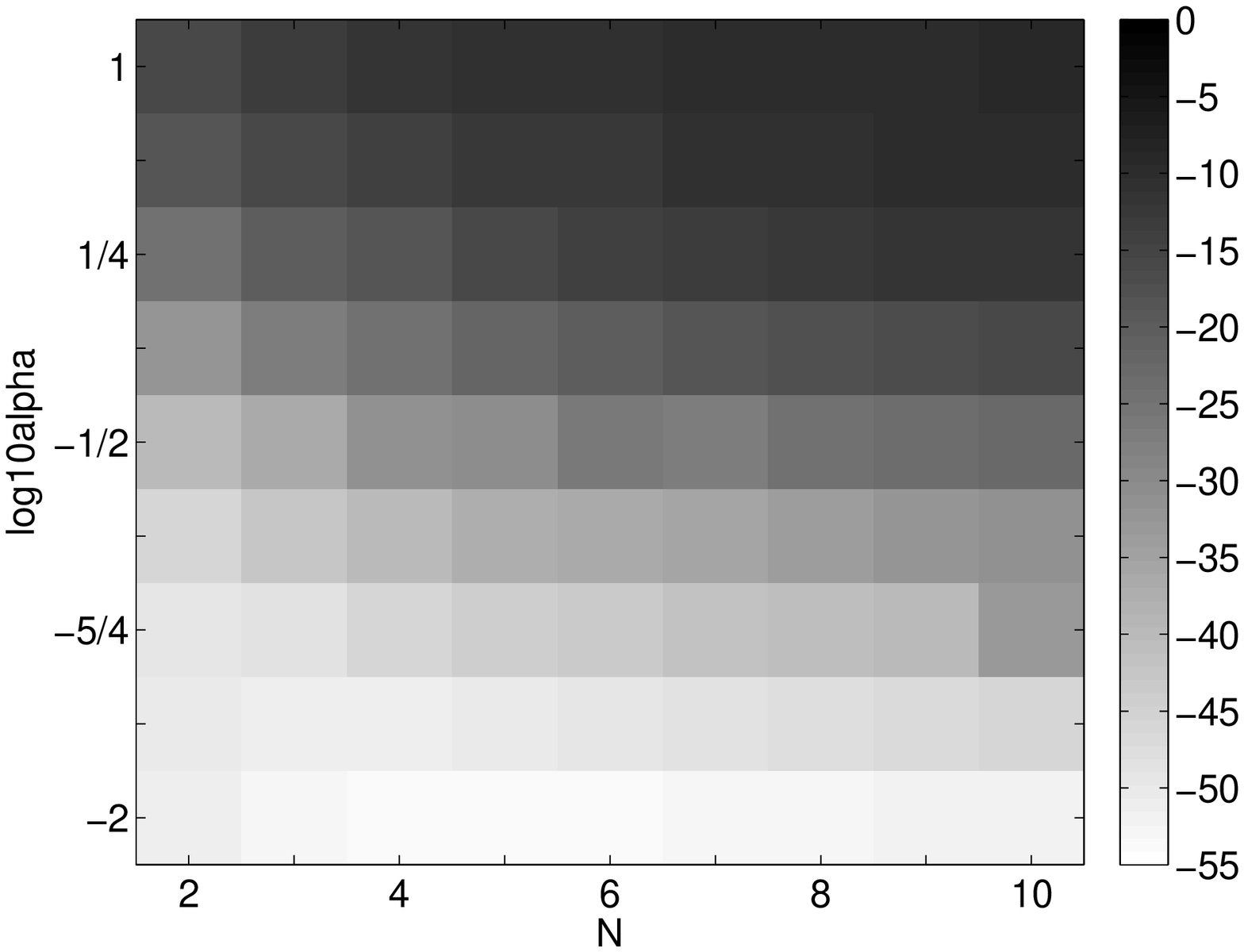}
  \end{minipage}
    \begin{minipage}{\pwid}
        \newcommand{\sz}{0.8}
 		\psfrag{log10alpha}[][][\sz]{\sf concentration $\log_{10}\alpha$}
 		\psfrag{N}[t][][\sz]{\sf \# materials $N$}
    \hspace{9 mm} (d)  L$\tfrac{1}{2}$NMF\\
    \includegraphics[width=\fwid]{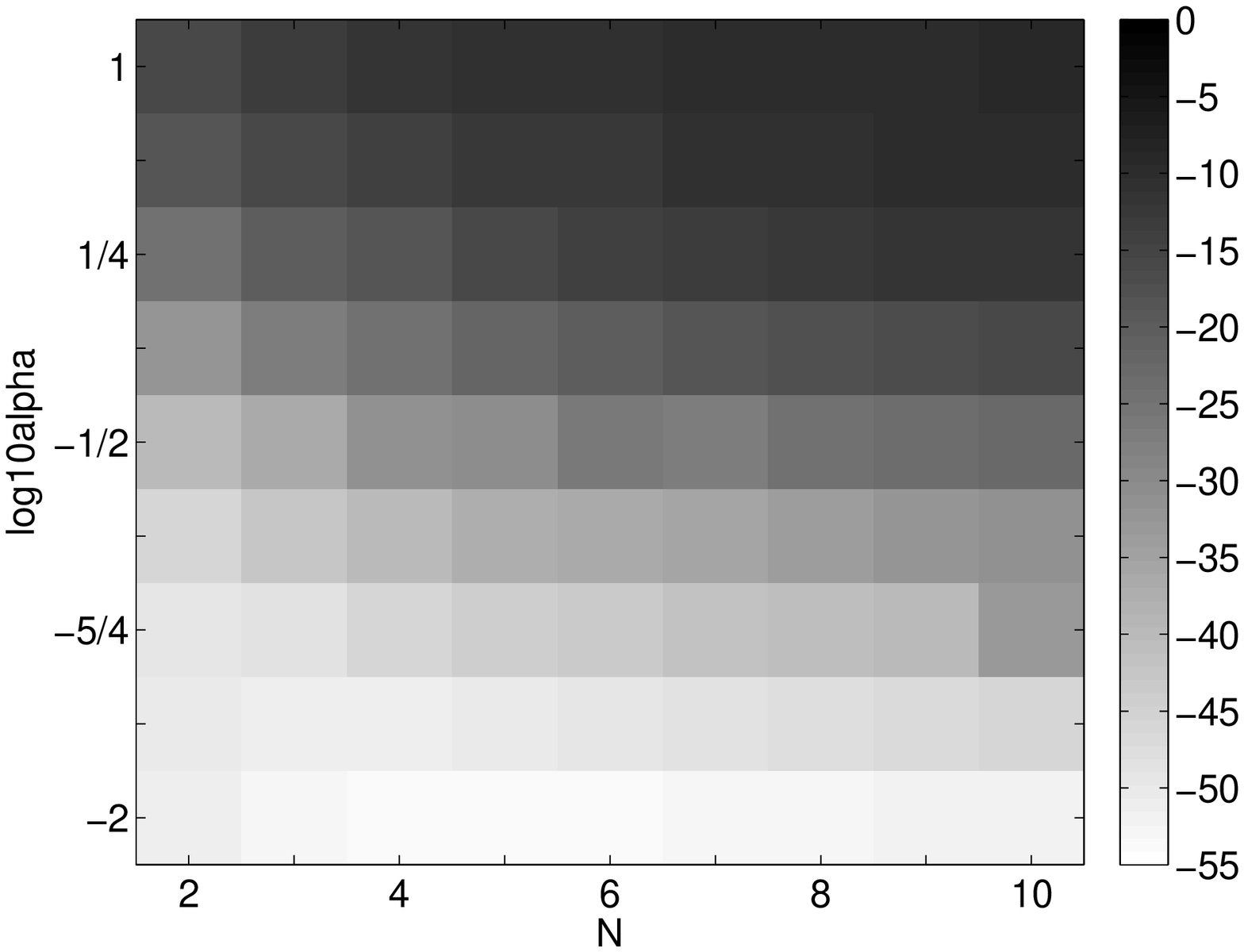}
  \end{minipage} \\[2 mm]
  \caption{Third experiment: Average $\NMSE_{\vec{S}}$ for recovery of $N$ USGS endmembers $\vec{S}$ with abundances $\vec{A}$ drawn from Dirichlet distribution with concentration $\alpha$ using (a) HUT-AMP, (b) MVSA, (c) VCA, and (d) L$\tfrac{1}{2}$NMF.}
  \label{fig:DIRPTC}
\end{figure}

Finally, we perform a variation on the previous experiment that again uses randomly selected USGS endmembers. 
But rather than using pure and/or $K$-sparse abundance vectors, it uses fully mixed abundances whose $N$ coefficients were generated from a Dirichlet distribution with concentration parameter $\alpha$ (recall \eqref{Dirichlet}).
Recall that larger values of $\alpha$ correspond to more dense mixing.
\Figref{DIRPTC} reports the average $\NMSE_{\vec{S}}$ of HUT-AMP, MVSA, VCA, and L$\frac{1}{2}$NMF versus both the number of materials, $N$, and the concentration parameter, $\alpha$, at $\SNR=40$ dB. 
The figure shows that HUT-AMP, VCA, and L$\frac{1}{2}$NMF gave similar performance overall, with small advantages to HUT-AMP when $N\leq 8$ and $\alpha\leq 10^{-13/8}$.
Relative to the other algorithms, MVSA tolerated higher values of $\alpha$, but was more sensitive to larger numbers of materials, $N$, when $\alpha$ was small.

\subsection{Pure-Pixel Synthetic Abundances} \label{sec:pure}

The second experiment uses synthetic pure-pixel abundances $\vec{A}$ with endmembers $\vec{S}$ chosen from the USGS Digital Spectral Library. 
To construct the data, we partitioned a scene of $T\!=\! 50\!\times\! 50$ pixels into $N\!=\!5$ equally sized vertical strips, each containing a single pure material.
We then selected endmembers corresponding to the materials Grossular, Alunite, well crystallized (wxl) Kaolinite, Hydroxyl-Apatite, and Amphibole, noting that similar results were obtained in experiments we conducted with other materials.
\Figref{HSI_pure_color_scaled} shows a false-color image constructed from the noiseless measurements $\vec{Z}$.
We then vary the $\SNR$ on a grid from $15$~dB to $35$dB by adding white Gaussian noise.

\putFrag{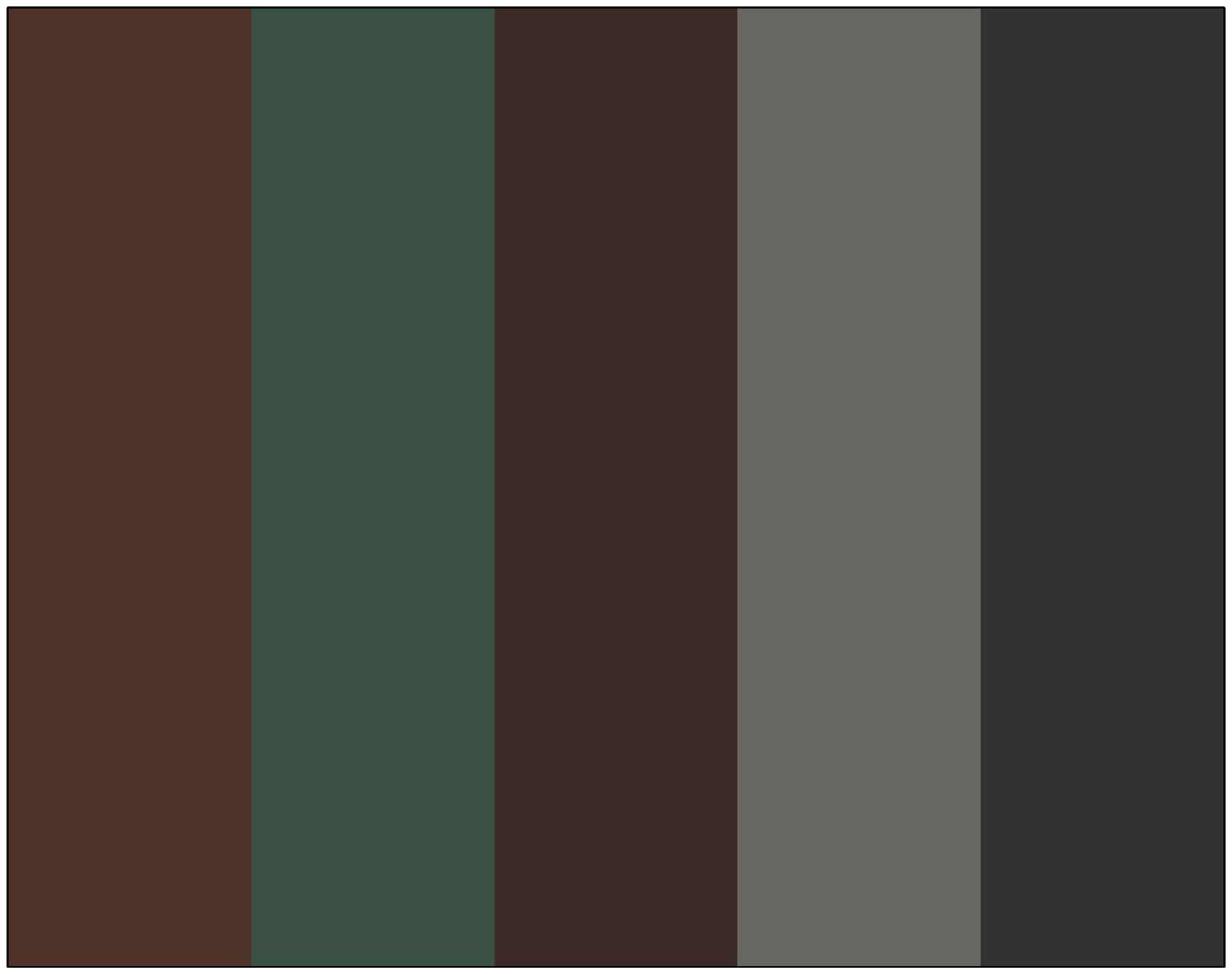}
{False-color image of the noiseless measurements $\vec{Z}$ used for the second experiment.  Since the pixels are pure, each strip shows the false color of one of the $N \!=\! 5$ materials. They are, in order from left to right: Grossular, Alunite, wxl Kaolinite, Hydroxyl-Apatite, and Amphibole.}
{2}
{}
 
Averaging over $r = 50$ realizations, Figures~\ref{fig:NMSES_pureSNR}~and~\ref{fig:NMSEA_pureSNR} show the normalized mean-squared error of the estimated endmembers and abundances, respectively (i.e., $\NMSE_{\vec{S}}$ and $\NMSE_{\vec{A}} \defn \norm{\vec{A} - \hat{\vec{A}}}^2_F/\norm{\vec{A}}^2_F$), while \figref{time_pureSNR} shows the average runtime versus $\SNR$.
For this pure-pixel dataset, these figures show HUT-AMP dominating the other algorithms in both endmember and abundance estimation accuracy at all $\SNR$s.
In particular, HUT-AMP outperformed the best competing techniques by $4$ to $12$~dB in $\NMSE_{\vec{S}}$ and as much as $90$~dB in $\NMSE_{\vec{A}}$.
We note that the biggest gains in $\NMSE_{\vec{A}}$ occurred when $\SNR \geq 24$~dB. 

\putFrag{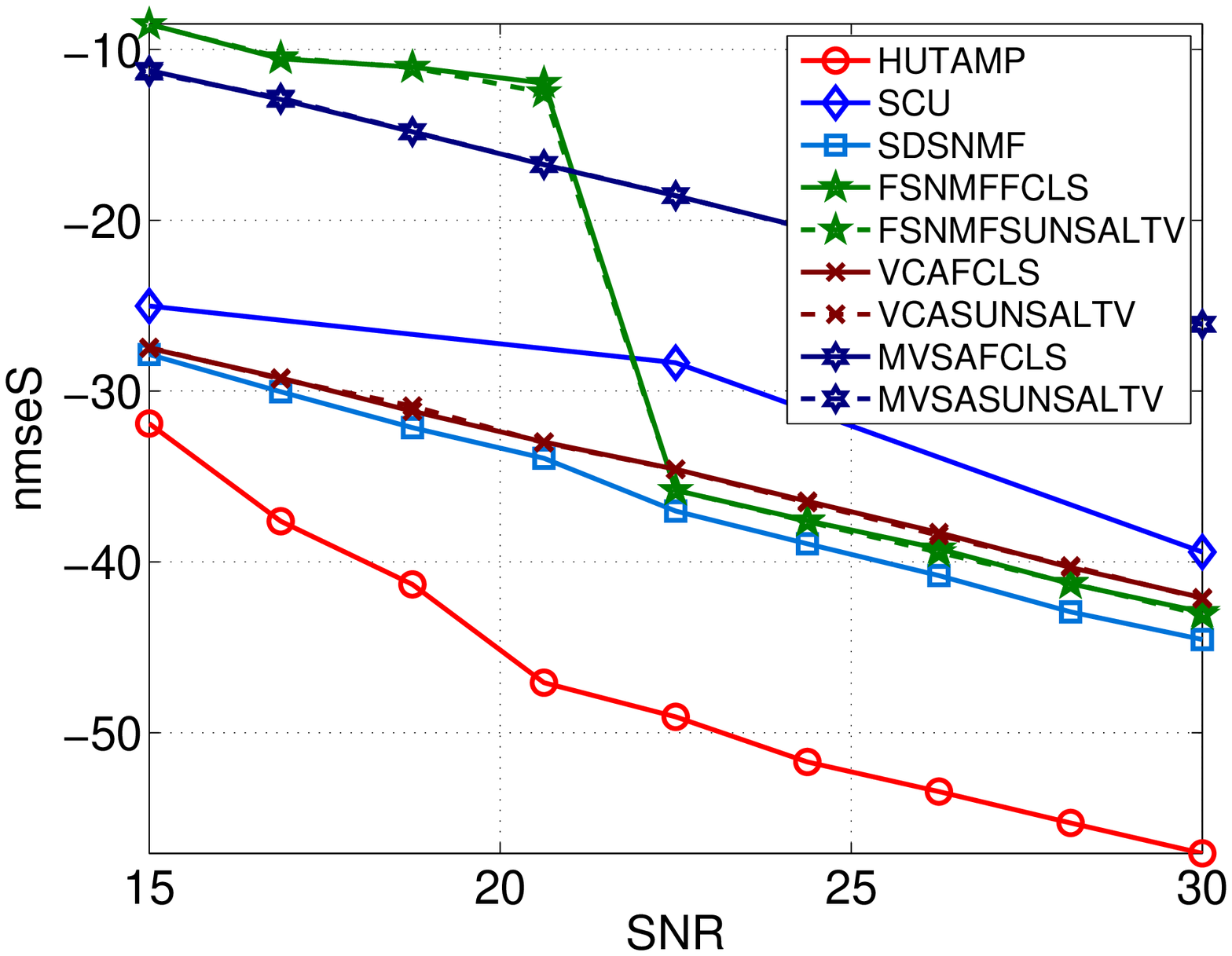}
{$\NMSE_{\vec{S}}$ vs.\ $\SNR$ for the synthetic pure pixel dataset.
}
{\figsize}
{
\newcommand{\sz}{0.65}
\newcommand{\szz}{0.9}
\psfrag{nmseS}[c][c][\szz]{\sf$\NMSE_{\vec{S}}$ [dB]}
\psfrag{SNR}[c][c][\szz]{\sf$\SNR$~ [dB]}
\psfrag{HUTAMP}[l][l][\sz]{\sf HUT-AMP}
\psfrag{VCAFCLS}[l][l][\sz]{\sf VCA+FCLS}
\psfrag{VCASUNSALTV}[l][l][\sz]{\sf VCA+SUnSAL-TV}
\psfrag{FSNMFFCLS}[l][l][\sz]{\sf FSNMF+FCLS}
\psfrag{FSNMFSUNSALTV}[l][l][\sz]{\sf FSNMF+SUnSAL-TV}
\psfrag{MVSAFCLS}[l][l][\sz]{\sf MVSA+FCLS}
\psfrag{MVSASUNSALTV}[l][l][\sz]{\sf MVSA+SUnSAL-TV}
\psfrag{SDSNMF}[l][l][\sz]{\sf SDSNMF}
\psfrag{SCU}[l][l][\sz]{\sf SCU}
}

\putFrag{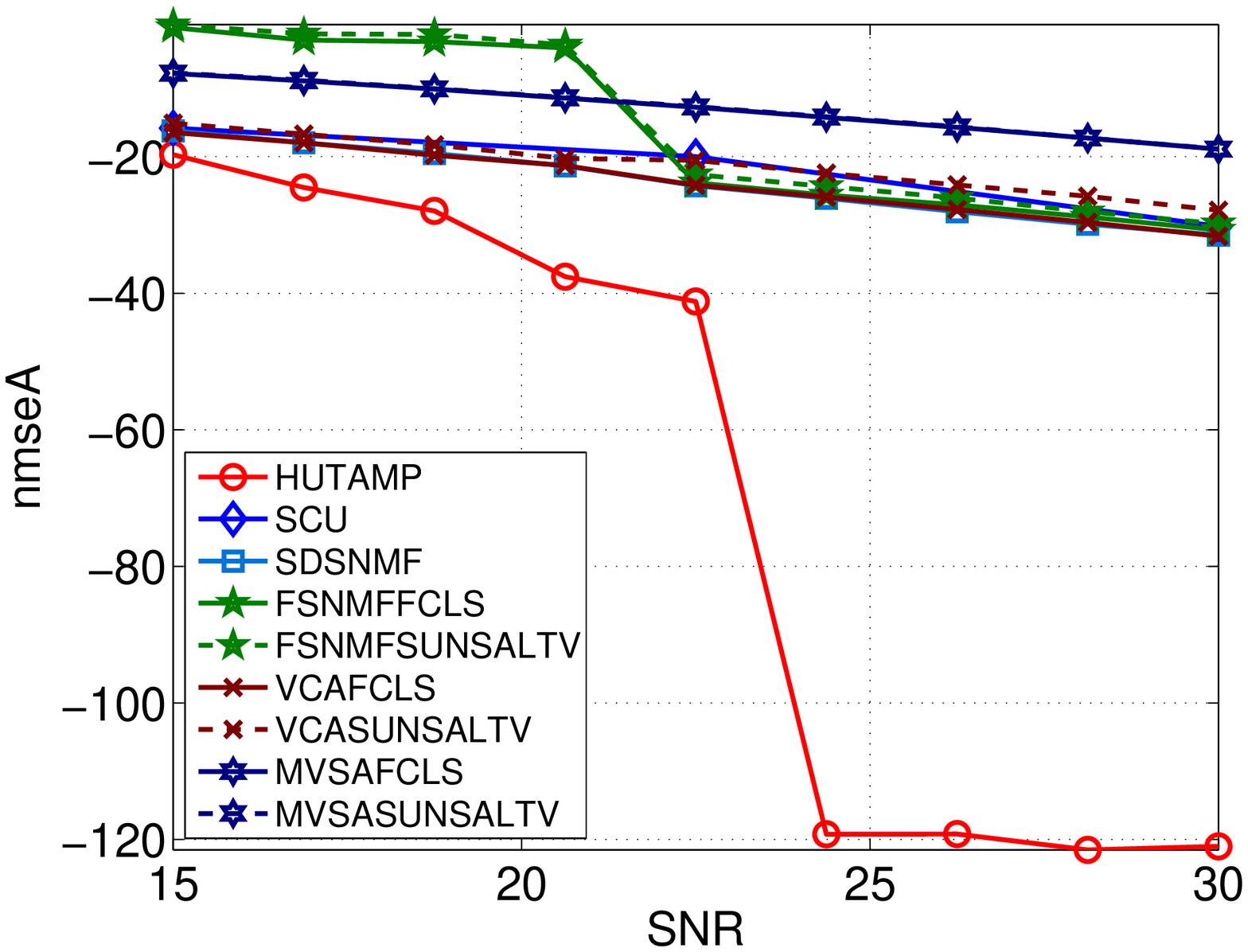}
{$\NMSE_{\vec{A}}$ vs.\ $\SNR$ for the synthetic pure pixel dataset.
}
{\figsize}
{
\newcommand{\sz}{0.65}
\newcommand{\szz}{0.9}
\psfrag{nmseA}[c][c][\szz]{\sf$\NMSE_{\vec{A}}$ [dB]}
\psfrag{SNR}[c][c][\szz]{\sf$\SNR$~ [dB]}
\psfrag{HUTAMP}[l][l][\sz]{\sf HUT-AMP}
\psfrag{VCAFCLS}[l][l][\sz]{\sf VCA+FCLS}
\psfrag{VCASUNSALTV}[l][l][\sz]{\sf VCA+SUnSAL-TV}
\psfrag{FSNMFFCLS}[l][l][\sz]{\sf FSNMF+FCLS}
\psfrag{FSNMFSUNSALTV}[l][l][\sz]{\sf FSNMF+SUnSAL-TV}
\psfrag{MVSAFCLS}[l][l][\sz]{\sf MVSA+FCLS}
\psfrag{MVSASUNSALTV}[l][l][\sz]{\sf MVSA+SUnSAL-TV}
\psfrag{SDSNMF}[l][l][\sz]{\sf SDSNMF}
\psfrag{SCU}[l][l][\sz]{\sf SCU}
}

\putFrag{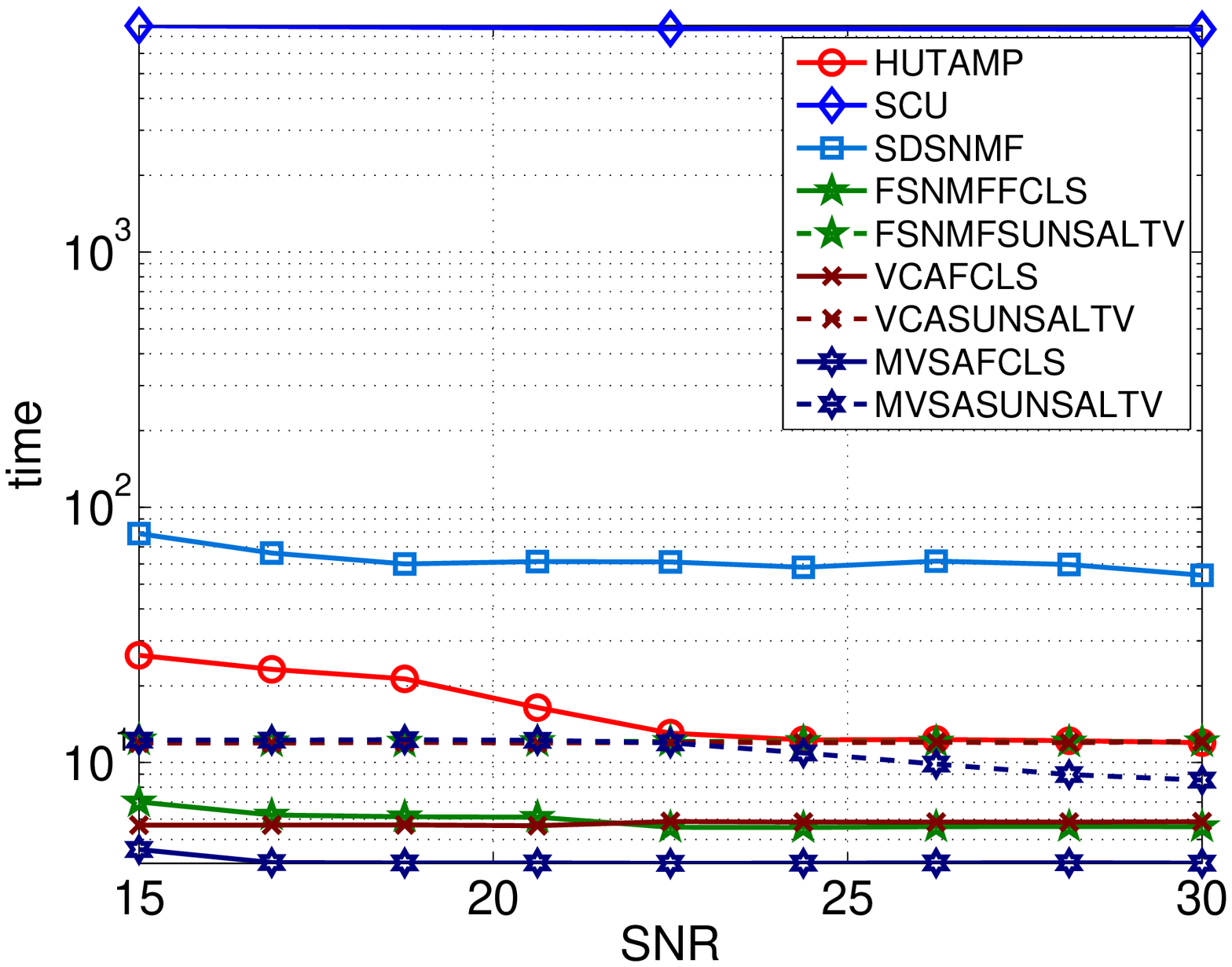}
{Runtime vs. $\SNR$ for the synthetic pure pixel dataset.
}
{\figsize}
{
\newcommand{\sz}{0.65}
\newcommand{\szz}{0.9}
\psfrag{time}[c][c][\szz]{\sf runtime [sec]}
\psfrag{SNR}[c][c][\szz]{\sf$\SNR$~ [dB]}
\psfrag{HUTAMP}[l][l][\sz]{\sf HUT-AMP}
\psfrag{VCAFCLS}[l][l][\sz]{\sf VCA+FCLS}
\psfrag{VCASUNSALTV}[l][l][\sz]{\sf VCA+SUnSAL-TV}
\psfrag{FSNMFFCLS}[l][l][\sz]{\sf FSNMF+FCLS}
\psfrag{FSNMFSUNSALTV}[l][l][\sz]{\sf FSNMF+SUnSAL-TV}
\psfrag{MVSAFCLS}[l][l][\sz]{\sf MVSA+FCLS}
\psfrag{MVSASUNSALTV}[l][l][\sz]{\sf MVSA+SUnSAL-TV}
\psfrag{SDSNMF}[l][l][\sz]{\sf SDSNMF}
\psfrag{SCU}[l][l][\sz]{\sf SCU}
}

We attribute HUT-AMP's excellent $\NMSE$ to several factors.  
First, it has the ability to \emph{jointly} estimate endmembers and abundances, to exploit spectral coherence in the endmembers, and to exploit both spatial coherence and sparsity in the abundances (of which there is plenty in this experiment).
Furthermore, due to the presence of pure-pixels throughout the scene, the ``active'' distribution $\zeta_n(\cdot)$ in \eqref{nu} is simply a Bernoulli distribution, which HUT-AMP is able to learn (via EM) and exploit (via BiG-AMP) for improved performance.

\Figref{time_pureSNR} shows that the runtime of HUT-AMP was approximately $3$ times slower than the EE-and-inversion techniques, 
approximately $4$ times faster than the spatial-coherence exploiting SDSNMF algorithm,
and approximately $200$ times faster than that of the Bayesian SCU algorithm.
We conjecture that the relatively slow runtime of SCU is due to its use of Gibbs sampling.

Although not shown in the above figures, we also ran HUT-AMP-MOS on this dataset at $\SNR \!=\! 30$~dB.
The result was that HUT-AMP-MOS correctly estimated the number of materials (i.e., $N\!=\!5$) on every realization and thus gave identical $\NMSE_{\vec{S}}$ and $\NMSE_{\vec{A}}$ as HUT-AMP.
The total runtime of HUT-AMP-MOS at this SNR was $94.27$~seconds, which was about $9$ times slower than HUT-AMP but still $75$ times faster than SCU.

\subsection{SHARE 2012 Avon Dataset} \label{sec:SHARE}

Next, we evaluated algorithm performance on the SHARE~2012 Avon dataset\footnote{The SHARE~2012 Avon dataset can be obtained from \url{http://www.rit.edu/cos/share2012/}.}\cite{Giannandrea:SPIE:13}, which uses $M\!=\!360$ spectral bands, corresponding to wavelengths between $400$ and $2450$~nm, over a large rural area. 
To do this, we first cropped down the full image to the scene shown in \figref{RIT1_rgb}, which is known to consist of $N\!=\!4$ materials: grass, dry sand, black felt, and white TyVek \cite{Canham:SPIE:13}.
This scene was explicitly constructed for use in hyperspectral unmixing experiments, as efforts were made to ensure that the vast majority of the pixels were pure. 
Also, the data was collected on a nearly cloudless day, implying that shadowing effects were minimal.
To construct ground-truth endmembers,\footnote{In practical HU data, ground truth is difficult to obtain, since lab-measured reflectivity can differ dramatically from received radiance at the sensor.  In this experiment, we circumvent these problems by exploiting the known purity of the pixels and by minimizing noise effects through averaging.} 
we averaged a $4\times 4$ pixel grid of the received spectra in a ``pure'' region for each material.  
We then computed \SAD\
between each ground-truth endmember $\vec{s}_n$ and the estimate $\hvec{s}_n$ produced by each algorithm. 

\begin{figure}[t]
\newcommand{\pwid}{3.4in}
\newcommand{\fwid}{3.4 in}
  \begin{minipage}{\pwid}
    (a) HUT-AMP (average runtime $= 14.68$ sec):\\
    \includegraphics[width=\fwid]{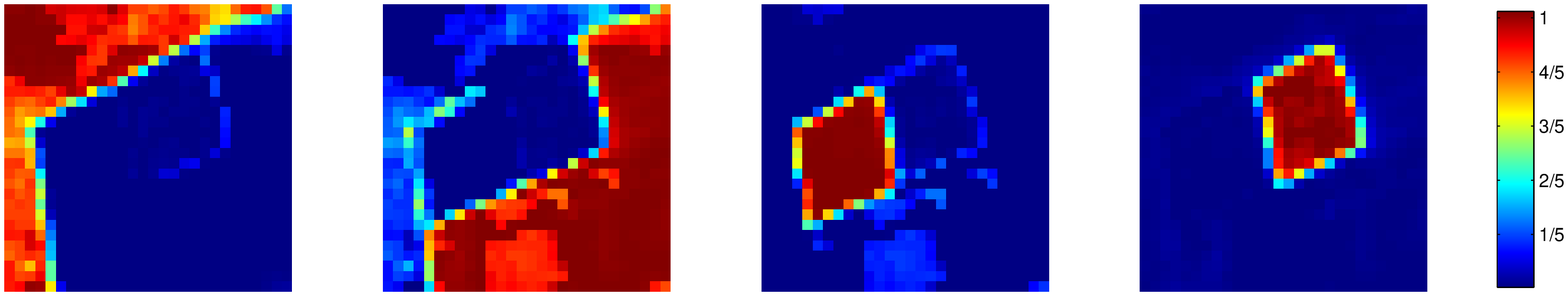}
  \end{minipage} \\[2mm]
  \begin{minipage}{\pwid}
    (b) VCA+FCLS (average runtime $= 2.50$ sec):\\
    \includegraphics[width=\fwid]{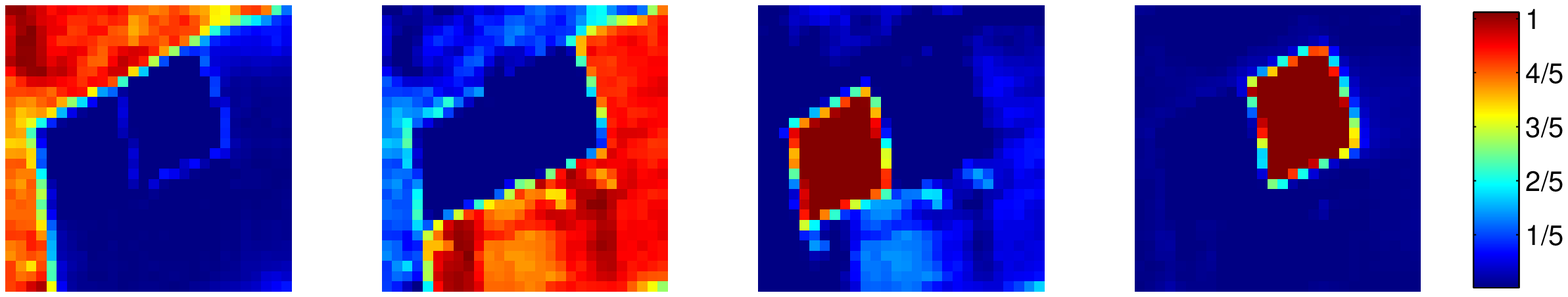}
  \end{minipage} \\[2mm]
  \begin{minipage}{\pwid}
    (c) VCA+SUnSAL-TV (average runtime $= 4.13$ sec):\\
    \includegraphics[width=\fwid]{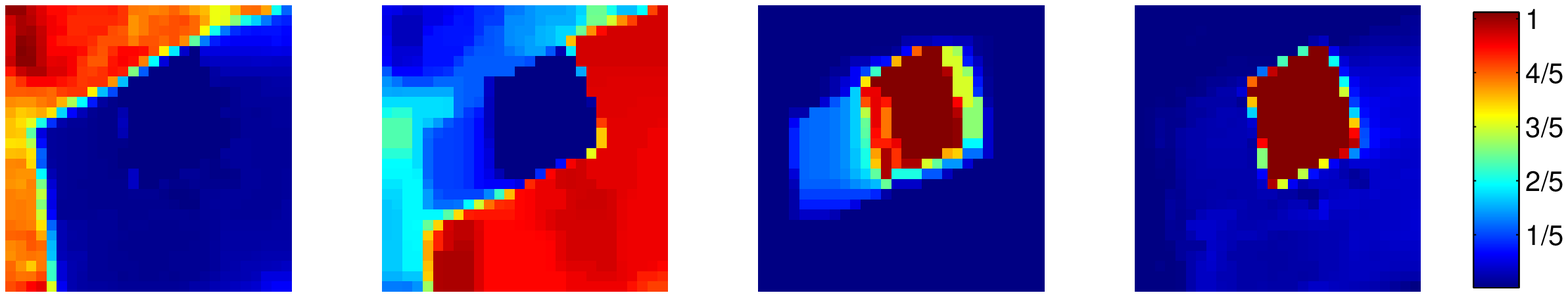}
  \end{minipage} \\[2mm]
    \begin{minipage}{\pwid}
    (d) FSNMF+FCLS (average runtime $= 1.67$ sec):\\
    \includegraphics[width=\fwid]{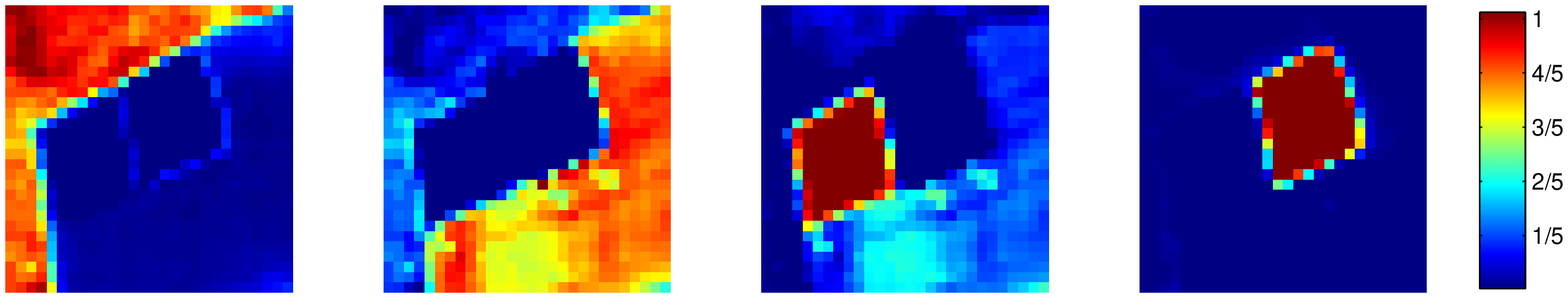}
  \end{minipage} \\[2mm]
  \begin{minipage}{\pwid}
    (e) FSNMF+SUnSAL-TV (average runtime $= 3.36$ sec):\\
    \includegraphics[width=\fwid]{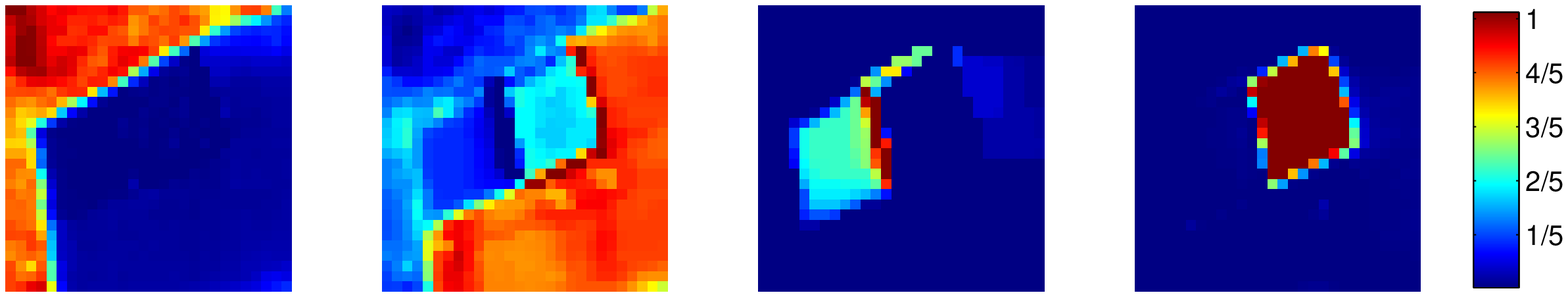}
  \end{minipage} \\[2mm]
  \begin{minipage}{\pwid}
    (f) MVSA+FCLS (average runtime $= 1.82$ sec):\\
    \includegraphics[width=\fwid]{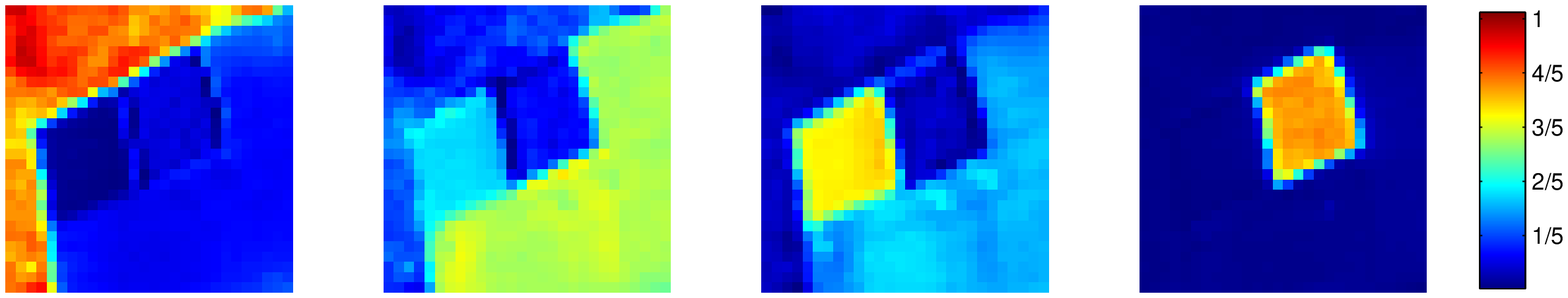}
  \end{minipage} \\[2mm]
    \begin{minipage}{\pwid}
    (g) MVSA+SUnSAL-TV (average runtime $= 4.08$ sec):\\
    \includegraphics[width=\fwid]{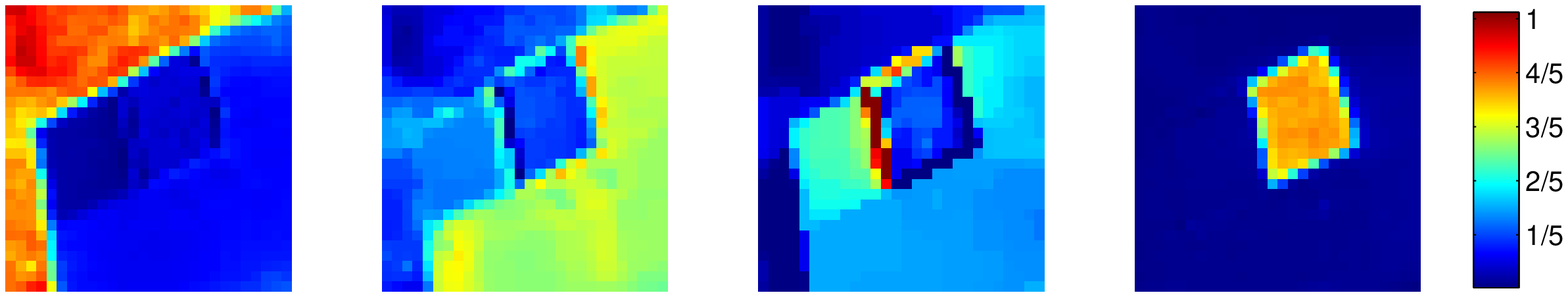}
  \end{minipage} \\[2mm]
    \begin{minipage}{\pwid}
    (h) SCU (average runtime $= 2438$ sec):\\
    \includegraphics[width=\fwid]{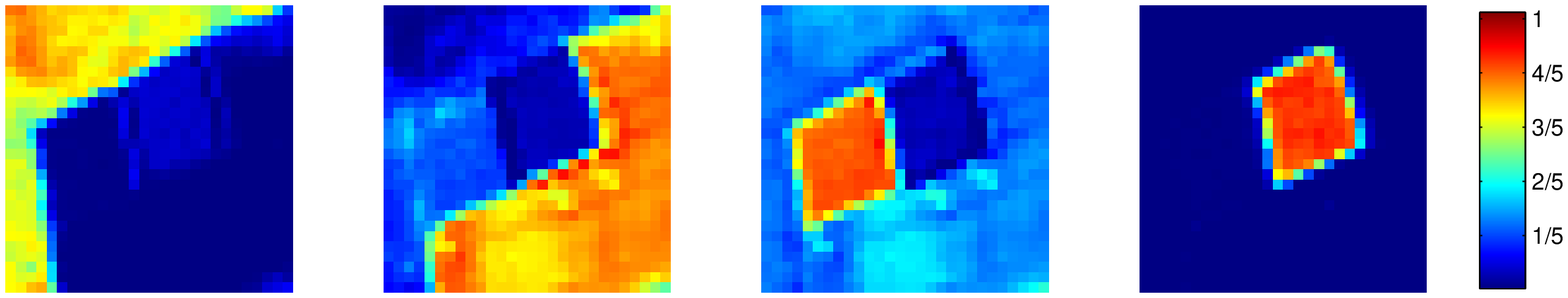}
  \end{minipage} \\ [2mm]
    \begin{minipage}{\pwid}
    (i) SDSNMF (average runtime $= 642.8$ sec):\\
    \includegraphics[width=\fwid]{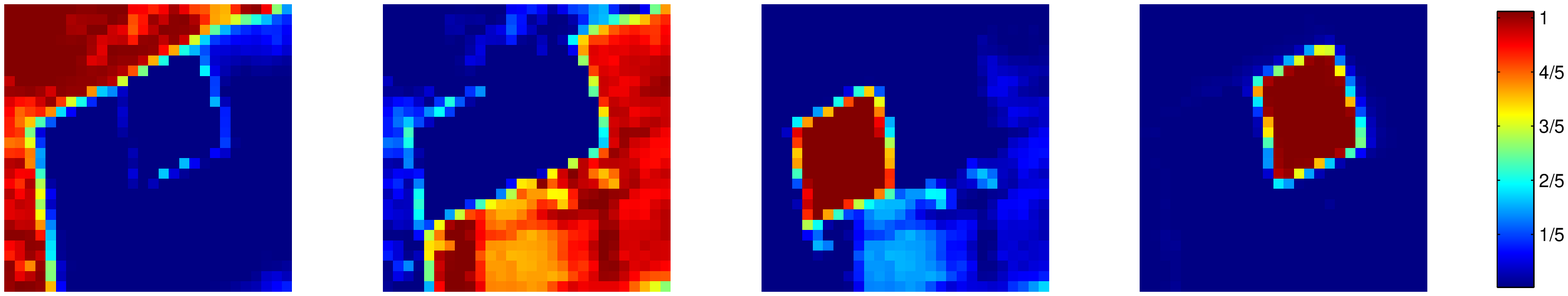}
  \end{minipage} \\
  \caption{Average runtimes and examples of recovered abundance maps for the SHARE 2012 experiment. From left to right, the materials are: grass, dry sand, black felt, and white TyVek.}
  \label{fig:abun}
\vspace{-4 mm}
\end{figure}

\tabref{SAD} shows median \SAD\ over $50$ realizations, using the original dataset and one with white Gaussian noise added to achieve $\SNR = 25$~dB. 
In the noiseless case, the table shows that HUT-AMP recovered the grass and white TyVek materials with the highest accuracy and recovered the dry sand and black felt materials with the second highest accuracy.
Meanwhile, in the noisy case, HUT-AMP recovered the TyVek material with the highest accuracy (in a tie with SDSNMF) and recovered the dry sand and black felt materials with the second highest accuracy.
Looking at the material-averaged SAD scores in the table, it is evident that the accuracies achieved by HUT-AMP are close to those attained by the most accurate algorithm, SDSNMF, and significantly better than those attained by any of the competing algorithms, in both the noiseless and noisy cases. 
Although SDSNMF offers slightly more accurate endmember recoveries, \figref{abun} shows that its runtime is $44$ times slower than that of HUT-AMP.
Therefore we conclude that HUT-AMP offers an excellent combination of endmember recovery accuracy and runtime.

For visual comparison, \figref{SHARE_endmembers} shows an example of the extracted and ground-truth endmembers in the noiseless case.
The figure shows HUT-AMP's estimates closely matching the ground-truth for all materials; by contrast, 
MVSA is mismatched in the case of grass,
MVSA and FSNMF are mismatched in the case of dry sand,
MVSA, VCA, and FSNMF, are mismatched in the case of black felt, 
and
MVSA, VCA, SCU, FSNMF, and MVSA are all mismatched in the case of white TyVek.
\Figref{SHARE_endmembers} reveals that MVSA does not always yield non-negative endmembers estimates, which may account for its relatively poor performance in all but our first experiment from \secref{mixed}.

\putFrag{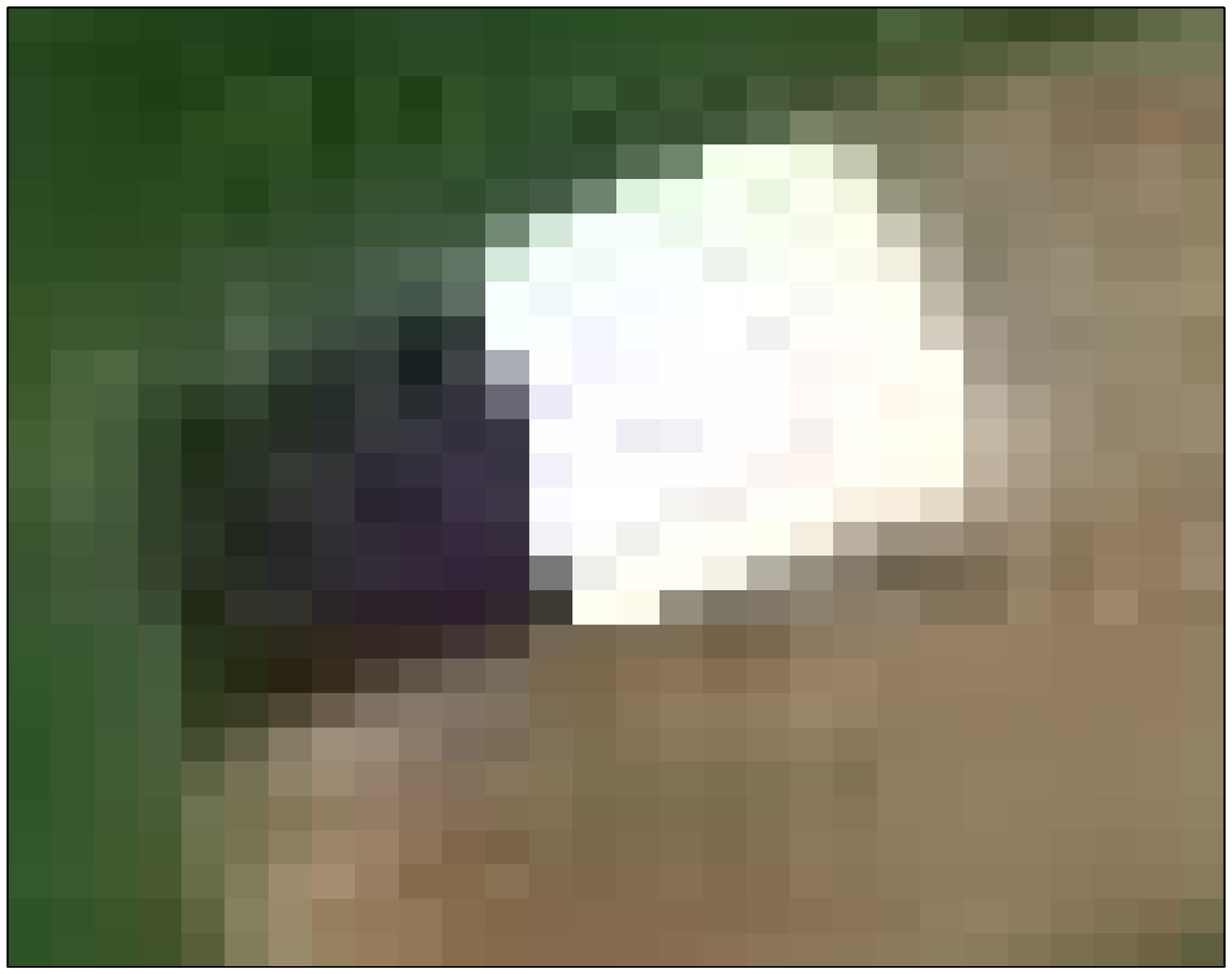}
{False-color image of the cropped scene of the SHARE~2012 dataset\cite{Giannandrea:SPIE:13}.}
{2.3}
{}

\putTable{SAD}{Median spectral angle distance (in degrees) between recovered and ground-truth endmembers in the SHARE~2012 experiment.}
{
\bgroup
\def\arraystretch{1.2}
\begin{tabular}{|c|l|c|@{\,}c@{\,}|@{\,}c@{\,}|@{\,}c@{\,}|c|}
\cline{3-7}
\multicolumn{2}{c|}{} &grass & dry sand &black felt & white TyVek &avg. \\ \hline
\multirow{6}{*}{\begin{sideways}{noiseless~~}\end{sideways}} 
&HUT-AMP &\textbf{1.54} &1.13 &3.53 &\textbf{0.39} &1.65 \\ \cline{2-7}
&VCA &1.58 &2.20 &11.01 &2.09 &4.22 \\ \cline{2-7}
&FSNMF &1.65 &1.68 &7.36 &1.46 &3.03 \\ \cline{2-7}
&MVSA &4.57 &10.42 &45.47 &1.60 &15.52 \\ \cline{2-7}
&SCU &2.69 &2.48 &32.10 &1.19 &9.61 \\ \cline{2-7}
&SDSNMF &1.86 &\textbf{0.71} &\textbf{2.85} &0.40 & \textbf{1.45} \\ \hline \hline
\multirow{6}{*}{\begin{sideways}{\sf $\SNR = 25$ dB}\end{sideways}} 
&HUT-AMP &1.60 &1.03 &3.68 &\textbf{0.44} &1.69 \\ \cline{2-7}
&VCA &\textbf{1.53} &1.77 &11.72 &2.17 &4.30 \\ \cline{2-7}
&FSNMF &1.58 &14.47 &4.39 &1.65 &5.52 \\ \cline{2-7}
&MVSA &4.52 &11.16 &48.05 &1.58 &16.33 \\ \cline{2-7}
&SCU &1.82 &2.03 &8.24 &1.73 &3.47 \\ \cline{2-7}
&SDSNMF &1.99 &\textbf{0.81} &\textbf{3.37} &\textbf{0.44} &\textbf{1.65} \\ \hline
\end{tabular}
\egroup
}

\putFragSpace{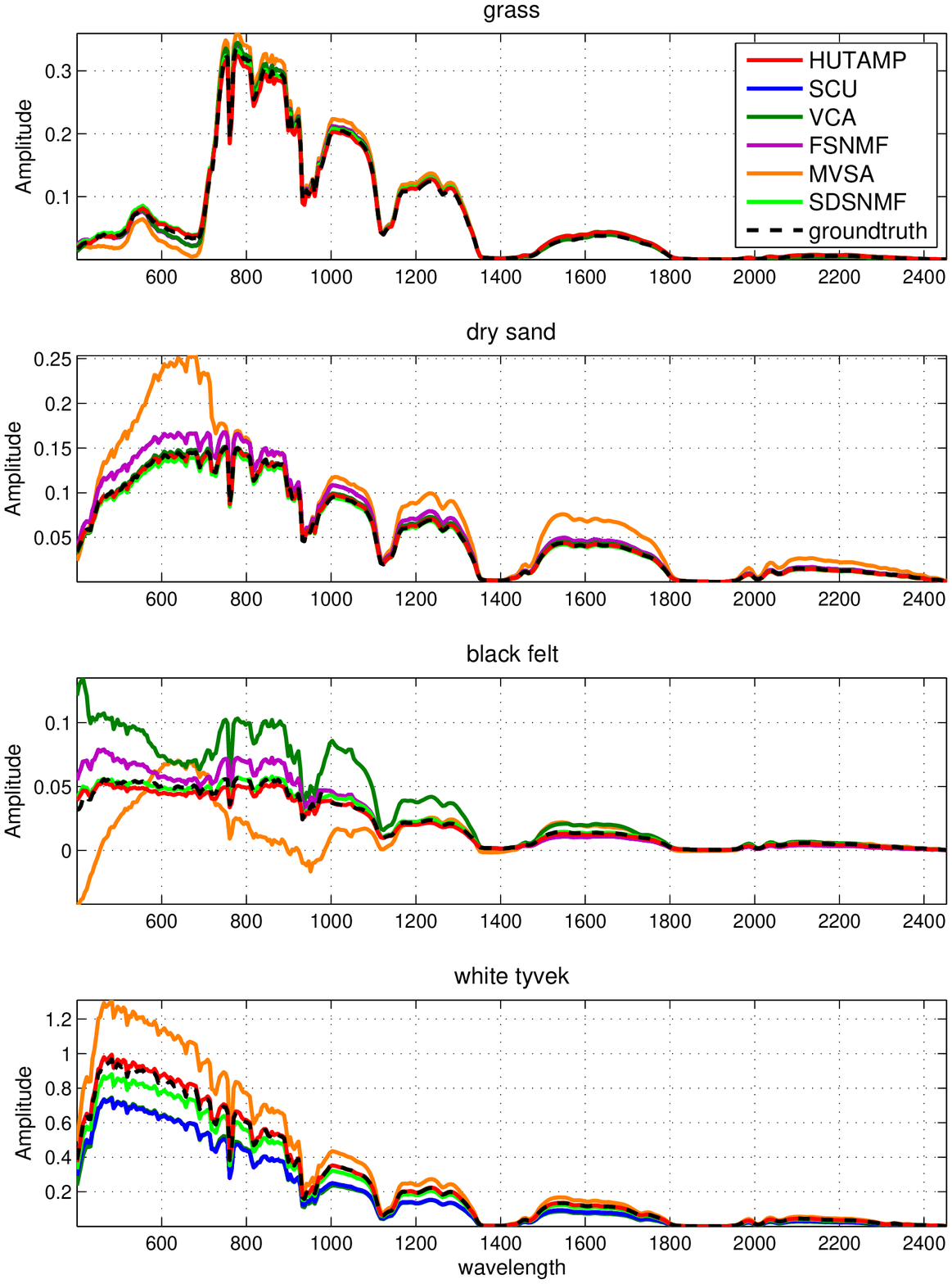}{Examples of recovered and ground-truth endmembers for the SHARE~2012 experiment.}
{\figsize}
{ \newcommand{\sz}{0.72}
    \psfrag{wavelength}[t][T][\sz]{\sf Wavelength [nm]}
    \psfrag{Amplitude}[B][B][\sz]{\sf Amplitude}
    \psfrag{grass}[B][B][\sz]{\sf grass}
    \psfrag{dry sand}[B][B][\sz]{\sf dry sand}
    \psfrag{black felt}[B][B][\sz]{\sf black felt}
    \psfrag{white tyvek}[B][B][\sz]{\sf white TyVek}
    \psfrag{HUTAMP}[l][l][\sz]{\sf HUT-AMP}
    \psfrag{SCU}[l][l][\sz]{\sf SCU}
    \psfrag{VCA}[l][l][\sz]{\sf VCA}
    \psfrag{FSNMF}[l][l][\sz]{\sf FSNMF}
    \psfrag{MVSA}[l][l][\sz]{\sf MVSA}
    \psfrag{SDSNMF}[l][l][\sz]{\sf SDSNMF}
    \psfrag{groundtruth}[l][l][0.62]{\sf ground-truth}
}

As another visual comparison, \figref{abun} shows an example of the recovered abundance maps in the noiseless case.
We reason that the best recoveries are the ones that are the most pure within the green, tan, black, and white regions of \figref{RIT1_rgb}, given that great care was taken during data collection to keep each region occupied by a single material.
\Figref{abun} shows that, in the case of dry sand and black felt, the abundances recovered by HUT-AMP were the most pure and, in the case of grass and Tyvek, the abundances recovered by HUT-AMP were among the most pure.
The other Bayesian approach, SCU, yielded abundance estimates with much less purity, and we conjecture that was due to its priors being less well-matched to this highly sparse scene.
Meanwhile, SUnSAL-TV (using both EE techniques) failed to recover the black felt material, which we attribute to its lack of a sum-to-one constraint.

Average runtimes are also reported next to each algorithm in \figref{abun}.
There we see that HUT-AMP's runtime was $4$-$9$ times slower than the EE-and-inversion techniques but $44$ times faster than SDSNMF and $166$ times faster than SCU, the other Bayesian technique.

We also ran HUT-AMP-MOS on the SHARE~2012 dataset and found that it correctly estimated the presence of $N\!=\!4$ materials, thus yielding identical recovery performance to HUT-AMP.   
HUT-AMP-MOS's runtime was $36.54$ seconds, which was $2.5$ times slower than HUT-AMP but still $67$ times faster than SCU.

\subsection{AVIRIS Cuprite Dataset} \label{sec:Cuprite}

Our final numerical experiment was performed on the well known AVIRIS Cuprite dataset. 
Although the original dataset consisted of $M \!=\! 224$ spectral bands, ranging from $0.4$ to $2.5 \ \mu$m,
we aimed to replicate the setup in \cite{Mittelman:TSP:12}, which removed bands $1$-$10$, $108$-$113$, $153$-$168$, and $223$-$224$ to avoid water-absorption effects, resulting in $M \!=\! 189$ spectral measurements per pixel.
And, like \cite{Mittelman:TSP:12}, we considered only the $80\!\times\!80$ pixel scene identified by the black square in \figref{Cuprite_map} and we assumed $N\!=\!5$ materials. 
According to the tricorder classification map in \figref{Cuprite_map}, this scene contains the materials Montmorillonite, Alunite, well crystallized (wxl) Kaolinite, and partially crystallized (pxl) Kaolinite.
Although \cite{Mittelman:TSP:12} conjectured that this area also contains Sphene, none of the algorithms produced endmember estimates that were close to Sphene, and is Sphene is not listed in \figref{Cuprite_map}. 
Thus, we did not consider Sphene as a ground-truth material.
Also, like in \cite{Mittelman:TSP:12}, we considered both noiseless and white-Gaussian-noise corrupted measurements (at $\SNR\!=\!30$~dB).

\putFrag{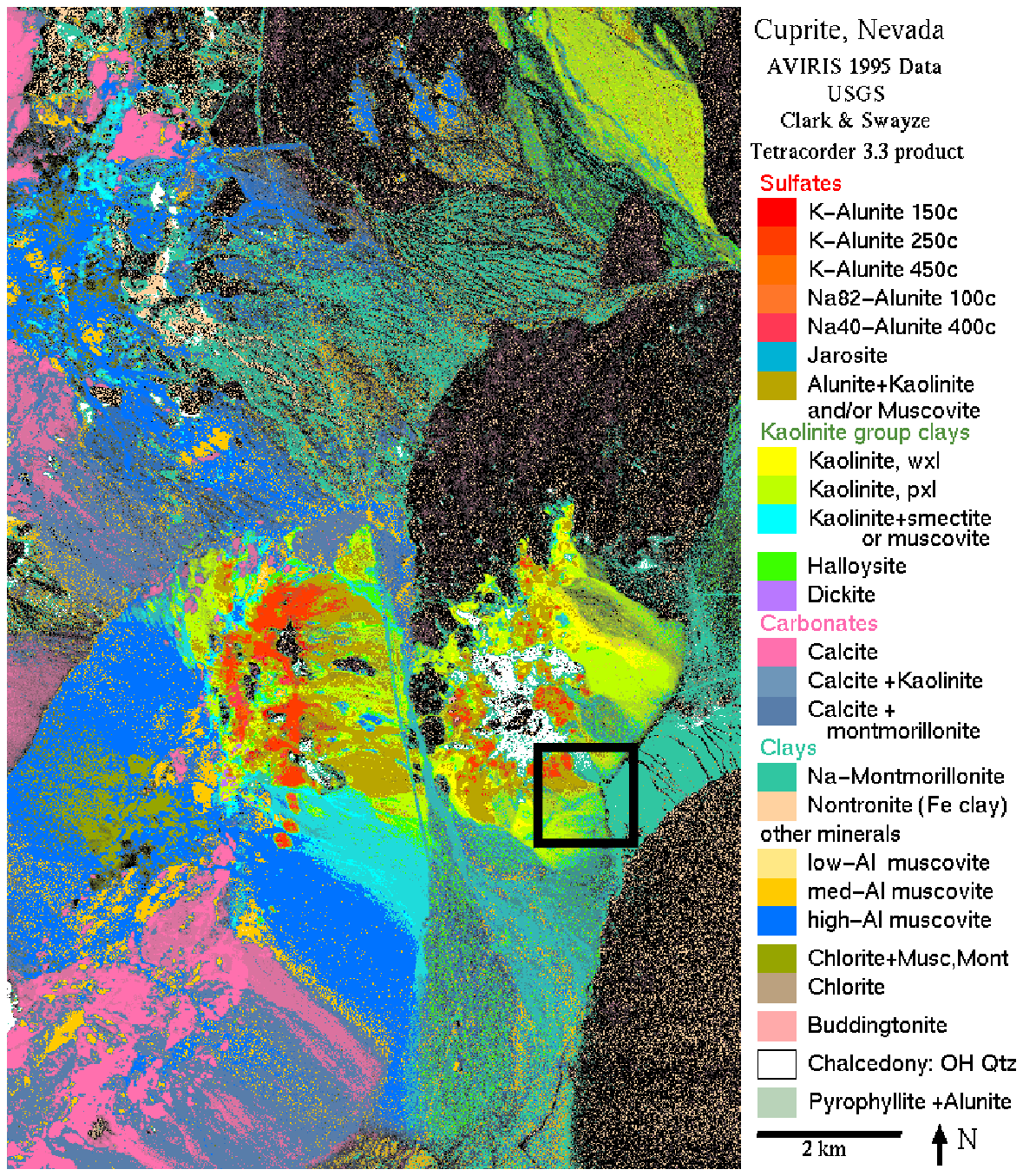}
{Mineral classification mapping of the Cuprite Dataset using the Tricorder 3.3 product\cite{Clark:JGEO:03}.  We used the scene cropped by the black rectangle.}
{\figsize}
{}

\tabref{cuprite} shows the median \SAD\ achieved during endmember extraction over $50$ realizations.
From the table, we see that, in the noiseless case, HUT-AMP achieved the best material-averaged \SAD\, as well as the best \SAD\ for two specific materials.
In the noisy case, HUT-AMP achieved the second-best material-averaged \SAD\, as well as the best \SAD\ for one material.
Meanwhile, the \SAD s produced by VCA, FSNMF, and SDSNMF were of a similar magnitude, while those produced by SCU and MVSA were noticeably larger.
These \SAD\ values should be interpreted with caution, however, since i) the ground-truth endmembers are laboratory-measured reflectance spectra from the 2006 USGS library as ground-truth, whereas the Cuprite dataset itself uses reflectance units obtained via atmospheric correction of radiance data,\footnote{The reflectance and radiance versions of the Cuprite dataset can be found at \url{http://aviris.jpl.nasa.gov/html/aviris.freedata.html}} and ii) it is not clear exactly which materials are truly present in the scene. 
The fact that the \SAD s reported here are so much larger than those reported in our SHARE experiment suggests that the Cuprite ground-truth may not be fully accurate.

\putTable{cuprite}{Median spectral angle distance (in degrees) for the Cuprite experiment.}
{%
\bgroup
\def\arraystretch{1.2}
\begin{tabular}[t]{|c|c|@{\,}c@{\,}|@{\,}c@{\,}|@{\,}c@{\,}|@{\,}c@{\,}|@{\,}c@{\,}|@{\,}c@{\,}|}
\hline
\multirow{2}{*}{} & \multirow{2}{*}{} & 
  \multicolumn{6}{c|}{\SAD\ [degrees]}\\
  \cline{3-8}
\multirow{2}{*}{} & \multirow{2}{*}{} & 
  HUT-AMP &SCU &VCA &FSNMF &MVSA &SDSNMF\\
\hline
\hline
\multirow{6}{*}{\begin{sideways}{\hspace{2mm} noiseless}\end{sideways}} 
&Montmor. 
&\textbf{3.42} &3.95 &3.91 &3.54 &6.03 &3.47\\ \cline{2-8}
&wxl Kaolinite
&\textbf{10.23} &13.14 &10.45 &10.86 &15.42 &11.46\\ \cline{2-8}
&pxl Kaolinite 
&9.10 &11.45 &9.22 &9.38 &10.51 &\textbf{9.09}\\ \cline{2-8}
&Alunite 
&{7.27} &6.62 &6.55 &\textbf{6.40} &7.11 &7.87\\ \cline{2-8}
&Average
&\textbf{7.50} &9.12 &7.53 &7.55 &9.77 &7.97
\\ \hline \hline
\multirow{6}{*}{\begin{sideways}{\hspace{2mm} $\SNR = 30$ dB}\end{sideways}} 
&Montmor. 
&3.53 &3.80 &3.79 &3.57 &5.64 &\textbf{3.48}\\ \cline{2-8}
&wxl Kaolinite 
&10.72 &12.46 &\textbf{10.62} &12.93 &15.59 &11.20\\ \cline{2-8}
&pxl Kaolinite
&\textbf{9.10} &11.47 &9.32 &10.49 &11.55 &9.43\\ \cline{2-8}
&Alunite 
&7.45 &7.94 &6.60 &\textbf{6.39} &7.16 &6.67\\ \cline{2-8}
& Average
&7.70 &8.92 &\textbf{7.59} &8.34 &9.98 &7.70
\\ \hline 
\end{tabular}
\egroup
}

\begin{figure*}[t!]
\newcommand{\pwid}{6.6 in}
\newcommand{\fwid}{6.6 in}
 {\scriptsize \hspace{10 mm} Montmorillonite \hspace{16 mm} Kaolin, wxl \hspace{18 mm} Kaolin, pxl \hspace{19.5 mm} Unknown \hspace{21 mm} Alunite} \\[2 mm]
 \begin{minipage}[t][0.9 in]{\pwid}
   \hspace{1 mm}\rotatebox{90}{\hspace{6.5 mm} \scriptsize HUT-AMP}
   \rotatebox{90}{\hspace{1 mm} \scriptsize avg runtime: 67.52}
   \includegraphics[trim=0 0.2in 0 0.3in, clip, width=\fwid]{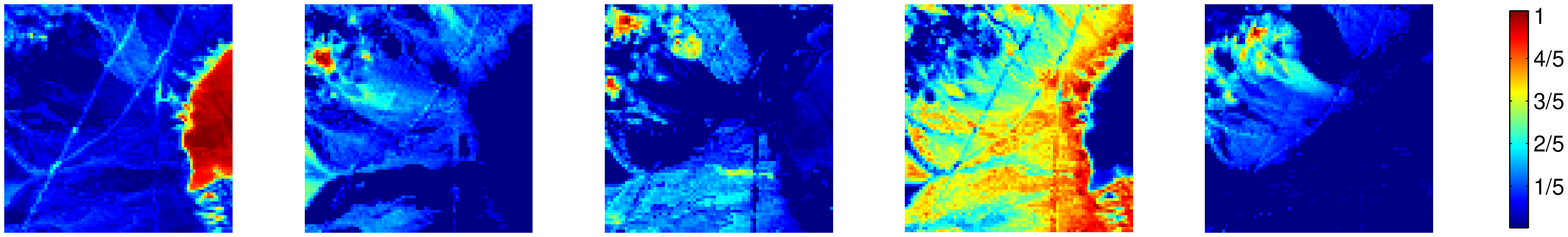}
 \end{minipage}\\
  \begin{minipage}[t][0.9 in]{\pwid}
   \hspace{1 mm}\rotatebox{90}{\hspace{6.5 mm} \scriptsize SDSNMF}
   \rotatebox{90}{\hspace{1 mm} \scriptsize avg runtime: 14\,382}
   \includegraphics[trim=0 0.2in 0 0.3in, clip, width=\fwid]{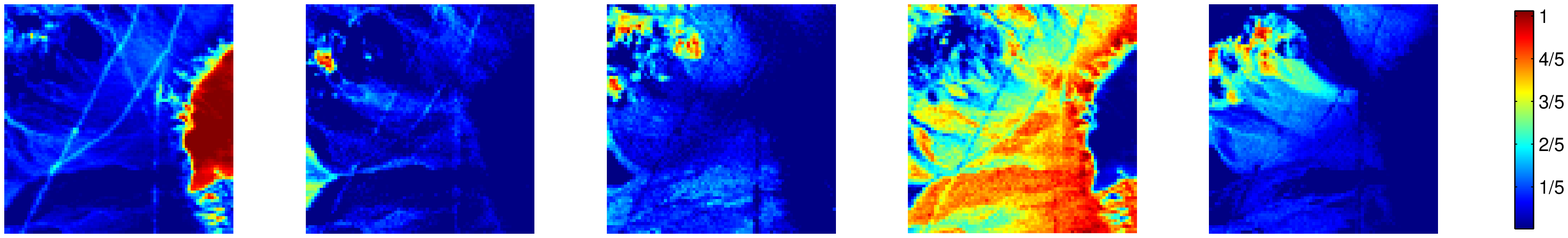}
 \end{minipage}\\
 \begin{minipage}[t][0.9 in]{\pwid}
  \hspace{1 mm}\rotatebox{90}{\hspace{4 mm} \scriptsize FSNMF+FCLS}
  \rotatebox{90}{\hspace{1.5 mm} \scriptsize avg runtime: 11.91}
   \includegraphics[trim=0 0.2in 0 0.3in, clip, width=\fwid]{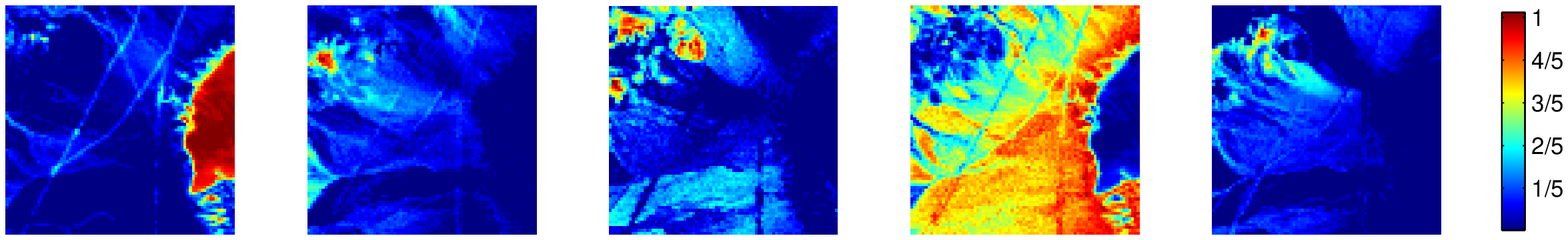}
 \end{minipage}\\
 \begin{minipage}[t][0.9 in]{\pwid}
 \hspace{1 mm}\rotatebox{90}{\hspace{6 mm} \scriptsize VCA+FCLS}
 \rotatebox{90}{\hspace{1.5 mm} \scriptsize avg runtime: 10.82}
   \includegraphics[trim=0 0.2in 0 0.3in, clip, width=\fwid]{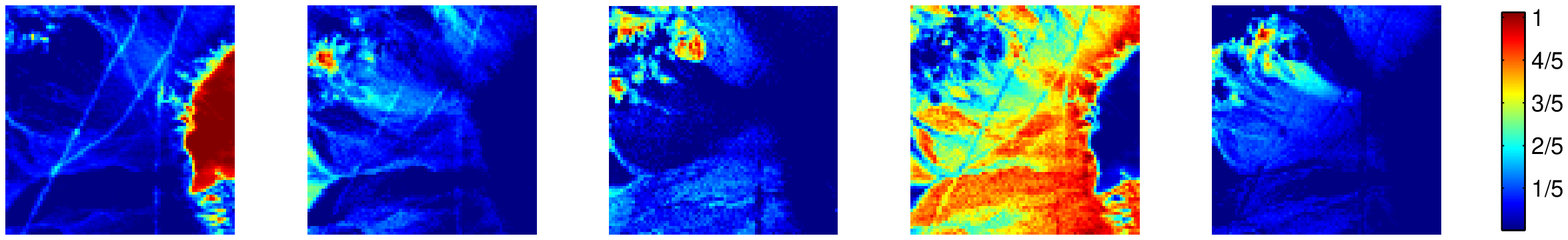}
 \end{minipage}\\
 \begin{minipage}[t][0.9 in]{\pwid}
 \hspace{1 mm}\rotatebox{90}{\hspace{1 mm} \scriptsize FSNMF+SUnSAL-TV}
 \rotatebox{90}{\hspace{1.5 mm} \scriptsize avg runtime: 11.43}
   \includegraphics[trim=0 0.2in 0 0.3in, clip, width=\fwid]{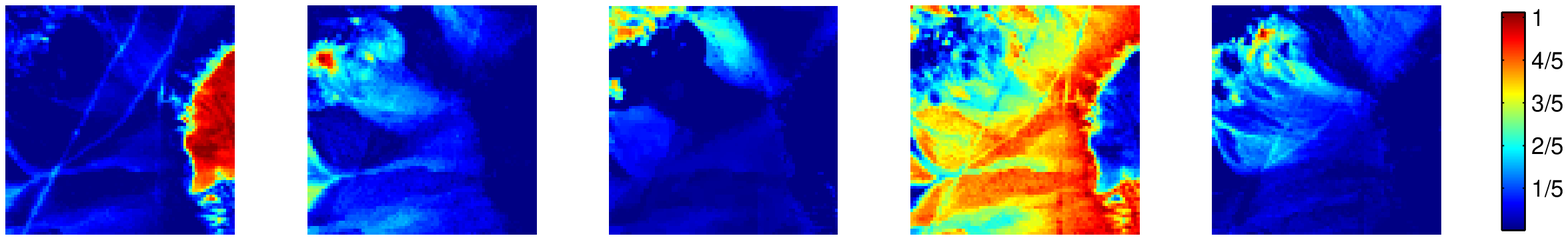}
 \end{minipage}\\
 \begin{minipage}[t][0.9 in]{\pwid}
 \hspace{1 mm}\rotatebox{90}{\hspace{1.5 mm} \scriptsize VCA+SUnSAL-TV}
 \rotatebox{90}{\hspace{1.5 mm} \scriptsize avg runtime: 15.22}
   \includegraphics[trim=0 0.2in 0 0.3in, clip, width=\fwid]{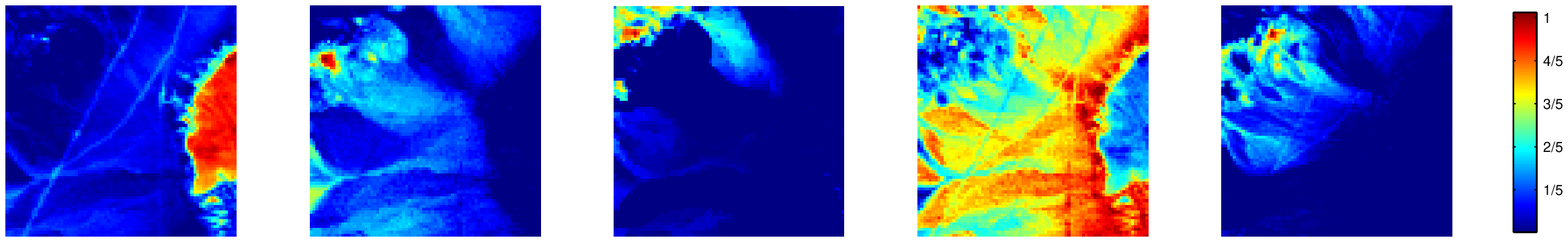}
 \end{minipage}\\
 \begin{minipage}[t][0.9 in]{\pwid}
  \hspace{1 mm}\rotatebox{90}{\hspace{11 mm} \scriptsize SCU}
  \rotatebox{90}{\hspace{2 mm} \scriptsize avg runtime: 5\,662}
   \includegraphics[trim=0 0.2in 0 0.3in, clip, width=\fwid]{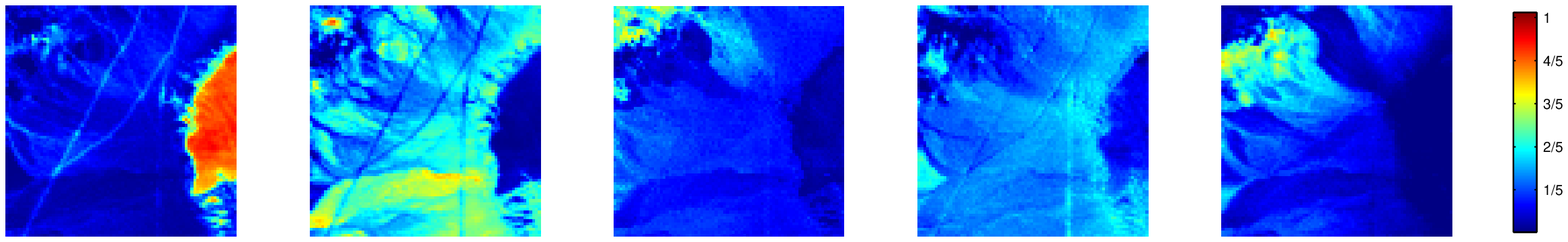}
 \end{minipage}
 \begin{minipage}[t][0.9 in]{\pwid}
  \hspace{1 mm}\rotatebox{90}{\hspace{4 mm} \scriptsize MVSA+FCLS}
  \rotatebox{90}{\hspace{2 mm} \scriptsize avg runtime: 8.71}
   \includegraphics[trim=0 0.2in 0 0.3in, clip, width=\fwid]{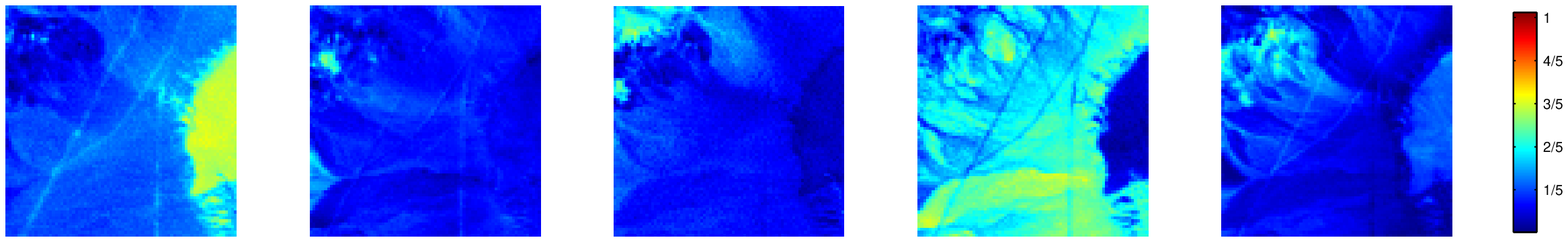}
 \end{minipage}\\
 \begin{minipage}[t][0.9 in]{\pwid}
 \hspace{1 mm}\rotatebox{90}{\hspace{1.0 mm} \scriptsize MVSA+SUnSAL-TV}
 \rotatebox{90}{\hspace{2 mm} \scriptsize avg runtime: 8.34}
   \includegraphics[trim=0 0.2in 0 0.3in, clip, width=\fwid]{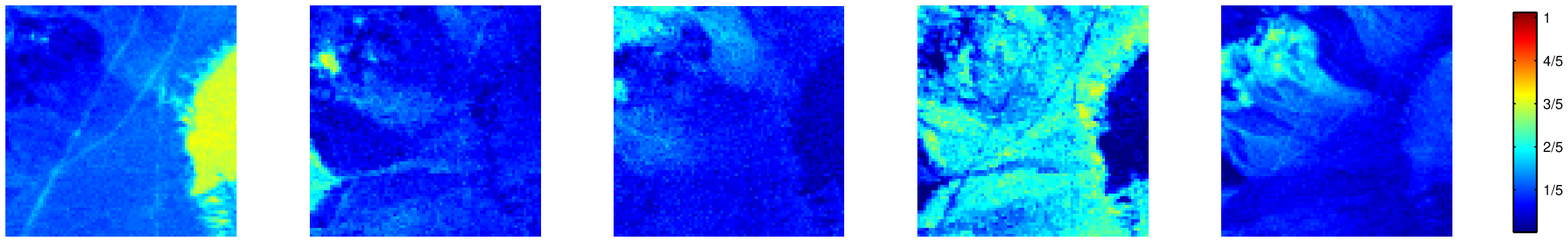}
 \end{minipage}\\
 \vspace{-6mm}
 \caption{Examples of recovered abundance maps in the noiseless Cuprite experiment.  Each row corresponds to an algorithm and each column corresponds to a material.  Average runtimes (in seconds) are also listed on the left.}
 \label{fig:cupabun}
\end{figure*}

For visual comparison, we plot examples of the abundance maps recovered in the noiseless experiment in \figref{cupabun}.
The figure shows that the abundance maps returned by HUT-AMP, SDSNMF, FSNMF+FCLS, VCA+FCLS, FSNMF+SUnSAL-TV, and VCA+SUnSAL-TV have the highest contrast, suggesting that if certain pixels are truly pure then these algorithms are accurately estimating those pixels.
The maps produced by SUnSAL-TV appear more ``blurred,'' probably as an artifact of TV regularization.
The abundances returned by SCU, MVSA+FCLS, and MVSA+SUnSAL-TV were of much lower contrast and suggest different material placements than the maps generated by the other algorithms.
For example, SCU suggests a significant wxl-Kaolin presence throughout the lower half of the scene, in contrast to other algorithms.
However, \tabref{cuprite} shows that SCU gave the worst \SAD\ for wxl-Kaolin.

\Figref{cupabun} also shows the total runtimes of the various algorithms.
There we see that HUT-AMP was $6$-$8$ times slower than the typical EE-and-inversion approach, but more than $80$ times faster than SCU and 
more than $200$ times faster than SDSNMF.

We also ran HUT-AMP-MOS on the Cuprite data and found that, in both the noiseless and noisy cases, it estimated the presence of $N\!=\!5$ materials, and thus returned identical estimates to HUT-AMP. 
Meanwhile, HUT-AMP-MOS gave an average runtime of $191.49$ seconds, which was $30$ times faster than SCU and $75$ times faster than SDSNMF.

\section{Conclusions}\label{sec:sigmodel}

In this paper, we proposed a novel empirical-Bayesian hyperspectral-unmixing algorithm that jointly estimates endmembers and abundance maps while exploiting the practical features of spectral and spatial coherence, as well as abundance sparsity.
Inference is performed using the ``turbo'' approach proposed in \cite{Schniter:CISS:10}, which breaks up the factor graph into three subgraphs, performs (loopy) BP individually on each subgraph, and then exchanges beliefs between subgraphs.
For the spectral and spatial coherence subgraphs, standard Gauss-Markov and discrete-Markov methods \cite{Bouman:ICIP:95,Li:Book:09}, respectively, are used, while for the non-negative bilinear-mixing subgraph, the recently proposed BiG-AMP method from \cite{Parker:TSP:14a} is used, which exploits the approximate message passing framework from \cite{Donoho:PNAS:09,Donoho:ITW:10a}.
Furthermore, the statistical parameters of all distributions are learned using expectation-maximization \cite{Vila:TSP:14}, and the number of materials in the scene is estimated using penalized log-likelihood maximization.
On the whole, the proposed HUT-AMP-MOS algorithm performs approximate MMSE inference that exploits spectral and spatial coherence, in addition to simplex constraints, while avoiding the need for the specification of any tuning parameters.

Through a detailed numerical study, we demonstrated that our proposed HUT-AMP algorithm yields accurate recoveries of both endmembers and abundances on both synthetic and real-world datasets.
In particular, we found that HUT-AMP gives recoveries that are close to---if not more accurate than---state-of-the-art unmixing algorithms like SDSNMF.
Meanwhile, the runtime required for HUT-AMP is much less than sophisticated spatial-coherence exploiting approaches like SDSNMF and SCU---often by several orders of magnitude---while within an order of magnitude of the fastest EE-and-inversion approach.  
Our experiments also demonstrated that our model-order selection technique was able to correctly estimate the number of materials in several synthetic and real-world datasets, without requiring a very large increase in runtime.

\appendices

\section{Mean removal} \label{app:mean}
We can see that $\uvec{S}$ from \eqref{LMMmr} is approximately zero-mean via 
\begin{align}
0 
&= \frac{1}{MT}\sum_{m=1}^M\sum_{t=1}^T \underline{y}_{mt} 
	\label{eq:zero1}\\
&= \underbrace{ 
	\frac{1}{MT}\sum_{m=1}^M\sum_{t=1}^T \sum_{n=1}^N\underline{s}_{mn}a_{nt} 
	}_{\displaystyle O(1)}
	+ \underbrace{ \frac{1}{MT}\sum_{m=1}^M\sum_{t=1}^T w_{mt} }_{\displaystyle O(1/N)}
	\label{eq:zero2}\\
&\approx 
\sum_{n=1}^N 
\underbrace{ \frac{1}{T}\sum_{t=1}^Ta_{nt} }_{\displaystyle  \defn \mu_n^a }
\frac{1}{M} \sum_{m=1}^M \underline{s}_{mn}
	\label{eq:zero3} ,
\end{align}
where \eqref{zero1} follows from the definitions \eqref{mu}-\eqref{LMMmr0}.
The underbraces in \eqref{zero2} show the scaling on each term in the large-system limit (i.e., as $N\rightarrow\infty$).
These particular scalings follow from our assumption that the noise is both zero-mean and white and the convention\cite{Parker:TSP:14a} that both $y_{mt}$ and the noise variance $\vec{\psi}$ scale as $O(1)$. 
Recalling that $\sum_{n=1}^N\mu_n^a=1$ due to the simplex constraint, expression \eqref{zero3} shows that a weighted average of elements in $\uvec{S}$ is approximately zero, where the approximation becomes exact in the large-system limit.

\bibliographystyle{IEEEtran}
\bibliography{macros_abbrev,books,misc,sparse,machine,hsi,comm,multicarrier}

\end{document}